\documentclass[12pt,british,pointlessnumbers,bibtotoc]{scrbook}
\usepackage[]{fontenc}
\usepackage[latin1]{inputenc}
\usepackage{geometry}
\usepackage{fancyhdr}
\usepackage{array}
\usepackage{subfigure}
\usepackage{float}
\usepackage{amsmath}
\usepackage{graphicx}
\usepackage{amssymb}

\makeatletter

\providecommand{\tabularnewline}{\\}

\usepackage{ae,aecompl}

\usepackage{hyperref}

\usepackage{babel}
\makeatother
\begin{document}
\renewcommand{\textrm}{\text}
%fancy aktivieren
\fancypagestyle{plain}{
\fancyhf{}
\renewcommand{\headrulewidth}{0.0pt}
\renewcommand{\footrulewidth}{0.0pt}
}
\fancyhead{}
\fancyfoot{}
\fancyhead[LE,RO]{\thepage}
\fancyhead[RE]{\sf{\nouppercase{\leftmark}}}
\fancyhead[LO]{\sf{\nouppercase{\rightmark}}}
%\lhead{\sf{\nouppercase{\leftmark}}}
%\chead{} 
%\rhead{\sf{\today}}
%\lhead{}
%\lfoot{\today} \cfoot{} \rfoot{Seite \thepage/\pageref{LastPage}} %\renewcommand{\headrulewidth}{0.0pt}
\newcommand{\kl}[1]{{\textstyle #1}}

\pagestyle{empty}

\setcounter{page}{1}
%\begin{minipage}{7cm}
%\begin{center}
%{\bf Technische Universit\"{a}t M\"{u}nchen} \\
%Physik Department \\
%Institut f\"{u}r Theoretische Physik T30d \\
%Univ.-Prof. Dr. M. Lindner \\
%\end{center}
%\end{minipage}
%\vspace*{-3.5cm}

\noindent \begin{center}\textbf{Technische Universität München}\\
Physik Department\\
Institut für Theoretische Physik T30d\\
\par\end{center}

\vspace*{2cm}

\begin{center}\textbf{\Large The Cosmological Constant and Discrete
Space-Times}\par\end{center}{\Large \par}

\begin{center}\vspace{1cm}{\large Dipl.-Phys. Univ. Florian Bauer}\vspace{1cm}\par\end{center}

Vollständiger Abdruck der von der Fakultät für Physik der Technischen
Universität München zur Erlangung des akademischen Grades eines

\smallskip{}

\begin{center}\textsl{Doktors der Naturwissenschaften (Dr. rer. nat.)}\par\end{center}

\smallskip{}

\begin{flushleft}\vspace{-0.4cm}genehmigten Dissertation.\vspace{1cm}\par\end{flushleft}

\begin{center}\begin{tabular}{ll>{\raggedright}p{10cm}}
Vorsitzender:&
&
Univ.-Prof. Dr. Lothar Oberauer\tabularnewline
&
&
\tabularnewline
Prüfer der Dissertation:&
1.&
\parbox[t]{10cm}{Prof. Dr. Manfred Lindner,\\
Ruprecht-Karls-Universität Heidelberg}\tabularnewline
&
&
\tabularnewline
&
2.&
Hon.-Prof. Dr. Wolfgang Hillebrandt\tabularnewline
\end{tabular}\par\end{center}

\vspace{1cm}

Die Dissertation wurde am 31.07.2006 bei der Technischen Universität
München einge\-reicht und durch die Fakultät für Physik am 23.08.2006
angenommen.

\newpage

\pagestyle{fancy}

\vspace*{2cm}

\textsf{\textbf{\large Abstract}}{\large \par}

In this thesis the cosmological constant is investigated from two
points of view. First, we study the influence of a time-dependent
cosmological constant on the late-time expansion of the universe.
Thereby, we consider several combinations of scaling laws motivated
by renormalisation group running and different choices for the interpretation
of the renormalisation scale. Apart from well known solutions like
de~Sitter final states we also observe the appearance of future singularities.
As the second topic we explore vacuum energy in the context of discrete
extra dimensions, and we calculate the Casimir energy density as a
contribution to the cosmological constant. The results are applied
in a deconstruction scenario, where we propose a method to determine
the zero-point energy of quantum fields in four dimensions. In a related
way we find a lower bound on the size of a discrete gravitational
extra dimension, and finally we discuss the graviton and fermion mass
spectra in a scenario, where the extra dimensions form a discrete
curved disk. 

\vspace{2cm}

\textsf{\textbf{\large Zusammenfassung}}{\large \par}

In dieser Arbeit wird die kosmologische Konstante aus zwei verschiedenen
Blickwinkeln untersucht. Als erstes behandeln wir den Einfluß einer
zeitabhängigen kosmologischen Konstante auf die Entwicklung des Universums
zu späten Zeiten. Dabei betrachten wir mehrere Skalengesetze, die
vom Renormierungsgruppenlaufen herrühren, und außerdem verschiedene
Möglichkeiten, die Renormierungsskala festzulegen. Neben bekannten
Lösungen wie dem de~Sitter Kosmos beobachten wir auch das Auftreten
von Singularitäten in endlicher Zukunft. Der zweite Schwerpunkt dieser
Arbeit stellt Vakuumenergie in diskreten extra Dimensionen dar, wo
die Casimirenergiedichte als Beitrag zur kosmologischen Konstante
berechnet wird. Im Rahmen von Deconstruction verwenden wir die Ergebnisse,
um eine Möglichkeit zu finden, die Nullpunktsenergie von Quantenfeldern
in vier Dimensionen zu bestimmen. Ebenso leiten wir eine untere Schranke
für die Größe einer diskreten gravitativen extra Dimension her. Und
schließlich diskutieren wir die Massenspektren von Gravitonen und
Fermionen in einem Modell, bei dem die extra Dimensionen eine diskrete
gekrümmte Scheibe bilden.

%\clearpage{\pagestyle{empty}\cleardoublepage}

\renewcommand{\baselinestretch}{1.5}\tableofcontents{}\renewcommand{\baselinestretch}{1.0}

%\newpage

\chapter{Introduction\label{cha:Introduction}}

Several years after the discovery of the accelerated cosmological
expansion, the question what drives this behaviour is still open.
Since general relativity with well known matter sources like dust
and radiation always exhibits a decelerating universe, we have to
expect something new that explains the observed behaviour, which has
gained great support by recent observations~\cite{Riess:1998cb,Perlmutter:1998np,Tegmark:2003ud,Boughn:2004zm,Hannestad:2005fg,Spergel:2006hy}.
The name {}``dark energy'' has become common to describe all energy
forms that are able to yield the acceleration, although other origins
are often put into the same category. The amount of possible frameworks
and models that have been proposed to describe DE has grown quite
huge as can be seen in recent reviews about this topic~\cite{Peebles:2002gy,Sahni:2004ai,paddy,Perivolaropoulos:2006ce,Copeland:2006wr}.
Interestingly, the first dark energy candidate was already introduced
by Albert Einstein almost a century ago. He considered a positive
cosmological constant (CC) in order to realise a static cosmos, where
the CC compensates the dust matter. However, it turned out that this
space-time was unstable and additionally not consistent with the afterwards
observed cosmological expansion. These days, Einstein's CC has come
back as the simplest candidate for dark energy since it is just a
constant in the action for classical general relativity. Furthermore,
in the context of quantum field theory the vacuum or zero-point energy
of quantum fields contributes also to the CC. But this quantum origin
also involves the so-called cosmological constant problem~\cite{CCProblem},
which is generally considered to be one of the biggest mysteries in
physics. The core of this problem is the fact that in quantum field
theory the absolute value of the CC has not been determined yet, because
the corresponding calculation leads to an infinite value. Furthermore,
naive estimations using energy cutoffs are many orders of magnitude
above its measured value. Due to this extreme discrepancy and the
fact that dark energy has become dominant very lately in the cosmological
evolution, other explanations have been proposed. Apart from new energy
forms like, e.g., scalar field condensates (quintessence)~\cite{quintessence},
that contribute to the energy content of the universe, modifications
of the theory of gravity or even of the space-time structure have
become subject of intensive investigation. Instead of discussing all
these possibilities, we will in this thesis concentrate on two subjects
in the context of vacuum energy as a major source of dark energy.
First, we will study a CC that becomes time-dependent due to quantum
effects. In this framework we discuss the late time cosmological evolution
and the occurrence of future singularities. The second topic of investigation
are space-times with discrete compact extra dimensions, where the
appearance of zero-point energy and the properties of fields will
be discussed.

To introduce the first subject let us start on the classical level,
where the CC is just a constant term in the action for general relativity,
implying that it remains constant in Einstein's equations, too. Assume
for the moment that the current cosmological acceleration is just
due to a positive CC in a universe with only cold dark matter ($\Lambda$CDM
cosmology), then one finds that our space-time is approaching an empty
and static de~Sitter universe. However, other dark energy sources
might change the cosmological fate significantly in comparison to
the $\Lambda$CDM case. For instance, they could cause a final domination
of matter or even future singularities~\cite{Future-Sing}, where
space-time collapses to a big crunch or gets torn apart in a big rip
event~\cite{Caldwell:2003vq}. Especially the big rip singularity
would require some very unusual energy forms like scalar fields with
negative kinetic terms and other {}``exotic'' things. But as we
will show in this work, such extreme final states might also appear
with the CC once it is allowed to be time-dependent or more generally
scale-dependent. Constants that become scale-dependent occur quite
generally in quantum theories like quantum electrodynamics, where
the fine-structure constant becomes energy-dependent by renormalisation
group effects. The corresponding renormalisation group equations (RGE)
describe the {}``running'' of the constant as a function of the
renormalisation scale~$\mu$ and masses of quantum fields that are
involved in the theory. Similarly, one obtains the RGEs for the CC
and also for Newton's constant from the effective action of quantum
fields on curved space-times~\cite{QFTonCurve-1,RGE-LamG,RGE-LamG-2,QFTonCurve-2}.
For free fields the RGEs can be calculated exactly thereby leading
to a scale dependence of the CC as we will see later. Also certain
theories of quantum gravity~\cite{QEG,time-LamG-QG-1,time-LamG-QG-2}
can be a source for a scale-dependent CC or Newton's constant. Unlike
running coupling constants emerging from interacting quantum fields,
where the renormalisation scale can be easily interpreted as external
momentum or temperature, the identification of the scale for the CC
is not always given by the theory since there is no external momentum.
To study the RG running in a cosmological context we will therefore
explore as candidates for the renormalisation scale several scales
that are characteristic for the cosmological evolution. This will
be the Hubble scale and the sizes of the particle and event horizons.
In addition, the consequences of several RGEs following from different
frameworks and motivations will be discussed as well. For each combination
of RGEs and renormalisation scale identifications we will finally
solve Einstein's equations and determine the possible final states
of the universe. Apart from the pure curiosity of the researcher,
the results of this analysis might be valuable also for deciding which
of the above combinations are reasonable after one has accepted or
rejected the idea of future singularities.

The second major part of this thesis deals with fact that vacuum energy
is also sensitive to external conditions, which might come from a
non-trivial space-time structure. The Casimir effect~\cite{Casimir:dh}
is a famous example, where the quantum field zero-point energy depends
on the boundary conditions that are imposed onto the quantum fields.
In the original setup the boundary conditions follow from two parallel
conducting plates in four dimensions, where the Casimir effect implies
a force on the plates. While in this case the corresponding vacuum
energy usually cannot be identified directly with contributions to
the CC, a space-time with compact extra dimensions (ED) might lead
to an effective four dimensional (4D) CC that depends on the properties
of the EDs~\cite{Appelquist:1982zs}. Generally, one can say that
the resulting Casimir energy of massless fields scales inversely with
the size of the EDs. Here, one encounters a severe problem because
small EDs produce large contributions to the effective 4D CC in contrast
to its observed tiny value. Too large EDs, on the other hand, could
be easily observed by fifth force experiments and other methods~\cite{Adelberger:2003zx,Cullen:1999hc,Hannestad:2001xi}.
One has therefore to cope with stringent bounds on the size of the
EDs. Instead of working with large EDs one can obtain small CC contributions
also by considering massive quantum fields, which yield a suppression
of the Casimir energy. This property allows to keep both the ED and
the Casimir energy small. In this thesis we will investigate the Casimir
effect and the behaviour of quantum fields more deeply in the context
of discretised EDs. In contrast to continuous EDs the discrete structure
implies some interesting features like a finite number of Kaluza-Klein
modes or typical lattice theory attributes. More motivation comes
from the fact that a discretised space-time structure exhibits a minimal
physical length scale that could in principle regularise space-time
singularities in general relativity and respectively ultra-violet~(UV)
divergences in quantum theories. It therefore represents a useful
concept for quantum gravity~\cite{Discrete-QuGr}. By the way, as
EDs have not been observed yet, the property of being discretised
represents a reasonable possibility. One should just imagine some
kind of higher dimensional solid state physics. Furthermore, it was
recently found that discretised EDs can be described within a fully
4D context called {}``deconstruction''~\cite{Arkani-Hamed:2001ca,Hill:2000mu}.
This model-building approach avoids problems that emerge when extra-dimensional
field momenta reach the fundamental Planck scale, which can be considerably
below the 4D Planck scale. Coming back to the CC problem, the vacuum
energy of 4D quantum fields in a deconstruction setup is, as expected,
divergent, which just leads to the renormalisation group running effects
mentioned above. Thus the overall CC scale is undetermined and has
to fixed by observations. However, in this work we propose a well
defined prescription to assign a finite value to the CC by employing
the correspondence between deconstruction and discretised higher dimensions.
We will demonstrate this idea within a specific deconstruction model,
where also the suppression of the Casimir energy by massive fields
will be applied.

Going one step further by discretising gravity in the EDs one approaches
new effects in the form of strongly interacting massive 4D gravitons~\cite{Arkani-Hamed:2002sp,Arkani-Hamed:2003vb}.
This strong coupling behaviour leads to an upper limit for masses
and energy scales appearing in the theory. Interestingly, the limit
might depend also on the size of the ED. As a consequence, the ability
to suppress the Casimir energies by large field masses is reduced
since the masses have to lie below the strong coupling scale. In this
context we will derive some limits on the ED size by requiring that
the resulting CC does not exceed its observed value. Finally, we consider
on a more formal level a six dimensional (6D) model, which involves
a discretised curved disk. In this scenario we discuss the implementation
of fermions and gravitons, and furthermore investigate the effect
of the curvature and its influence on the mass spectra of the resulting
4D fields.

The structure of this thesis is given as follows: in Chap.~\ref{cha:RGE1-TimeDepCC}
we introduce a number of scaling laws for the CC together with some
renormalisation scale identifications. Afterwards we explain how to
solve Einstein's equations with a time-dependent CC and Newton's constant
and subsequently discuss cosmological late-time solutions and the
corresponding fates of the universe. Chap.~\ref{cha:DEED} is devoted
the Casimir effect in continuous higher dimensions, where the vacuum
energy for a five dimensional (5D) setup is derived. In Chap.~\ref{cha:DiED}
we introduce the discretisation of the fifth dimension and explicitly
show how to calculate the Casimir energy in the corresponding scenario.
Several effects like bulk masses and lattice artefacts are discussed,
too. As an application we apply in Chap.~\ref{cha:VacD} our results
to a deconstruction model and propose a way to determine the zero-point
energy of the 4D fields. Discretised gravitational EDs are the main
subject of Chap.~\ref{cha:DiGr}, where we derive some bounds on
the size of the ED. In Chap.~\ref{cha:Disk} we investigate a 6D
scenario with two curved and discretised EDs. We calculate the corresponding
mass spectra of gravitons and fermions and close the chapter with
a little application. Finally, we summarise the results of this work
and present our conclusions.

\minisec{Conventions}

Throughout this work we use the Einstein sum convention, and we set
the speed of light~$c$, Planck's constant~$\hbar$ and Boltzmann's
constant~$k_{\textrm{B}}$ to unity. Other conventions are given
in the text, where necessary.

\chapter{Time-dependent Cosmological Constant\label{cha:RGE1-TimeDepCC}}

\section{Quantum Effects and the Cosmological Constant\label{sec:RGE1-Quantum-CC}}

In this chapter we will investigate a positive CC as the most prominent
candidate for dark energy. As a component in Einstein's equations
of classical general relativity it can be treated as a perfect fluid
with a constant energy density~$\Lambda>0$ and an equation of state~$\omega=p/\Lambda=-1$
that corresponds to a negative pressure~$p=-\Lambda$. On the quantum
level the CC emerges as vacuum energy of quantum fields with the same
equation of state as the classical CC. It is therefore an unavoidable
constituent of the matter content of the universe. Unfortunately,
one does not know how to calculate its value in a unique way, because
it can be written in the form of a quartically divergent momentum~($p$)
integral like~$\int\textrm{d}^{3}p\cdot p$. Respectively, for compact
dimensions one obtains the infinite sum of zero-point energies~$\sum\frac{1}{2}p$.
In Minkowski space one usually eliminates the infinite vacuum energy
by normal ordering in QFT since it has no influence on flat space-time
physics. Gravity, on the other hand, is sensitive to all forms of
energy and matter, and we thus have to deal with vacuum energy in
cosmology. The naive assumption of an UV cutoff to regularise the
infinite integral at some known energy scale~$M_{\textrm{x}}$ leads
to an unobserved high value of the CC, which illustrates the old CC
problem~\cite{CCProblem}. For example, let~$M_{\textrm{x}}\sim10^{3}\,\textrm{GeV}$
be the energy scale, where supersymmetry is assumed to be broken.
This leads via\begin{equation}
\int_{0}^{M_{\textrm{x}}}\textrm{d}^{3}p\cdot p\sim M_{\textrm{x}}^{4}\gg\rho_{\textrm{obs}}\sim10^{-47}\,\textrm{GeV}^{4}\label{eq:RGE1-Cutoff}\end{equation}
to a mismatch of about~$60$ orders of magnitude. Without extreme
fine-tuning in the theory the naive cutoff method for determining
the CC obviously does not work and should be rejected. Fortunately,
the procedure of renormalisation in QFT can handle infinities, thereby
leading to a dependence of the renormalised constants on some energy
scale~$\mu$. In many cases, this renormalisation scale can be identified
with an external momentum, or at least with some characteristic scale
(e.g., the temperature) of the environment. Studying QFT on curved
space-time~\cite{QFTonCurve-1,QFTonCurve-2} leads to infinities
in the effective action or in the vacuum expectation values (VEV)
of the energy-momentum tensors of the fields. This can be treated
by renormalisation to yield a scale-dependent or running CC and a
running Newton constant. However, the absolute values are still not
calculable, but the change with respect to the renormalisation scale
can be calculated via RGEs. In the following we will investigate the
influence of RGEs originating in QFT~\cite{RGE-LamG,Bauer:2005rp,Bauer:2005tu}
and quantum gravity~\cite{QEG}. Unlike the running coupling constants
in the standard model of particles, here, the physical meaning of
the scale is not given by the theory since there is no external momentum.
In the cosmological context that we consider, it is reasonable to
identify the renormalisation scale with some characteristic scales
in cosmology, which will be done in Sec.~\ref{sec:RGE1-Choice-scale}.
For phenomenological reasons, we should request that the scale~$\mu$
does not change too much over cosmological time scales. Once we have
fixed the combination of RGE and scale identification we are able
to discuss in Sec.~\ref{sec:RGE1-Late-time} the late-time behaviour
and the fate of the universe with a running CC. In addition to a running
CC we will also consider a running of Newton's constant for one case
in Sec.~\ref{sec:RGE1-Ldot-Gdot}. Please note, that in contrast
to dark energy scenarios with a time-dependent equation of state as
in Refs.~\cite{quintessence}, here, the equation of state of the
CC is still exactly~$\omega=-1$. Nevertheless, it is possible to
obtain an effective time-dependent equation of state~\cite{Eff-EOS}
due to a non-standard scaling of the matter energy density with the
cosmological time. This non-standard scaling is explained in Sec.~\ref{sec:RGE1-Time-Einstein},
where we discuss how to solve Einstein's equations on a Robertson-Walker
background when the CC and Newton's constant depend on the cosmological
time.

\section{Scaling Laws\label{sec:RGE1-Scaling-laws}}

According to QFT on curved space-times the CC and Newton's constant~$G$
are subject to renormalisation group running like the running of the
fine-structure constant in quantum electrodynamics. The corresponding
scaling laws of these constants depend crucially on the considered
quantum fields and their masses. In the following we consider two
different RGEs emerging in this framework and in addition one scaling
law that emerges in the context of {}``quantum Einstein gravity''~\cite{QEG}.
Throughout the text we will identify the CC with the corresponding
vacuum energy density~$\Lambda$. 

Let us first start with non-interacting quantum fields on a curved
space-time, namely a Friedmann-Robertson-Walker universe with a positive
CC. For one fermionic and one bosonic degree of freedom with masses~$m_{\textrm{F}}$
and~$m_{\textrm{B}}$, respectively, the $1$-loop effective action
can be written in the form~\cite{QFTonCurve-1}\begin{eqnarray}
S_{\textrm{eff}} & = & \int\textrm{d}^{4}x\sqrt{-g}\left[\frac{\textrm{Ric}}{16\pi G}-\Lambda+\left(D+\ln\frac{m_{\textrm{F/B}}}{\mu}\right)\right.\nonumber \\
 &  & \left.\times\left(\frac{m_{\textrm{F}}^{4}-m_{\textrm{B}}^{4}}{32\pi^{2}}-\frac{\textrm{Ric}}{16\pi^{2}}\left[(\xi-\kl{\frac{1}{6}})m_{\textrm{B}}^{2}-\kl{\frac{1}{12}}m_{\textrm{F}}^{2}\right]\right)\right]+C\label{eq:RGE1-Seff}\end{eqnarray}
where~$D=\kl{\frac{1}{2}}\gamma_{\textrm{Euler}}+\lim_{n\rightarrow4}(n-4)^{-1}$
is a divergent term, which does not depend on the renormalisation
scale~$\mu$. Furthermore,~$\xi$ is a coupling constant%
\footnote{In the action~$S=\int\textrm{d}^{4}x\sqrt{-g}\left[\frac{\textrm{Ric}-2\lambda}{16\pi G}+\frac{1}{2}\phi_{;\alpha}\phi^{;\alpha}-\frac{1}{2}[m^{2}+\xi\textrm{Ric}]\phi^{2}\right]$
of a scalar field~$\phi$ on a curved space-time, the constant~$\xi$
occurs in the coupling term~$\xi\cdot\textrm{Ric}\cdot\phi^{2}$
between the scalar field and the Ricci scalar~$\textrm{Ric}$.%
}, and the variable~$C$ represents all further terms in the effective
action, that are neither proportional to the Ricci scalar~$\textrm{Ric}$
nor to the vacuum energy density\[
\Lambda:=\frac{\lambda}{8\pi G}.\]

The relevant $\beta$-functions in the $\overline{\textrm{MS}}$-scheme
for the vacuum energy density~$\Lambda$ and Newton's constant~$G$
are obtained by the requirement that the effective action~$S_{\textrm{eff}}$
must not depend on the renormalisation scale~$\mu$,\[
\mu\frac{\textrm{d}S_{\textrm{eff}}}{\textrm{d}\mu}=0.\]
Because of this condition, $\Lambda$ and~$G$ have to be treated
as $\mu$-dependent functions in Eq.~(\ref{eq:RGE1-Seff}), which
consequently have to obey the RGEs given by\[
\mu\frac{\textrm{d}\Lambda}{\textrm{d}\mu}=-\frac{m_{\textrm{F}}^{4}-m_{\textrm{B}}^{4}}{32\pi^{2}},\,\,\,\,\,\mu\frac{\textrm{d}}{\textrm{d}\mu}\left(\frac{1}{G}\right)=-\frac{1}{\pi}\left[(\xi-\kl{\frac{1}{6}})m_{\textrm{B}}^{2}-\kl{\frac{1}{12}}m_{\textrm{F}}^{2}\right].\]
Note, that the divergent term~$D$ has dropped out, leaving over
just the masses~$m_{\textrm{F/B}}$ and~$\xi$. Assuming constant
masses, the RGEs can be integrated, hence, the equation for the vacuum
energy density reads\begin{equation}
\Lambda(\mu)=\Lambda_{0}(1-q_{1}\ln\kl{\frac{\mu}{\mu_{0}}})\,,\,\,\,\,\Lambda_{0}:=\Lambda(\mu_{0}),\label{eq:RGE1-RGE1}\end{equation}
where~$\Lambda_{0}$ denotes the vacuum energy density today, when
the renormalisation scale~$\mu$ has the value~$\mu_{0}$. Moreover,
the sign of the parameter\begin{equation}
q_{1}:=\frac{1}{32\pi^{2}\Lambda_{0}}\left(m_{\textrm{F}}^{4}-m_{\textrm{B}}^{4}\right)\label{eq:RGE1-q1Def}\end{equation}
depends on whether bosons or fermions dominate. In this context, a
real scalar field counts as one bosonic degree of freedom, and a Dirac
field as four fermionic ones. The generalisation to more than one
quantum field in the RGE, can be achieved by summing over the fourth
powers of their masses. For Newton's constant~$G$ we obtain the
RGE in the integrated form\begin{equation}
G(\mu)=\frac{G_{0}}{1-q_{2}\ln\frac{\mu}{\mu_{0}}}\,,\,\,\,\, G_{0}:=G(\mu_{0}).\label{eq:RGE1-RGE1-G}\end{equation}
Again, we omit the generalisation to more fields, that follows from
summing over the squared masses of the fields. For one bosonic and
one fermionic degree of freedom the mass parameter~$q_{2}$ is given
by\begin{equation}
q_{2}:=\frac{G_{0}}{\pi}\left[(\xi-\kl{\frac{1}{6}})m_{\textrm{B}}^{2}-\kl{\frac{1}{12}}m_{\textrm{F}}^{2}\right].\label{eq:RGE1-q2Def}\end{equation}
Finally, we remark, that Eq.~(\ref{eq:RGE1-RGE1}) for the running
vacuum energy density~$\Lambda(\mu)$ was derived in a renormalisation
scheme, which is usually associated with the high energy regime. It
is therefore not known to what extend it can be applied at late times
in cosmology. In addition, to avoid conflicts with observations, the
field content has to be fine-tuned to obtain $|q_{1}|\le O(1)$, which
we assume for the rest of this work. Unfortunately, the corresponding
covariantly derived equations for the low energy sector are not known
yet~\cite{DecLamG}. Therefore, we prefer to work with the above
RGEs, which were derived in a covariant way, and study the consequences
and the constraints on the mass parameters~$q_{1}$ and~$q_{2}$.

The second scaling law follows from a RGE that shows a decoupling
behaviour. Considering only the most dominant terms at low energy,
the corresponding $\beta$-function for~$\Lambda$ is given by $\mu\frac{\textrm{d}\Lambda}{\textrm{d}\mu}=A_{1}\mu^{2}$,
where $A_{1}\sim\pm M^{2}$ is set by the masses~$M$ and the spins
of the fields. RGEs of this kind have been studied extensively in
the literature, see Refs.~\cite{RGE-LamG}. Assuming constant masses
and $\mu_{0}$ to be of the order of today's Hubble scale~$H_{0}$
one obtains\begin{equation}
\frac{\Lambda(\mu)}{\Lambda_{0}}=L_{0}+L_{1}\frac{\mu^{2}}{\mu_{0}^{2}},\,\,\,\, L_{1}\sim\pm\frac{M^{2}}{M_{\textrm{P}}^{2}},\label{eq:RGE1-RGE2}\end{equation}
where~$L_{0}:=1-L_{1}$ and~$M_{\textrm{P}}$ denotes the Planck
mass today. Here, the running of~$\Lambda$ is suppressed since~$|L_{1}|\ll1$
for sub-Planckian masses~$M$.

The last scaling laws come from the RGEs in quantum Einstein gravity~\cite{QEG,time-LamG-QG-1}.
In this framework the effective gravitational action becomes dependent
on a renormalisation scale~$\mu$, which leads to RGEs for~$\Lambda$
and~$G$. An interesting feature is the occurrence of an UV fixed-point~\cite{QEG-Fixpoint}
in the renormalisation group flow of the dimensionless quantities%
\footnote{In this work~$\Lambda$ denotes the vacuum energy density corresponding
to a CC, whereas in many articles about quantum gravity~$\Lambda$
is the CC~$\lambda$ itself.%
}~$\Lambda\mu^{-4}$ and~$G\mu^{2}$ at very early times in cosmology
corresponding to~$\mu\rightarrow\infty$. Motivated by strong infrared~(IR)
effects in quantum gravity one has proposed that there might also
exist an IR fixed-point in quantum Einstein gravity~\cite{QEG-IR}
leading to significant changes in cosmology at late times%
\footnote{However, it was argued in Ref.~\cite{QEG-NoIR} that the region of
strong IR effects can never be reached.%
}, where $\mu\rightarrow0$. If this were true, one would obtain the
RGEs\begin{equation}
\frac{\Lambda}{\Lambda_{0}}=\frac{\mu^{4}}{\mu_{0}^{4}},\,\,\,\,\frac{G}{G_{0}}=\frac{\mu_{0}^{2}}{\mu^{2}},\label{eq:RGE1-RGE3}\end{equation}
 where $\mu_{0}$ corresponds to $\Lambda=\Lambda_{0}$ and~$G=G_{0}$.
In the epoch between the UV and IR fixed-points,~$\Lambda$ and~$G$
vary very slowly with~$\mu$, and we will treat them as constants
in this region. Therefore, we assume that the scaling laws~(\ref{eq:RGE1-RGE3})
are valid from today until the end of the universe.

\section{Renormalisation Scales\label{sec:RGE1-Choice-scale}}

In order to study the effects on the cosmological expansion due to
the scaling laws of Sec.~\ref{sec:RGE1-Scaling-laws}, we have to
define the physical meaning of the renormalisation scale~$\mu$.
Unfortunately, the theories underlying the scaling laws often do not
determine the scale explicitly%
\footnote{In the framework of Refs.~\cite{G-von-H}, Newton's constant was
found to depend explicitly on the Hubble scale.%
}, apart from the usual interpretations as an (IR) cutoff or a scale
characterising the physical environment (temperature, external momenta).
In our cosmological setting we will investigate three different choices
for~$\mu$, given by the Hubble scale~$H$, the inverse radius~$R^{-1}$
of the cosmological event horizon, and finally the inverse radius~$T^{-1}$
of the particle horizon.

Let us consider our universe on sufficiently large scales, where it
is well described by the Robertson-Walker metric given by\begin{equation}
\textrm{d}s^{2}=\textrm{d}t^{2}-a^{2}(t)\left(\frac{\textrm{d}r^{2}}{1-kr^{2}}+r^{2}\textrm{d}\Omega^{2}\right)\!,\label{eq:RGE1-RWwithK}\end{equation}
where~$k$ fixes the spatial curvature. In accordance with recent
observation we will ignore the spatial curvature and consider in the
following the spatially flat ($k=0$) metric\begin{equation}
\textrm{d}s^{2}=\textrm{d}t^{2}-a^{2}(t)\textrm{d}\vec{x}^{2},\label{eq:RGE1-RWflat}\end{equation}
where the scale factor~$a(t)$ depends only on the cosmological time~$t$
and not on the spatial coordinates~$\vec{x}$. 

Our first candidate for the renormalisation scale~$\mu$ is the Hubble
scale\begin{equation}
H(t):=\frac{\dot{a}(t)}{a(t)},\label{eq:RGE1-Scale1}\end{equation}
 which describes the actual expansion rate of the universe. On the
other hand, the horizon scales~$R$ and~$T$ describe the cosmological
evolution of the future and the past, respectively. 

In universes with late-time acceleration like the $\Lambda$CDM model
there usually exists a cosmological event horizon. Its proper radius~$R$
corresponds to the proper distance that a (light) signal can travel
when it is emitted by a comoving observer at the time~$t$: \begin{equation}
R(t):=a(t)\int_{t}^{\infty}\frac{\textrm{d}t^{\prime}}{a(t^{\prime})}.\label{eq:RGE1-Scale2}\end{equation}
In the case that the universe comes to an end within finite time the
upper limit of the integral has to be replaced by this time. Similar
to the event horizon of a black hole, the cosmological event horizon
exhibits thermodynamical properties like the emission of radiation
with the Gibbons-Hawking temperature~\cite{horizon-thermo}.

\begin{figure}
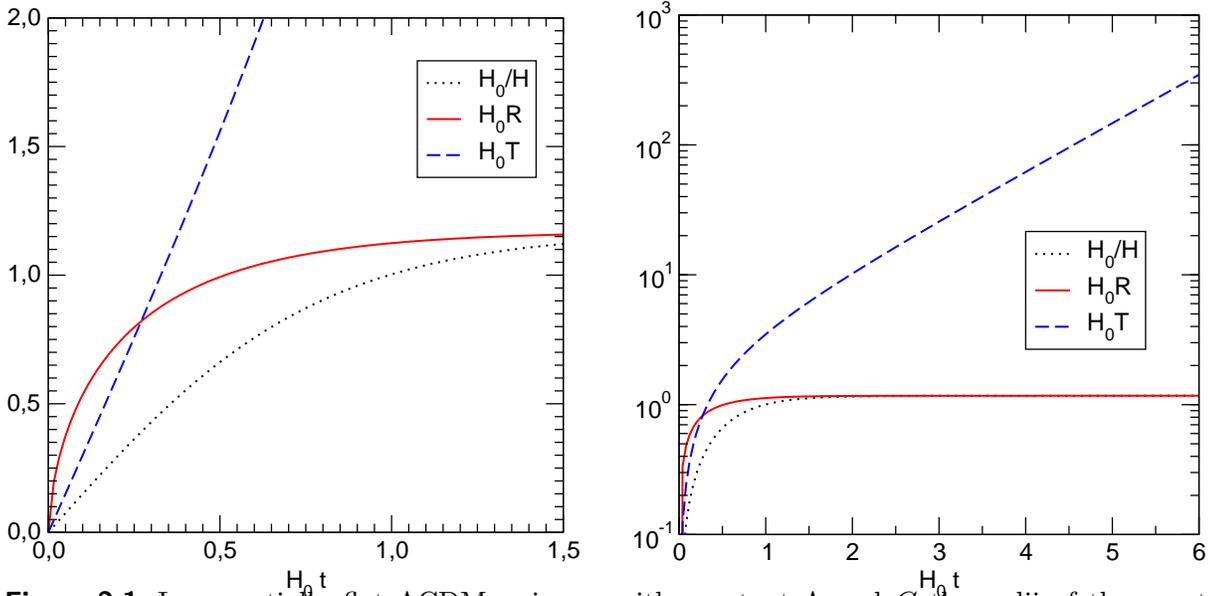

\begin{centering}\includegraphics[clip,width=0.48\columnwidth,keepaspectratio]{LCDM-Horizons-1}\hfill{}\includegraphics[clip,width=0.48\columnwidth,keepaspectratio]{LCDM-Horizons-2}\par\end{centering}

\caption{\label{fig:RGE1-ScalesLCDM}In a spatially flat $\Lambda$CDM universe
with constant~$\Lambda$ and~$G$ the radii of the event horizon~$R$
and the Hubble horizon~$H^{-1}$ approach the same constant value
for~$t\rightarrow\infty$. At early times the particle horizon radius~$T$
grows linearly with time and exponentially with time for~$t\rightarrow\infty$.
At the time~$t_{0}=0,99H_{0}^{-1}$ (today) the relative vacuum energy
is given by~$\Omega_{\Lambda0}=0,73$. }
\end{figure}

The counterpart of~$R$ is the particle horizon radius~$T$, which
is given by the proper distance that a signal has travelled since
the beginning of the world ($t=0$):\begin{equation}
T(t):=a(t)\int_{0}^{t}\frac{\textrm{d}t^{\prime}}{a(t^{\prime})}.\label{eq:RGE1-Scale3}\end{equation}
In a simple cosmological model, where the universe begins at~$t=0$
and then evolves according to the decelerating scale factor~$a(t)\propto t^{n}$
with~$0<n<1$, the particle horizon radius reads $T(t)=t/(1-n)$.
Now we assume that at some time~$t_{0}$ a de~Sitter phase sets
in, which corresponds to a scale factor~$a(t)\propto\exp(H_{0}(t-t_{0}))$
for~$t>t_{0}$. This leads to a radius function~$T$ that grows
exponentially with~$t$ at late times:\begin{equation}
T(t>t_{0})=\left(T(t_{0})+\frac{1}{H_{0}}\right)\exp(H_{0}(t-t_{0}))-\frac{1}{H_{0}}.\label{eq:RGE1-T-dS}\end{equation}
Since the asymptotic behaviour of the scales~$H$,~$R$ and~$T$
plays a major role in this work, it is plotted in Fig.~\ref{fig:RGE1-ScalesLCDM}
for a universe with dust-like matter and~$\Lambda$ and~$G$ being
positive and constant.%
 Note that if one of the horizon radii diverges, one says that the
horizon does not exist. The geometric meaning of the scales given
here has been discussed, e.g., in Ref.~\cite{horizons}, and some
arguments for the event horizon in the context of the CC can be found
in Refs.~\cite{EventHor-3,EventHor-4}.

\section{General Relativity with Time-dependent Constants \label{sec:RGE1-Time-Einstein}}

In this section, we derive the evolution equation for the cosmic scale
factor~$a(t)$ in the framework of the spatially isotropic and homogeneous
Friedmann-Robertson-Walker universe with a time-dependent CC and Newton's
constant. On this background, radiation and pressureless matter (dust)
can both be described by a perfect fluid with energy density~$\rho$
and pressure~$p=\omega\rho$, where the constant~$\omega$ characterises
the equation of state. For instance, dust-like matter like CDM has
the equation of state~$\omega=0$ and incoherent radiation~$\omega=\frac{1}{3}$,
respectively. The corresponding energy-momentum tensor for these energy
forms reads\[
T^{\alpha\beta}=(\rho+p)u^{\alpha}u^{\beta}-pg^{\alpha\beta},\]
with~$u^{\alpha}$ being the four-velocity vector field of the fluid.
With our choices of the renormalisation scale,~$G$ and~$\Lambda$
depend only on the cosmic time~$t$. From Einstein's equations\[
G^{\alpha\beta}=8\pi G(\Lambda g^{\alpha\beta}+T^{\alpha\beta})\]
and from the contracted Bianchi identities~$G_{\,\,\,\,\,\,;\beta}^{\alpha\beta}=0$
for the Einstein tensor~$G^{\alpha\beta}$, we obtain the generalised
conservation equations\[
[G\Lambda g^{\alpha\beta}+GT^{\alpha\beta}]_{;\beta}=0,\]
whose~$\alpha=0$ component reads\[
\dot{G}(\Lambda+\rho)+G(\dot{\Lambda}+\dot{\rho}+3\kl{\frac{\dot{a}}{a}}\rho(1+\omega))=0.\]
Here the dot denotes the derivative with respect to the cosmological
time~$t$, and we do not assume~$T_{\,\,\,\,\,\,;\beta}^{\alpha\beta}=0$.
For constant~$\Lambda$ and~$G$ the last equation can be integrated
to yield the usual scaling law for the matter energy density~$\rho\propto a^{-2(Q+1)}$,
where we have for our convenience introduced the equation of state
parameter%
\footnote{For a dominant matter energy density~$\rho\gg\Lambda$ and flat spatial
curvature, the acceleration quantity~$q:=\frac{\ddot{a}a}{\dot{a}^{2}}$
is given by the negative value of the~$Q$.%
}\begin{equation}
Q:=\frac{1}{2}(1+3\omega).\label{eq:RGE1-QDef}\end{equation}
 Note that for non-constant~$\Lambda$ and~$G$ this simple scaling
rule for the matter content is not valid anymore, because it is now
possible to transfer energy between the matter and the vacuum, in
addition to~$\dot{G}\neq0$. At this stage we have to admit that
this energy transfer implies an effective interaction between the
gravitational sector ($\Lambda$,~$G$) and matter, which is not
part of the original Lagrangian. In this sense it should be compared
with gravitational particle production~\cite{Parker:1968mv} resulting
also from the interplay of gravity with quantum physics.

Since we have at this point no information about~$\rho$ as function
of time or the scale factor, we have to combine the Friedmann equations
for the Hubble scale~$H:=\frac{\dot{a}}{a}$ and the acceleration~$\frac{\ddot{a}}{a}$,\begin{eqnarray}
\left(\frac{\dot{a}}{a}\right)^{2}+\frac{k}{a^{2}} & = & \frac{8\pi}{3}G(t)(\Lambda(t)+\rho(t)),\label{eq:RGE1-Friedmann1}\\
\frac{\ddot{a}}{a} & = & \frac{8\pi}{3}G(t)(\Lambda(t)-Q\rho(t)),\label{eq:RGE1-Friedmann2}\end{eqnarray}
in order to eliminate the matter energy density~$\rho$. The left-hand
side of the resulting is abbreviated by~$F(t)$:\begin{equation}
F(t):=\frac{\ddot{a}}{a}+Q\left[\left(\frac{\dot{a}}{a}\right)^{2}+\frac{k}{a^{2}}\right]=\frac{8\pi}{3}G\Lambda\cdot(1+Q).\label{eq:RGE1-Friedmann3}\end{equation}
By introducing the constant\begin{equation}
K_{0}:=\frac{3}{8\pi G_{0}\Lambda_{0}(Q+1)}=\frac{H_{0}^{-2}}{\Omega_{\Lambda0}(Q+1)},\label{eq:RGE1-K0def}\end{equation}
with the relative vacuum energy density~$\Omega_{\Lambda0}=8\pi G_{0}\Lambda_{0}/(3H_{0}^{2})$
and the Hubble scale~$H_{0}$ at the time~$t=t_{0}$, we obtain
the main equation in the compact form\begin{equation}
K_{0}F(t)=\frac{\Lambda G}{\Lambda_{0}G_{0}}\label{eq:RGE1-MainEqu}\end{equation}
Here, we can now insert the RGEs for~$\Lambda$ and~$G$ from Sec.~\ref{sec:RGE1-Scaling-laws}
together with a choice of the renormalisation scale~$\mu$ from Sec.~\ref{sec:RGE1-Choice-scale}.
For the scale identifications~$\mu=R^{-1}$ and~$\mu=T^{-1}$ we
have to solve integro-differential equations, whereas~$\mu=H$ just
leads to an ordinary differential equation.

\section{Cosmological Evolution at Late Times\label{sec:RGE1-Late-time}}

In this section we study the late-time evolution of the universe with
variable~$\Lambda$ and~$G$, thereby assuming from today on the
validity of the scaling laws (\ref{eq:RGE1-RGE1})--(\ref{eq:RGE1-RGE3})
and the correct identification of the renormalisation scale~$\mu$
with the scales (\ref{eq:RGE1-Scale1})--(\ref{eq:RGE1-Scale3}).
Our aim is to determine in all nine cases the possible final states
of the universe. This depends, of course, on the choice of parameters,
but we restrict ourselves to parameter values which comply vaguely
to current observations. Today, at the cosmological time~$t_{0}=0,99H_{0}^{-1}$
we fix the initial values~$H_{0}$ and~$\Omega_{\Lambda0}=0,73$
by observations. Furthermore, the initial value~$T_{0}$ of the particle
horizon radius would be fixed if the past cosmological evolution was
known from the Big Bang on. In contrast to this, the value~$R_{0}$
of today's event horizon radius depends on the future cosmological
evolution and is treated here as a free parameter. In the following
we derive some properties of the solutions analytically, in particular,
we study the stability of (asymptotic) de~Sitter solutions and the
occurrence of future singularities, where the scale factor or one
of its derivatives diverge within finite time~\cite{Future-Sing,BigRip-Analysis}.
For simplicity we denote a big rip or a big crunch by the lowest order
divergent derivative that is positive or negative, respectively. Since
some combinations of scaling laws and renormalisation scales lead
to complicated equations we derive in some cases only approximate
or numerical statements. 

In order to solve Eq.~(\ref{eq:RGE1-MainEqu}) numerically, we have
to remove the integrals in the definitions (\ref{eq:RGE1-Scale2}),
(\ref{eq:RGE1-Scale3}) of~$R$ and~$T$ by a differentiation with
respect to~$t$. Using the relations \[
\dot{R}=RH-1,\,\,\,\,\dot{T}=TH+1\]
 an ordinary differential equation for the scale factor can be obtained.
Let us, for instance, consider~$\mu=R^{-1}$ in combination with
the RGEs (\ref{eq:RGE1-RGE1}) and (\ref{eq:RGE1-RGE1-G}). First,
we solve the main equation~(\ref{eq:RGE1-MainEqu}) for~$R$,\begin{equation}
\frac{\mu_{0}}{\mu(t)}=\frac{R(t)}{R_{0}}=\exp\left[\frac{K_{0}F(t)-1}{q_{1}-q_{2}K_{0}F(t)}\right]\!,\label{eq:RGE1-RvonF}\end{equation}
which has the time derivative\begin{equation}
\left[\frac{\dot{a}}{a}+\frac{(q_{2}-q_{1})K_{0}\dot{F}}{(q_{1}-q_{2}K_{0}F)^{2}}\right]\cdot\exp\left[\frac{K_{0}F-1}{q_{1}-q_{2}K_{0}F}\right]-\frac{1}{R_{0}}=0,\label{eq:RGE1-ODE-a}\end{equation}
where we have substituted~$R$ by Eq.~(\ref{eq:RGE1-RvonF}). This
equation can be integrated numerically. Afterwards we have to check
whether the functions~$R$ and~$T$, calculated from the numerical
solution of~$a(t)$, agree with~$R$ and~$T$ that follow directly
from Eq.~(\ref{eq:RGE1-MainEqu}). If they do not match, the numerical
solution has to be discarded. Some of these cases are illustrated
and discussed in Sec.~\ref{sec:RGE1-Ldot-Gdot}. Furthermore, solutions
involving a negative matter energy density are questionable on physical
grounds. This happens when the vacuum energy density~$\Lambda$ becomes
greater than the critical energy density~$\rho_{\textrm{c}}=3H^{2}/(8\pi G)$,
which follows from the first Friedmann equation~(\ref{eq:RGE1-Friedmann1}).

\subsection{\label{sub:RGE1-A1}$\boldsymbol{\Lambda=\Lambda_{0}(1-q_{1}\ln\frac{\mu}{\mu_{0}})}$,
$\boldsymbol{\mu=H}$}

\begin{figure}[t]
\begin{centering}\includegraphics[clip,width=0.4\columnwidth,keepaspectratio]{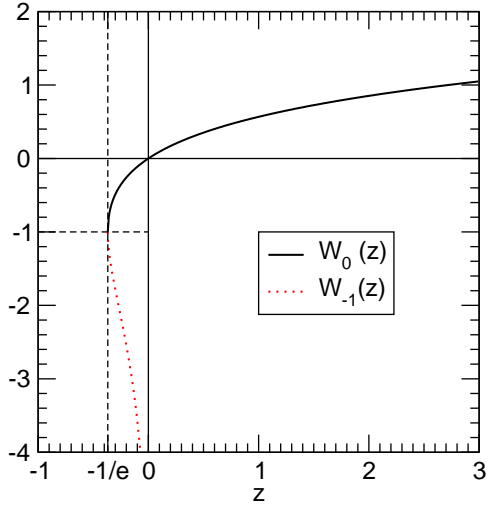}\par\end{centering}

\caption{\label{fig:RGE1-LambertW}Lambert's $W$-function is the inverse
function of~$z=xe^{x}$. For~$-1/e\le z<0$ it has two real-valued
brunches~$W_{0}(z)$ and~$W_{-1}(z)$, and for~$z\ge0$ only~$W_{0}(z)$
is real.}
\end{figure}

\begin{figure}[t]
\begin{centering}\includegraphics[clip,width=1\columnwidth,keepaspectratio]{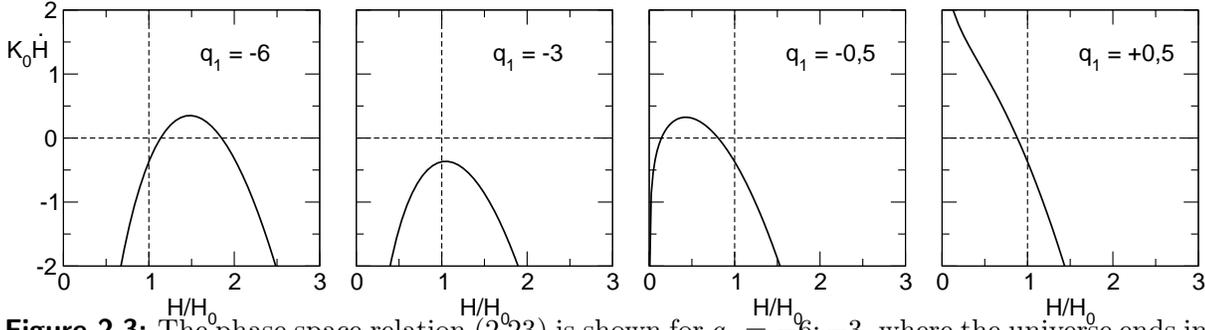}\par\end{centering}

\caption{\label{fig:RGE1-A1-HdotH}The phase space relation~(\ref{eq:RGE1-A1-Hdot})
is shown for~$q_{1}=-6;-3$, where the universe ends in a big crunch
singularity, and respectively for~$q_{1}=-\frac{1}{2};+\frac{1}{2}$,
which leads to a de~Sitter final state ($\dot{H}=0$). In all cases
the evolution begins at~$H=H_{0}$, where~$\Omega_{\Lambda0}=0,73$
and~$\dot{H}_{0}<0$.}
\end{figure}

For the given scaling law and scale choice Eq.~(\ref{eq:RGE1-MainEqu})
reads\begin{equation}
K_{0}(\dot{H}+(Q+1)H^{2})=1-q_{1}\ln\frac{H}{H_{0}}.\label{eq:RGE1-A1-Hdot}\end{equation}
 We first look for asymptotic de~Sitter solutions by applying $\dot{H}=0$
and~$H\rightarrow H_{\textrm{e}}$. Thus the final Hubble scale~$H_{\textrm{e}}$
is given by \begin{equation}
\frac{H_{\textrm{e}}^{2}}{H_{0}^{2}}=\frac{q_{1}}{2}\Omega_{\Lambda0}W_{u}\left(\frac{2}{q_{1}\Omega_{\Lambda0}}e^{2/q_{1}}\right)\!\!.\label{eq:RGE1-A1-He}\end{equation}
Here, $W_{u}(z)$ with~$u=0,-1$ denotes one of the two real-valued
branches of Lambert's \mbox{W-function}, which is the solution of~$z=xe^{x}$,
see Fig.~\ref{fig:RGE1-LambertW}.%
 For~$q_{1}>0$ there is always one solution for~$H_{\textrm{e}}$,
and for~$q_{1}<0$ two solutions exist if the argument of~$W_{u}$
in Eq.~(\ref{eq:RGE1-A1-He}) is greater than~$-e^{-1}$. This means
either~$q_{1}<n_{1}$ or~$q_{1}>n_{2}$ with\[
n_{1}:=\frac{2}{W_{0}(-\Omega_{\Lambda0}e^{-1})}\,\,\,\,\textrm{and}\,\,\,\, n_{2}:=\frac{2}{W_{-1}(-\Omega_{\Lambda0}e^{-1})}>n_{1}.\]
These solution are stable if\[
\left.\frac{\textrm{d}(K_{0}\dot{H})}{\textrm{d}H}\right|_{H\rightarrow H_{\textrm{e}}}=-\frac{q_{1}}{H_{\textrm{e}}}-2H_{\textrm{e}}K_{0}(Q+1)<0,\]
where we used Eq.~(\ref{eq:RGE1-A1-Hdot}). For positive~$q_{1}$
this condition is always fulfilled, whereas for negative~$q_{1}$
it means~$W_{u}(\frac{2}{q_{1}\Omega_{\Lambda0}}e^{2/q_{1}})<-1$,
which follows from Eq.~(\ref{eq:RGE1-A1-He}). Again, the argument
of~$W_{u}$ is constrained yielding~$q_{1}>n_{2}$. Therefore only
$q_{1}>n_{2}$, which includes~$q_{1}>0$, leads to stable de~Sitter
solutions. Using the phase space relation~(\ref{eq:RGE1-A1-Hdot}),
which is plotted in Fig.~\ref{fig:RGE1-A1-HdotH}, we conclude that
for other values of~$q_{1}$ the cosmological evolution will always
end within finite time in a Big Crunch singularity, where~$H\rightarrow0$
and~$\dot{H}\rightarrow-\infty$.%

\subsection{\label{sub:RGE1-B1}$\boldsymbol{\Lambda=\Lambda_{0}(L_{0}+L_{1}\frac{\mu^{2}}{\mu_{0}^{2}})}$,
$\boldsymbol{\mu=H}$}

Using this choice for~$\Lambda$ and~$\mu$ the main equation~(\ref{eq:RGE1-MainEqu})
becomes\[
\frac{\ddot{a}}{a}+H^{2}(Q-L_{1}(Q+1)\Omega_{\Lambda0})=H_{0}^{2}L_{0}\Omega_{\Lambda0}(Q+1),\]
which has the exact solution\[
\frac{a(t)}{a_{0}}=\left[\sinh([Q+1][\Omega_{\Lambda0}L_{0}(1-L_{1}\Omega_{\Lambda0})]^{\frac{1}{2}}H_{0}t)\right]^{n}\]
with the parameter\[
n:=[(Q+1)(1-L_{1}\Omega_{\Lambda0})]^{-1}.\]
At early times,~$H_{0}t\ll1$, it describes a power-law expansion
$a(t)\propto t^{n}$, whereas at late times it approaches the de~Sitter
expansion law $a(t)\propto\exp(H_{\textrm{e}}t)$ with the final Hubble
scale~$H_{\textrm{e}}$ given by\[
H_{\textrm{e}}=H_{0}\sqrt{\frac{\Omega_{\Lambda0}L_{0}}{1-L_{1}\Omega_{\Lambda0}}}.\]
Further aspects of this case have been studied, e.g., in Refs.~\cite{RGE-LamG}.

\subsection{\label{sub:RGE1-C1}$\boldsymbol{\Lambda G=\Lambda_{0}G_{0}\frac{\mu^{2}}{\mu_{0}^{2}}}$,
$\boldsymbol{\mu=H}$}

Following Sec.~\ref{sec:RGE1-Scaling-laws} we assume that this scaling
law is valid from today on. Here, Eq.~(\ref{eq:RGE1-MainEqu}) is
given by~$\frac{\ddot{a}}{a}+BH^{2}=0$ with~$B:=Q-(Q+1)\Omega_{\Lambda0}$.
It has an exact power-law solution\[
\frac{a(t)}{a_{0}}=[(1+B)(t-t_{1})]^{1/(B+1)},\,\,\,\, t_{1}=\textrm{const.},\]
that exhibits a constant acceleration~$\ddot{a}a/\dot{a}^{2}=-B>0$
as well as~$\Omega_{\Lambda}=\textrm{const}$. Note that a similar
solution was found in Refs.~\cite{QEG-IR}.

\subsection{\label{sub:RGE1-A2}$\boldsymbol{\Lambda=\Lambda_{0}(1-q_{1}\ln\frac{\mu}{\mu_{0}})}$,
$\boldsymbol{\mu=R^{-1}}$}

At first we look for de~Sitter solutions with $a(t)\propto\exp(H_{\textrm{e}}t)$,
where the inverse Hubble scale and the event horizon radius approach~$R\rightarrow R_{\textrm{e}}=H_{\textrm{e}}^{-1}$
in addition to~$\dot{H}=0$. Plugging this asymptotic form for~$a(t)$
into Eq.~(\ref{eq:RGE1-MainEqu}), one arrives at\[
\frac{K_{0}(1+Q)}{R_{\textrm{e}}^{2}}=q_{3}x^{-2}=1+q_{1}\ln x,\]
where the variables~$x:=R_{\textrm{e}}/R_{0}$ and~$q_{3}:=K_{0}(1+Q)/R_{0}^{2}>0$
have been introduced. The solutions for~$x$ are given by\begin{equation}
x=\frac{R_{\textrm{e}}}{R_{0}}=\sqrt{\frac{2q_{3}}{q_{1}\cdot W_{u}(\frac{2q_{3}}{q_{1}}e^{2/q_{1}})}},\label{eq:RGE1-A2-ReR0}\end{equation}
involving%
\begin{figure}[t]
\begin{centering}\includegraphics[clip,scale=0.8]{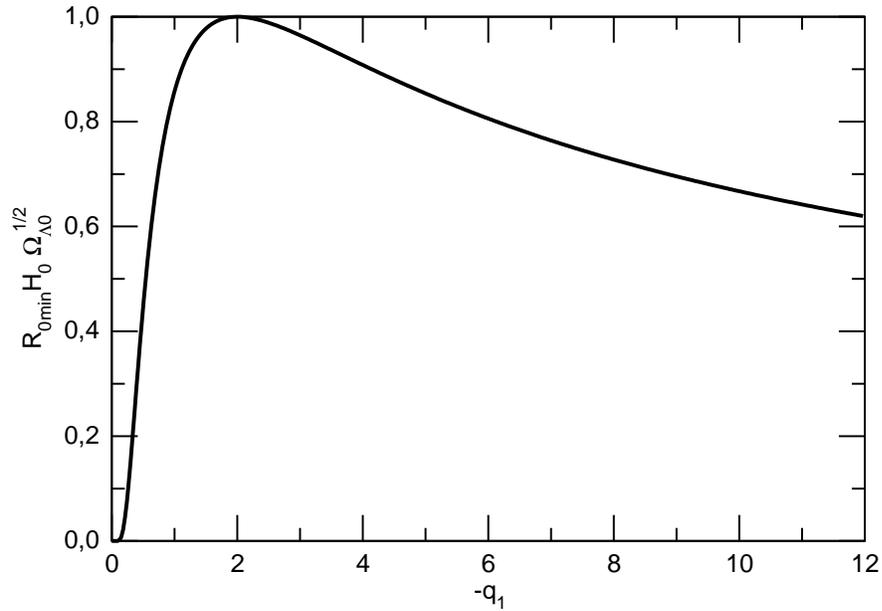}\par\end{centering}

\caption{\label{fig:RGE1-A2-R0min}The lower bound~$R_{0\textrm{min}}$ of
the initial event horizon radius~$R_{0}$ from Eq.~(\ref{eq:RGE1-A2-R0min})
for negative~$q_{1}$. For~$R_{0}<R_{0\textrm{min}}$ a final de~Sitter
state does not exist. }
\end{figure}
 Lambert's $W$-function~$W_{u}(z)$ with~$u=0,-1$ (Fig.~\ref{fig:RGE1-LambertW}).
Since the $W$-function is real only for arguments~$z\ge-e^{-1}$
and~$z<0$ for~$W_{-1}(z)$ we obtain for negative~$q_{1}$ the
constraint~$q_{3}\le-\frac{q_{1}}{2}\exp(-\frac{2}{q_{1}}-1)$, which
implies an lower bound for the initial value~$R_{0}$ of the horizon
radius as shown in Fig.~\ref{fig:RGE1-A2-R0min}:\begin{equation}
R_{0}\ge R_{0\textrm{min}}:=\sqrt{-\frac{2}{q_{1}\Omega_{\Lambda0}H_{0}^{2}}\exp\left(\frac{2}{q_{1}}+1\right)}.\label{eq:RGE1-A2-R0min}\end{equation}
If~$R_{0}$ is smaller than this minimal value, then Eq.~(\ref{eq:RGE1-A2-ReR0})
has no positive solutions and a final de~Sitter state does not exist.
For~$R_{0}=R_{0\textrm{min}}$ there is exactly one solution~$x=\exp(-\frac{1}{q_{1}}-\frac{1}{2})$,
for higher values~$R_{0}$ there are two solutions. %
\begin{figure}[t]
\begin{centering}\includegraphics[clip,width=0.9\textwidth,keepaspectratio]{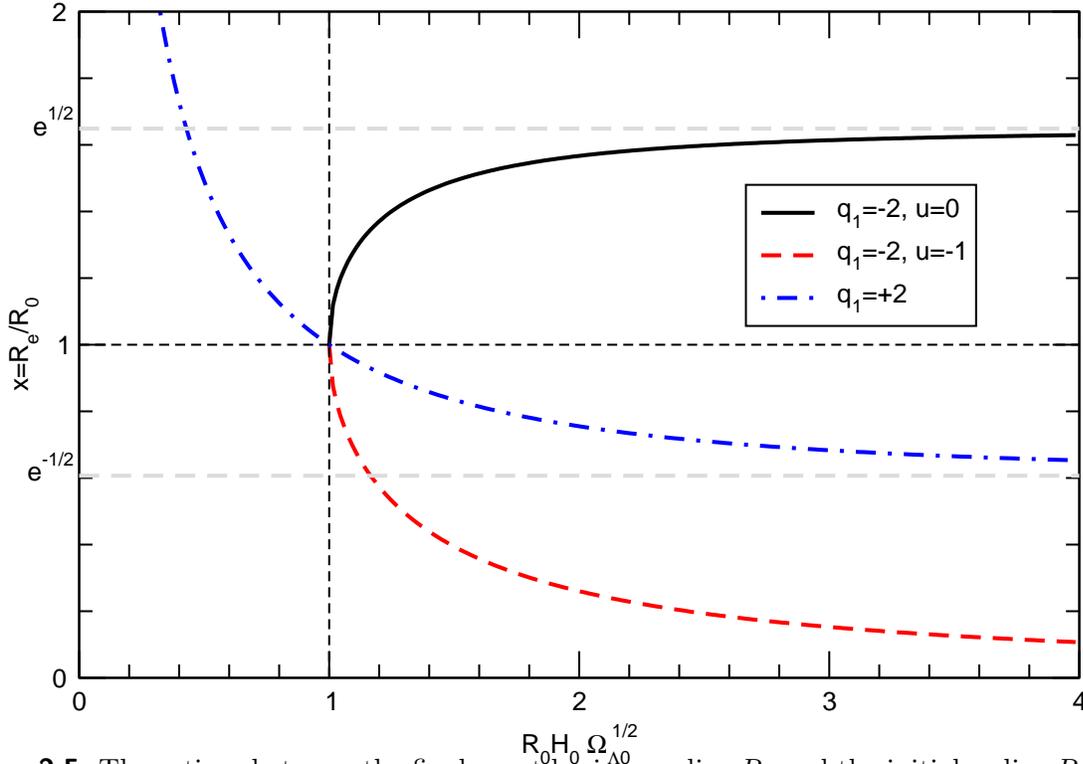}\par\end{centering}

\caption{\label{fig:RGE1-A2-ReR0}The ratio~$x$ between the final event
horizon radius~$R_{\textrm{e}}$ and the initial radius~$R_{0}$
as a function of~$R_{0}$, Eq.~(\ref{eq:RGE1-A2-ReR0}). Here~$q_{1}$
is set to~$\pm2$. All stable de~Sitter final states lie on the
curve~$u=0$.}
\end{figure}
In the case of a positive value of the parameter~$q_{1}$, the initial
value~$R_{0}$ must be smaller than~$1/\sqrt{H_{0}^{2}\Omega_{\Lambda0}}$.
Otherwise the final horizon radius~$R_{\textrm{e}}$ is smaller than
the initial one,~$x<1$. Both cases are plotted in Fig.~\ref{fig:RGE1-A2-ReR0}.

Since we have found several de~Sitter solutions, we now have to study
the stability of these final states. Therefore, we write~$K_{0}\dot{F}$
as a function of~$K_{0}F=K_{0}(\dot{H}+(Q+1)H^{2})$,\[
K_{0}\dot{F}=q_{1}\left[H-\frac{1}{R_{0}}\exp\left(\frac{1-K_{0}F}{q_{1}}\right)\right]\!,\]
where we used~$\dot{R}=RH-1$. In the final de~Sitter state we have~$K_{0}\dot{F}=0$
and~$R=R_{\textrm{e}}=1/H_{\textrm{e}}=\textrm{const}$. Near this
point we can neglect~$\dot{H}$ in the function~$F$ and replace~$H$
by \[
\sqrt{\frac{K_{0}F}{K_{0}(Q+1)}}.\]
 For a stable solution it is required that \begin{equation}
\left.\frac{\textrm{d}(K_{0}\dot{F})}{\textrm{d}(K_{0}F)}\right|_{\textrm{dS}}=\frac{q_{1}}{2}[K_{0}^{2}F(Q+1)]^{-\frac{1}{2}}+\frac{1}{R_{0}}\exp\left[\frac{1-K_{0}F}{q_{1}}\right]<0,\label{eq:RGE1-A2-dFdotdF}\end{equation}
in the final point, where~$K_{0}F=K_{0}(Q+1)H_{\textrm{e}}^{2}=q_{3}x^{-2}$.
With~$q_{3}$ and~$x$ from above, this yields the stability condition\[
\left[W_{u}\left(\frac{2q_{3}}{q_{1}}e^{2/q_{1}}\right)\right]^{-1}<-1,\]
implying that there are no stable de~Sitter solutions for positive
values of~$q_{1}$, because the $W$-function is positive. Thus a
big crunch singularity ($F,\dot{H}\rightarrow-\infty$, $R\rightarrow0$)
might happen for certain values of initial conditions and parameters,
or there exists no solution. A big rip singularity ($F,\dot{H}\rightarrow\infty$,
$R\rightarrow0$) does not exist because it is not compatible with~$R\rightarrow\infty$.

For negative~$q_{1}$ we get the condition\[
W_{u}\left(\frac{2q_{3}}{q_{1}}e^{2/q_{1}}\right)>-1,\]
which means that only the solution with $u=0$ is stable. This renders
the final event horizon radius~$R_{\textrm{e}}=R_{0}x$ unique.

\begin{figure}[t]
\begin{centering}\includegraphics[clip,width=1\columnwidth,keepaspectratio]{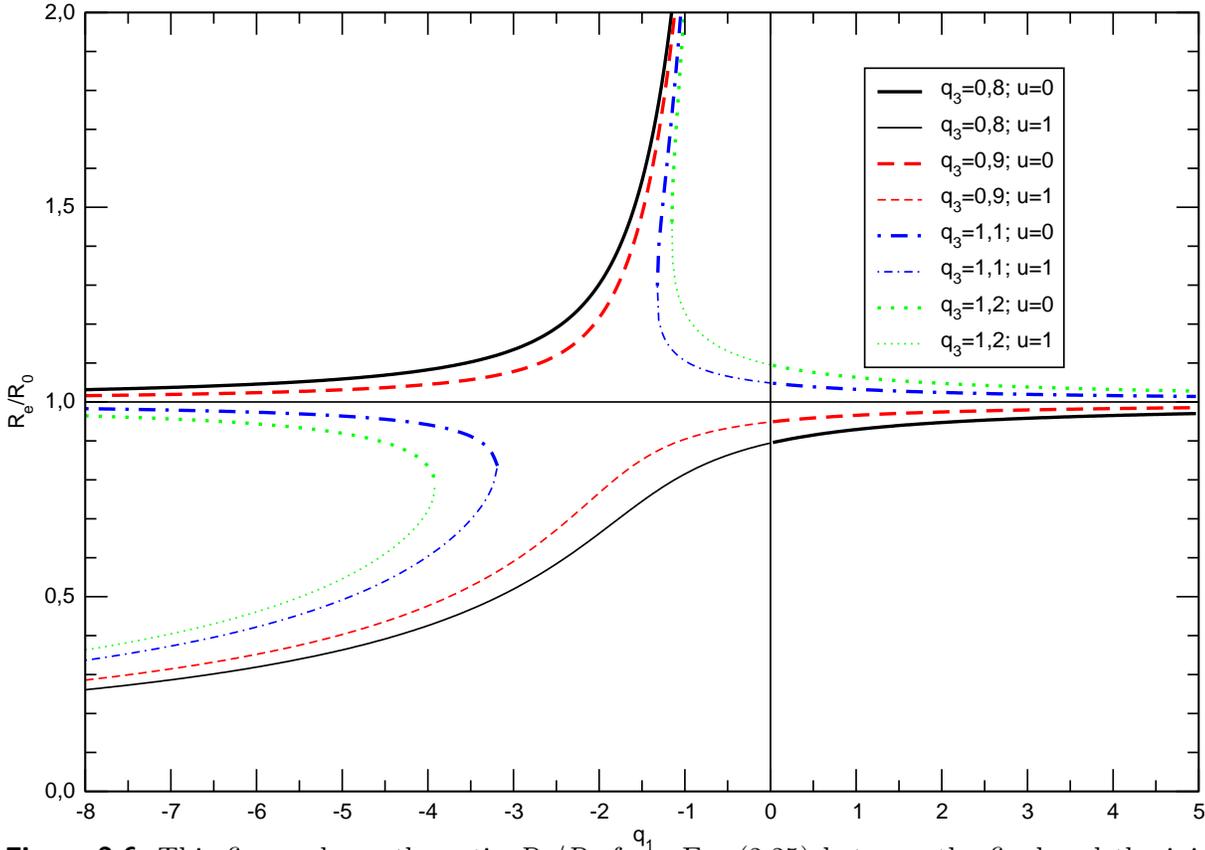}\par\end{centering}

\caption{\label{fig:RGE1-A2-ReR0vonq1}This figure shows the ratio~$R_{\textrm{e}}/R_{0}$
from Eq.~(\ref{eq:RGE1-A2-ReR0}) between the final and the initial
event horizon radius as a function of~$q_{1}$. We plotted four cases
for different values of~$q_{3}=1/(H_{0}^{2}R_{0}^{2}\Omega_{\Lambda0})$,
where the choice~$q_{3}=0,8;\,0,9$ corresponds to an initial radius~$R_{0}>1/(H_{0}\sqrt{\Omega_{\Lambda0}})$,
and~$q_{3}=1,1;\,1,2$ to~$R_{0}<1/(H_{0}\sqrt{\Omega_{\Lambda0}})$.
In the latter case there are no solutions for a certain range of values
of~$q_{1}$ as described by Eq.~(\ref{eq:RGE1-A2-q1Conditions}).
The thick lines show the ($u=0$)-branch of~$R_{\textrm{e}}/R_{0}$,
and the thin lines the ($u=-1$)-branch, respectively.}
\end{figure}

Finally, we take a closer look at the ratio~$R_{\textrm{e}}/R_{0}$
as a function of the mass parameter~$q_{1}$. For initial values~$R_{0}<1/(H_{0}\sqrt{\Omega_{\Lambda0}})$,
which means~$q_{3}>1$, there is a certain range of values of~$q_{1}$
where no solutions for~$R_{\textrm{e}}$ exist. This range is again
given by the requirement that the argument of the~$W$-function must
be greater than or equal to~$-e^{-1}$, leading to the conditions\begin{equation}
q_{1}\leq\frac{2}{W_{0}(-\frac{1}{eq_{3}})}\,\,\,\,\textrm{or}\,\,\,\, q_{1}\geq\frac{2}{W_{-1}(-\frac{1}{eq_{3}})}.\label{eq:RGE1-A2-q1Conditions}\end{equation}
In Fig.~\ref{fig:RGE1-A2-ReR0vonq1} the exclusion range for~$q_{1}$
is obvious for~$q_{3}>1$. In the case that~$q_{1}$ lies above
this range, the unstable solution for~$R_{\textrm{e}}$ is reached
first during the future cosmic evolution. For~$q_{1}$ below this
range, the stable solution is nearer to the initial value~$R_{0}$
than the unstable one, however, both solutions for~$R_{\textrm{e}}$
lie below~$R_{0}$. Initial values~$R_{0}>1/(H_{0}\sqrt{\Omega_{\Lambda0}})$
(i.e.~$q_{3}<1$) lead to stable final states with~$R_{\textrm{e}}>R_{0}$
for all negative values of~$q_{1}$. Moreover, all stable de~Sitter
solutions are realised except the last one corresponding to the parameter
region\[
R_{0}^{2}H_{0}^{2}\Omega_{\Lambda0}<1,\,\,\,\, q_{1}>\frac{2}{W_{0}[-1/(e^{1}q_{3})]}.\]
 Numerical solutions show that in this last case the universe will
end in a big rip singularity, where~$F,\dot{H}\rightarrow\infty$
and~$R\rightarrow0$. A big crunch ($F\rightarrow-\infty$) is not
possible for negative~$q_{1}$ because it contradicts~$R\rightarrow0$.
This can also be seen from the time-derivative of the function~$K_{0}F$
given by\[
K_{0}\dot{F}=q_{1}\frac{\dot{R}}{R}=q_{1}\left(H-\frac{1}{R}\right)\!,\,\,\, q_{1}<0.\]
This becomes positive for~$H<0$ (big crunch), thus preventing a
further decrease of~$K_{0}F$ and the final collapse of the scale
factor. 

More on the numerical solutions of this case can be found in Sec.~\ref{sec:RGE1-Ldot-Gdot}
and Ref.~\cite{Bauer:2005rp}, where additionally the running of
Newton's constant is taken into account.

\subsection{\label{sub:RGE1-B2}$\boldsymbol{\Lambda=\Lambda_{0}(L_{0}+L_{1}\frac{\mu^{2}}{\mu_{0}^{2}})}$,
$\boldsymbol{\mu=R^{-1}}$}

\begin{figure}[t]
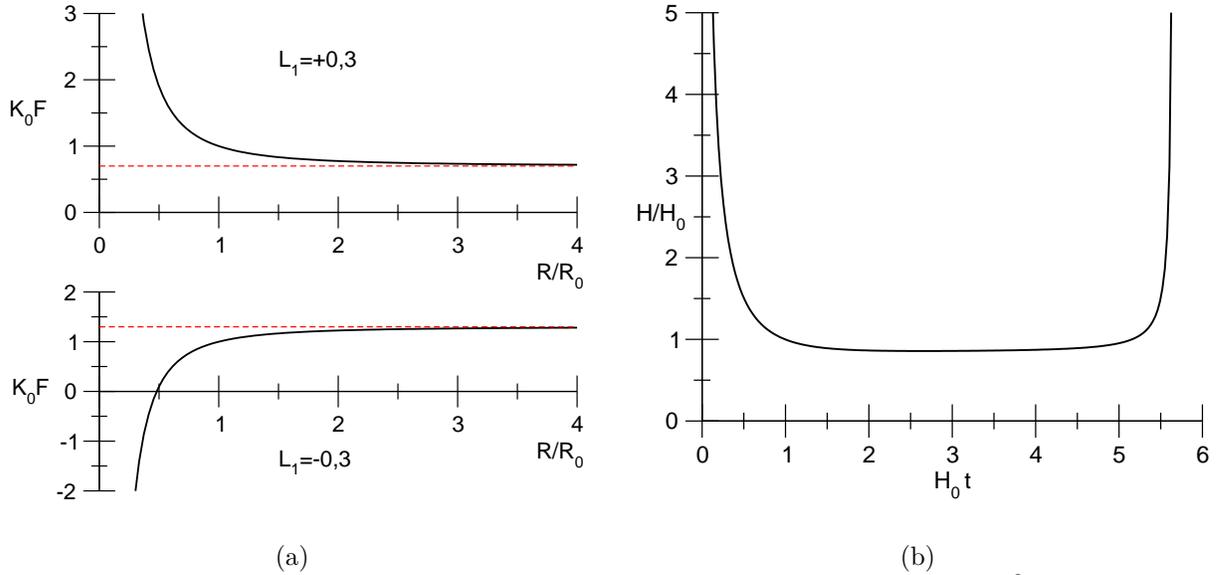

\begin{centering}\subfigure[]{\includegraphics[clip,width=0.48\columnwidth,keepaspectratio]{Scaling-B2-K0FvonR}}\hfill{}\subfigure[]{\includegraphics[clip,width=0.48\columnwidth,keepaspectratio]{Scaling-B2-BR-H}}\vspace*{-0.2cm}\par\end{centering}

\caption{\label{fig:RGE1-B2-KoF-BR}Corresponding to Sec.~\ref{sub:RGE1-B2}
we plot (a) the function $K_{0}F=L_{0}+L_{1}\frac{R_{0}^{2}}{R^{2}}$
for $L_{1}=\pm0,3$ and (b) the Hubble scale $H(t)$ for~$L_{1}=+0,1$;~$R_{0}=1,1$;~$\Omega_{\Lambda0}=0,73$
and~$Q=\frac{1}{2}$ (dust), where a big rip singularity occurs.}
\end{figure}

In the de~Sitter limit,~$\dot{H}=0$ and~$R\rightarrow R_{\textrm{e}}=H_{\textrm{e}}^{-1}$,
we find here the solution \[
R_{\textrm{e}}^{2}=(H_{0}^{-2}\Omega_{\Lambda0}^{-1}-L_{1}R_{0}^{2})/L_{0},\]
 but it is unstable for all possible values of~$L_{1}$, which follows
from \[
\left.\frac{\textrm{d}(K_{0}\dot{F})}{\textrm{d}(K_{0}F)}\right|_{\textrm{dS}}=R_{\textrm{e}}H_{0}^{2}L_{0}\Omega_{\Lambda0}>0.\]
 For~$L_{1}<0$ our numerical calculations did not yield any solution
that is compatible with Eqs.~(\ref{eq:RGE1-Scale2}) and~(\ref{eq:RGE1-MainEqu}).
This is also true in a certain parameter range for~$L_{1}>0$, where
otherwise a Big Rip singularity occurs, where $F,\dot{H}\rightarrow\infty$
and~$R\rightarrow0$, see Fig.~\ref{fig:RGE1-B2-KoF-BR}. A big
crunch is not possible since~$K_{0}F$ is bounded from below.%

\subsection{\label{sub:RGE1-C2}$\boldsymbol{\Lambda G=\Lambda_{0}G_{0}\frac{\mu^{2}}{\mu_{0}^{2}}}$,
$\boldsymbol{\mu=R^{-1}}$}

Here, one cannot find a prediction for~$R_{\textrm{e}}$ in the de~Sitter
limit ($\dot{H}=0$, $R\rightarrow R_{\textrm{e}}=H_{\textrm{e}}^{-1}$)
since Eq.~(\ref{eq:RGE1-MainEqu}) only leads to the constraint~$R_{0}^{2}=1/(H_{0}^{2}\Omega_{\Lambda0})$.
To find solutions corresponding to other values of~$R_{0}$ we impose
the ansatz~$a(t)=a_{0}(t-t_{1})^{n}$ with~$a_{0},n,t_{1}=\textrm{const}$.
The Hubble scale is given by~$H=n(t-t_{1})^{-1}$ which implies due
to~$H_{0}>0$ that~$t_{1}<t$ for~$n>0$ and~$t_{1}>t$ for~$n<0$,
respectively. For~$0<n<1$ the event horizon does not exist, in the
other cases~$R$ can be calculated exactly via Eq.~(\ref{eq:RGE1-Scale2}):\[
R(t)=a(t)\int_{t}^{t_{\textrm{E}}}\frac{\textrm{d}t^{\prime}}{a(t^{\prime})}=(t-t_{1})^{n}\cdot\frac{(t_{\textrm{E}}-t_{1})^{(1-n)}-(t-t_{1})^{(1-n)}}{1-n}.\]
 Note that for negative~$n$ there is a big rip singularity at~$t\rightarrow t_{1}$,
where~$a(t)$ diverges and the universe ends: $t_{\textrm{E}}=t_{1}$.
For~$n>1$ the scale factor~$a(t)$ describes power-law acceleration
and there is no future singularity which means~$t_{\textrm{E}}\rightarrow\infty$.
As a result, both cases yield \[
R(t)=\frac{(t-t_{1})}{n-1}.\]
Thus Eq.~(\ref{eq:RGE1-MainEqu}) can be solved exactly, which determines
the constant~$n$, see Fig.~\ref{fig:RGE1-C2-C3-n12Vonx}:%
\begin{figure}[t]
\begin{centering}\includegraphics[clip,width=0.4\columnwidth,keepaspectratio]{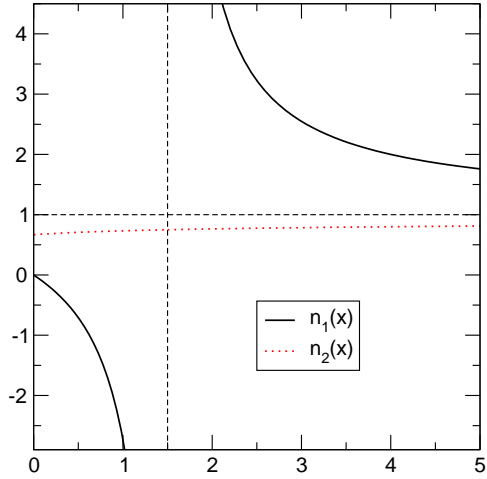}\par\end{centering}

\caption{\label{fig:RGE1-C2-C3-n12Vonx}The exponents~$n_{1,2}(x)$ from
Eq.~(\ref{eq:RGE1-C2-C3-n12Vonx}), which occur in the power-law
expansion scale factor~$a(t)=a_{0}(t-t_{1})^{n}$, are shown. In
Sec.~\ref{sub:RGE1-C2} we find $n=n_{1}(R_{0}^{2}/K_{0})$, whereas
$n=n_{2}(T_{0}^{2}/K_{0})$ in Sec.~\ref{sub:RGE1-C3}.}
\end{figure}
\begin{eqnarray}
n_{1,2}(x) & = & \frac{(x-\frac{1}{2})\pm\sqrt{(x-\frac{1}{2})^{2}-x(x-(Q+1))}}{x-(Q+1)}\label{eq:RGE1-C2-C3-n12Vonx}\\
\textrm{with}\,\, x & := & \frac{R_{0}^{2}}{K_{0}}=R_{0}^{2}H_{0}^{2}\Omega_{\Lambda0}(Q+1).\end{eqnarray}
For~$Q>0$ we find~$(Q+1)^{-1}<n_{2}<1$ for all positive values
of~$x$, therefore~$n_{2}$ can be dropped as the event horizon
does not exist. In the case~$R_{0}^{2}<1/(H_{0}^{2}\Omega_{\Lambda0})$
the constant~$n=n_{1}$ is negative leading to a big rip singularity
at the time~$t=t_{1}$, and~$R_{0}^{2}>1/(H_{0}^{2}\Omega_{\Lambda0})$
implies a positive~$n$ corresponding to power-law acceleration.

\subsection{\label{sub:RGE1-A3}$\boldsymbol{\Lambda=\Lambda_{0}(1-q_{1}\ln\frac{\mu}{\mu_{0}})}$,
$\boldsymbol{\mu=T^{-1}}$}

Solving Eq.~(\ref{eq:RGE1-MainEqu}) for this choice of~$\Lambda$
and~$\mu$ can be done approximately. First we propose the ansatz~\begin{equation}
a(t)=a_{0}\exp[u^{2}(t-t_{1})^{2}]\label{eq:RGE1-A3-exp-plus}\end{equation}
 for the scale factor with the constants~$a_{0}$,~$u$ and~$t_{1}$.
Therefore\[
H=2u^{2}(t-t_{1}),\,\,\,\,\frac{\ddot{a}}{a}=4u^{4}(t-t_{1})^{2}+2u^{2},\,\,\,\, F=2u^{2}+4u^{4}(Q+1)(t-t_{1})^{2},\]
and the particle horizon radius at late times reads\begin{eqnarray*}
T & = & \exp(u^{2}(t-t_{1})^{2})\int_{t_{x}}^{t}\exp(-u^{2}(t^{\prime}-t_{1})^{2})\textrm{d}t^{\prime},\,\,\,\, t>t_{x}=\textrm{const.}\\
 & = & \exp(u^{2}(t-t_{1})^{2})\cdot\left[\frac{\sqrt{\pi}}{2u}\textrm{erf}(u(t-t_{1}))\right]_{t_{x}}^{t}.\end{eqnarray*}
For~$t\rightarrow\infty$ the series expansion of the square brackets
reads~$C+\mathcal{O}(t^{-1})$ with~$C>0$ and thus~$T\approx C\exp(u^{2}(t-t_{1})^{2})$.
In this limit the given ansatz for~$a(t)$ solves~$K_{0}F=1-q_{1}\ln\frac{T_{0}}{T}$
exactly and determines the constant~$u^{2}=\frac{1}{4}q_{1}H_{0}^{2}\Omega_{\Lambda0}$.
Therefore we have found that Eq.~(\ref{eq:RGE1-A3-exp-plus}) represents
an approximate late-time solution in the case~$q_{1}>0$, which we
call super-exponential accelerated expansion.

\begin{figure}[t]
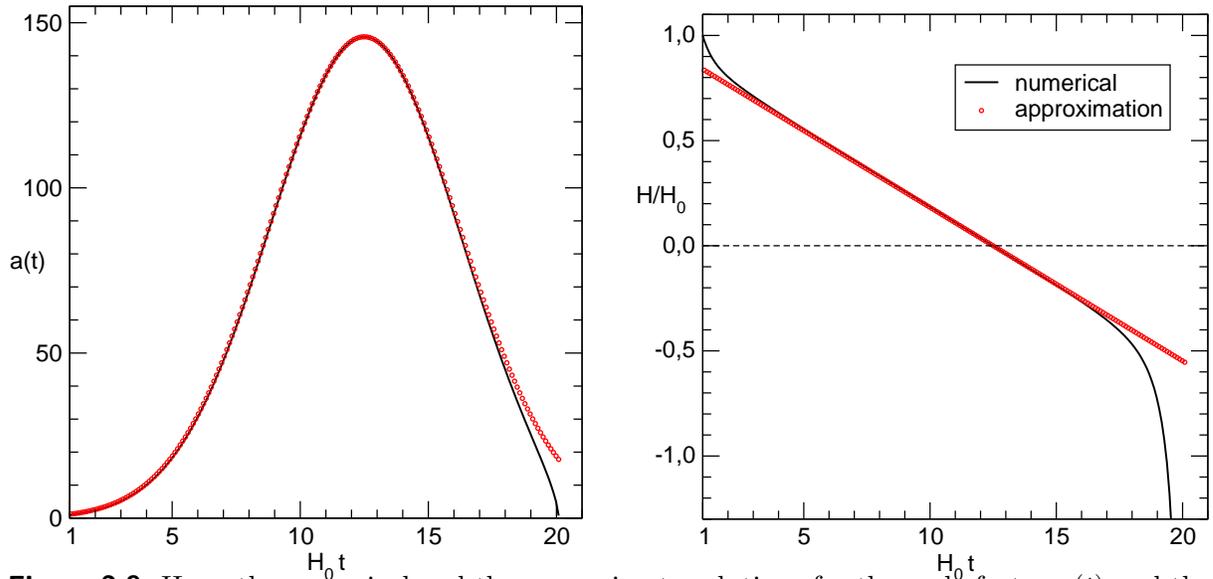

\begin{centering}\includegraphics[clip,width=0.48\columnwidth,keepaspectratio]{Scaling-A3-q1neg-a}\hfill{}\includegraphics[clip,width=0.48\columnwidth,keepaspectratio]{Scaling-A3-q1neg-H}\par\end{centering}

\caption{\label{fig:RGE1-A3-q1neg-a-H}Here, the numerical and the approximate
solutions for the scale factor~$a(t)$ and the Hubble scale~$H(t)$
from Sec.~\ref{sub:RGE1-A3} are shown for negative~$q_{1}=-0,2$.
Furthermore, we have used~$\Omega_{\Lambda0}=0,73$; $T_{0}H_{0}=3$
and~$Q=\frac{1}{2}$ (dust). In this case a big crunch singularity
will occur in the future. Note that the analytical approximation from
Eq.~(\ref{eq:RGE1-A3-exp-minus}) works very well except near the
future singularity.}
\end{figure}

The numerical solutions in the case~$q_{1}<0$ exhibits a future
big crunch singularity. However, at times sufficiently before this
event the scale factor can be well approximated by the ansatz~\begin{equation}
a(t)=a_{0}\exp[-u^{2}(t-t_{1})^{2}],\label{eq:RGE1-A3-exp-minus}\end{equation}
which implies~$F=-2u^{2}+4u^{4}(Q+1)(t-t_{1})^{2}$ and\[
T=\exp(-u^{2}(t-t_{1})^{2})\cdot\left[\frac{\sqrt{\pi}}{2iu}\textrm{erf}(iu(t-t_{1}))\right]_{t_{x}}^{t},\,\,\,\, t>t_{x}=\textrm{const}.\]
We expand the square bracket term around~$t\rightarrow t_{1}$, where
the scale factor is maximal, and find~$T\approx C\exp(-u^{2}(t-t_{1})^{2})$
with~$C>0$. Solving~$K_{0}F=1-q_{1}\ln\frac{T_{0}}{T}$ leads to~$u^{2}=-\frac{1}{4}q_{1}H_{0}^{2}\Omega_{\Lambda0}$
in accordance with~$q_{1}<0$. The numerical and approximate solutions
are shown in Fig.~\ref{fig:RGE1-A3-q1neg-a-H}.%

\subsection{\label{sub:RGE1-B3}$\boldsymbol{\Lambda=\Lambda_{0}(L_{0}+L_{1}\frac{\mu^{2}}{\mu_{0}^{2}})}$,
$\boldsymbol{\mu=T^{-1}}$}

In the late-time de~Sitter limit,~$a(t)\propto\exp(H_{\textrm{e}}t)$,
the particle horizon radius has the asymptotic form\[
T\propto\exp(H_{\textrm{e}}t)\rightarrow\infty\,\,\,\,\textrm{for}\,\,\,\, t\rightarrow\infty,\]
which follows from Eq.~(\ref{eq:RGE1-T-dS}). This means that~$K_{0}F\rightarrow L_{0}$
and thus determines~$H_{\textrm{e}}^{2}=H_{0}^{2}\Omega_{\Lambda0}L_{0}$.
To verify the stability of this solution we calculate\[
K_{0}\dot{F}=-2L_{1}\frac{\dot{T}}{T}\cdot\frac{T_{0}^{2}}{T^{2}}=-2L_{1}\frac{T_{0}^{2}}{T^{2}}\left(H+\frac{1}{T}\right)\!\!.\]
At the initial time (today)~$K_{0}\dot{F}$ is negative for~$L_{1}>0$
and $K_{0}F$ keeps on approaching the de~Sitter limit unless a sign
change in~$K_{0}\dot{F}$ would occur. But this requires a negative~$H$,
which is only possible if~$K_{0}F=\dot{H}+(Q+1)H^{2}<0$ at some
time. Since~$K_{0}F\ge L_{0}>0$ we conclude that the de~Sitter
limit will be reached always. For~$L_{1}<0$ the initial values of~$K_{0}F$
and~$K_{0}\dot{F}$ are positive and~$K_{0}F$ approaches the de~Sitter
limit. Again, a sign change in~$K_{0}\dot{F}$ is only possible for~$H<0$,
which requires~$K_{0}F<0$ at some time. These conditions cannot
be realised because~$K_{0}F>0$ as long as~$K_{0}\dot{F}>0$. In
summary all reasonable values of~$L_{1}$ lead to a stable de~Sitter
final state.

\subsection{\label{sub:RGE1-C3}$\boldsymbol{\Lambda G=\Lambda_{0}G_{0}\frac{\mu^{2}}{\mu_{0}^{2}}}$,
$\boldsymbol{\mu=T^{-1}}$}

By using at late times the power-law ansatz~$a(t)=a_{0}(t-t_{1})^{n}$
with~$a_{0},t,n=\textrm{const.}$, we find\[
F=(n^{2}(Q+1)-n)(t-t_{1})^{-2},\]
 and the particle horizon radius is given by\begin{equation}
T=a_{0}(t-t_{1})^{n}\left[\int_{0}^{t_{0}}\frac{\textrm{d}t^{\prime}}{a(t^{\prime})}+\int_{t_{0}}^{t}\frac{\textrm{d}t^{\prime}}{a(t^{\prime})}\right]=\frac{t-t_{1}}{1-n}+(t-t_{1})^{n}\left[T_{0}-\frac{(t_{0}-t_{1})^{1-n}}{1-n}\right]\!\!,\label{eq:RGE1-C3-T}\end{equation}
where~$T_{0}=T(t_{0})>0$ depends on the past evolution of the scale
factor. In the case~$n<0$ a positive Hubble scale~$H=n(t-t_{1})^{-1}$
requires~$t,t_{0}<t_{1}$, which implies~$T<0$ and thus the non-existence
of the particle horizon. For~$0<n<1$ the second term in Eq.~(\ref{eq:RGE1-C3-T})
can be neglected at late times, and Eq.~(\ref{eq:RGE1-MainEqu})
can be solved exactly, leading to \begin{eqnarray*}
n_{1,2}(x) & = & \frac{(x-\frac{1}{2})\pm\sqrt{(x-\frac{1}{2})^{2}-x(x-(Q+1))}}{x-(Q+1)}\\
\textrm{with}\,\, x & := & \frac{T_{0}^{2}}{K_{0}}=T_{0}^{2}H_{0}^{2}\Omega_{\Lambda0}(Q+1).\end{eqnarray*}
This is the same equation as in Sec.~\ref{sub:RGE1-C2} with~$R_{0}$
replaced by~$T_{0}$, see also Fig.~\ref{fig:RGE1-C2-C3-n12Vonx}.
Here, solution~$n_{2}$ is the physical one because of~$(Q+1)^{-1}<n_{2}<1$,
that corresponds to a decelerating universe for~$t\rightarrow\infty$.
In comparison with Sec.~\ref{sub:RGE1-C2} the identification of
the particle horizon radius as renormalisation scale yields a complementary
cosmological behaviour.

\section{Time-dependent Cosmological and Newton's Constant\label{sec:RGE1-Ldot-Gdot}}

In the previous sections we have discussed only the running of the
CC except for the scaling laws~(\ref{eq:RGE1-RGE3}), where the change
of Newton's constant is comparable with that of the CC. For the scaling
law~(\ref{eq:RGE1-RGE1}) we have required that the mass parameter~$q_{1}$
is small in order to obtain a viable phenomenology. Since the largest
known field masses are of the order of~$10^{2}\,\textrm{GeV}$ we
had to assume fine-tuning or some unknown suppression mechanism to
achieve the smallness of~$q_{1}$. What happens to Newton's constant
when it is controlled by the RGE~(\ref{eq:RGE1-RGE1-G})? As we can
see from Eq.~(\ref{eq:RGE1-q2Def}), the corresponding mass parameter~$q_{2}$
is of the order~$m^{2}G_{0}=m^{2}/M_{\textrm{Pl}}^{2}$ and therefore
strongly suppressed by the Planck mass~$M_{\textrm{Pl}}$ even for
larger masses~$m$. Obviously, the running of~$G$ is suppressed
from the beginning, which agrees with the strong bounds on the time-variation
of~$G$, see, e.g., Ref.~\cite{Gdot}. Additionally, this property
has the advantage, that today we are far away from the Landau pole
of~$G(\mu)$, where the function~$F$ diverges in Eq.~(\ref{eq:RGE1-MainEqu}).
Since the RGE for~$G$ follows directly from the effective action~(\ref{eq:RGE1-Seff}),
we are interested in its influence on the cosmological evolution. 

\begin{figure}[H]
\begin{centering}\includegraphics[clip,width=0.85\textwidth,height=19cm]{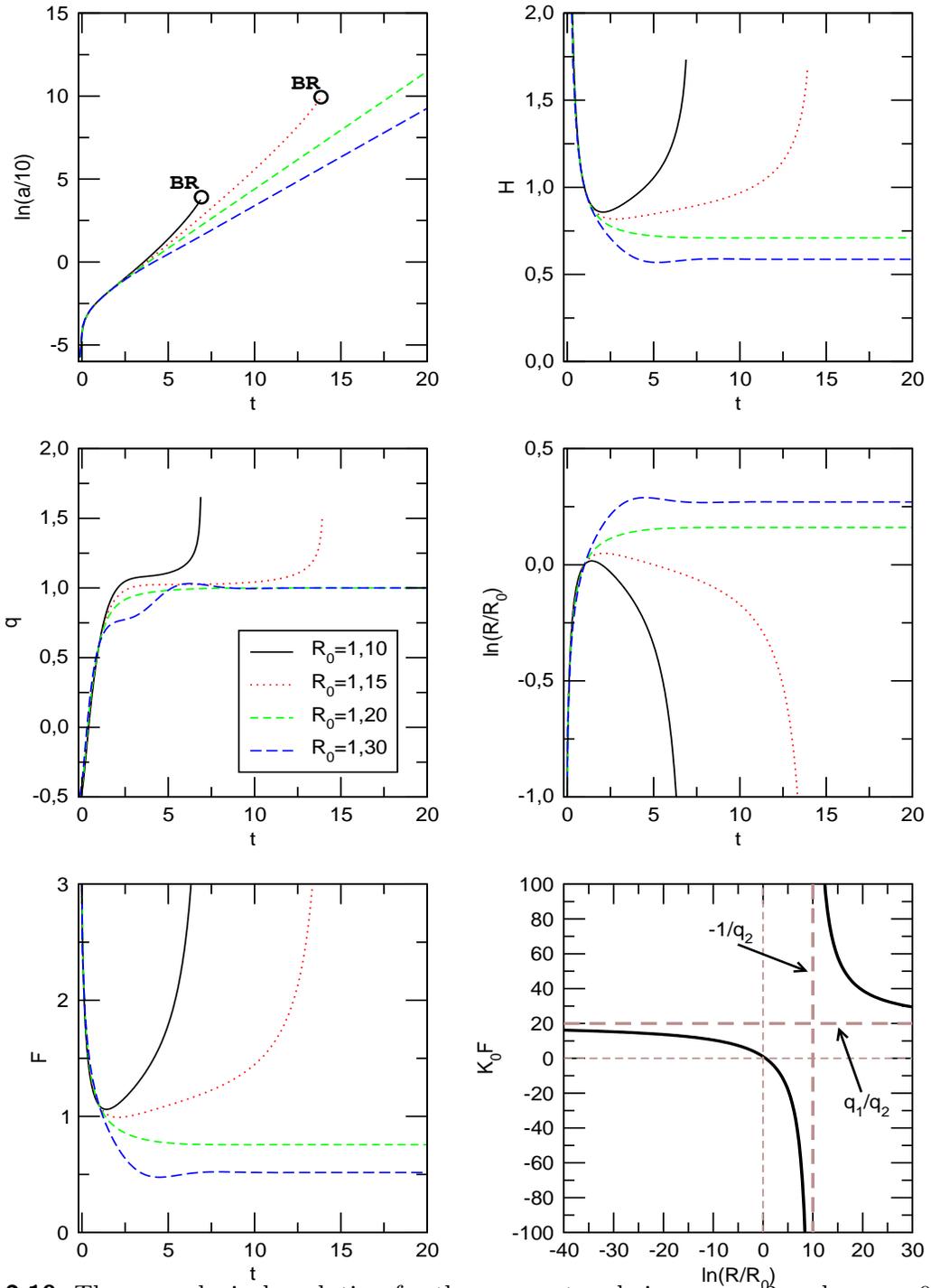}\par\end{centering}

\caption{\label{fig:RGE1-q1min2-q2min01}The cosmological evolution for the
parameter choice~$q_{1}=-2$ and~$q_{2}=-0,1$ and different values
of the initial event horizon radius~$R_{0}$. The fate of this universe
is either a stable de~Sitter state when choosing~$R_{0}=1,20;\,1,30$,
or a big rip~(\texttt{BR}) in the case~$R_{0}=1,10;\,1,15$. $K_{0}F$
is bounded from above. Nomenclature: Scale factor~$a$, Hubble scale~$H=\frac{\dot{a}}{a}$,
acceleration~$q=\frac{\ddot{a}a}{\dot{a}^{2}}$, event horizon radius~$R$
and its initial value~$R_{0}$. For the function~$K_{0}F$ see Eqs.~(\ref{eq:RGE1-K0FofR}),
and~(\ref{eq:RGE1-K0def}), for the mass parameters~$q_{1},q_{2}$
see Eqs.~(\ref{eq:RGE1-q1Def}) and~(\ref{eq:RGE1-q2Def}).}
\end{figure}

\begin{figure}[H]
\begin{centering}\includegraphics[clip,width=0.85\textwidth,height=19cm]{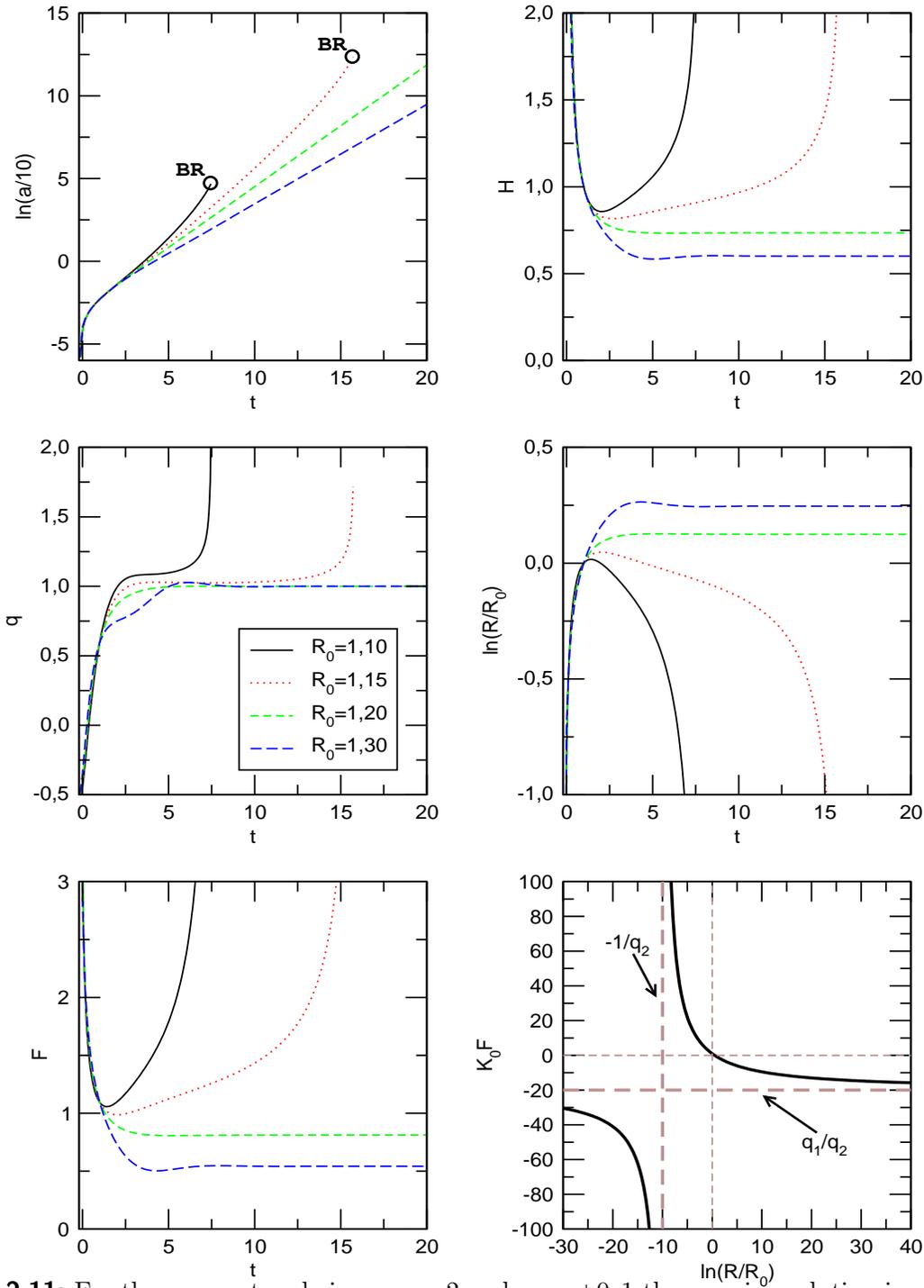}\par\end{centering}

\caption{\label{fig:RGE1-q1min2-q2plu01}For the parameter choice~$q_{1}=-2$
and~$q_{2}=+0,1$ the cosmic evolution is not very different from
the case~$q_{2}=-0,1$ (Fig.~\ref{fig:RGE1-q1min2-q2min01}). In
the future, there is either a stable de~Sitter state for~$R_{0}=1,20;\,1,30$,
or a big rip~(\texttt{BR}) when~$R_{0}=1,10;\,1,15$. $K_{0}F$
is bounded from below. Nomenclature: Scale factor~$a$, Hubble scale~$H=\frac{\dot{a}}{a}$,
acceleration~$q=\frac{\ddot{a}a}{\dot{a}^{2}}$, event horizon radius~$R$
and its initial value~$R_{0}$. For the function~$K_{0}F$ see Eqs.~(\ref{eq:RGE1-K0FofR})
and~(\ref{eq:RGE1-K0def}), for the mass parameters~$q_{1},q_{2}$
see Eqs.~(\ref{eq:RGE1-q1Def}) and~(\ref{eq:RGE1-q2Def}).}
\end{figure}

\begin{figure}[H]
\begin{centering}\includegraphics[clip,width=0.85\textwidth,height=19cm]{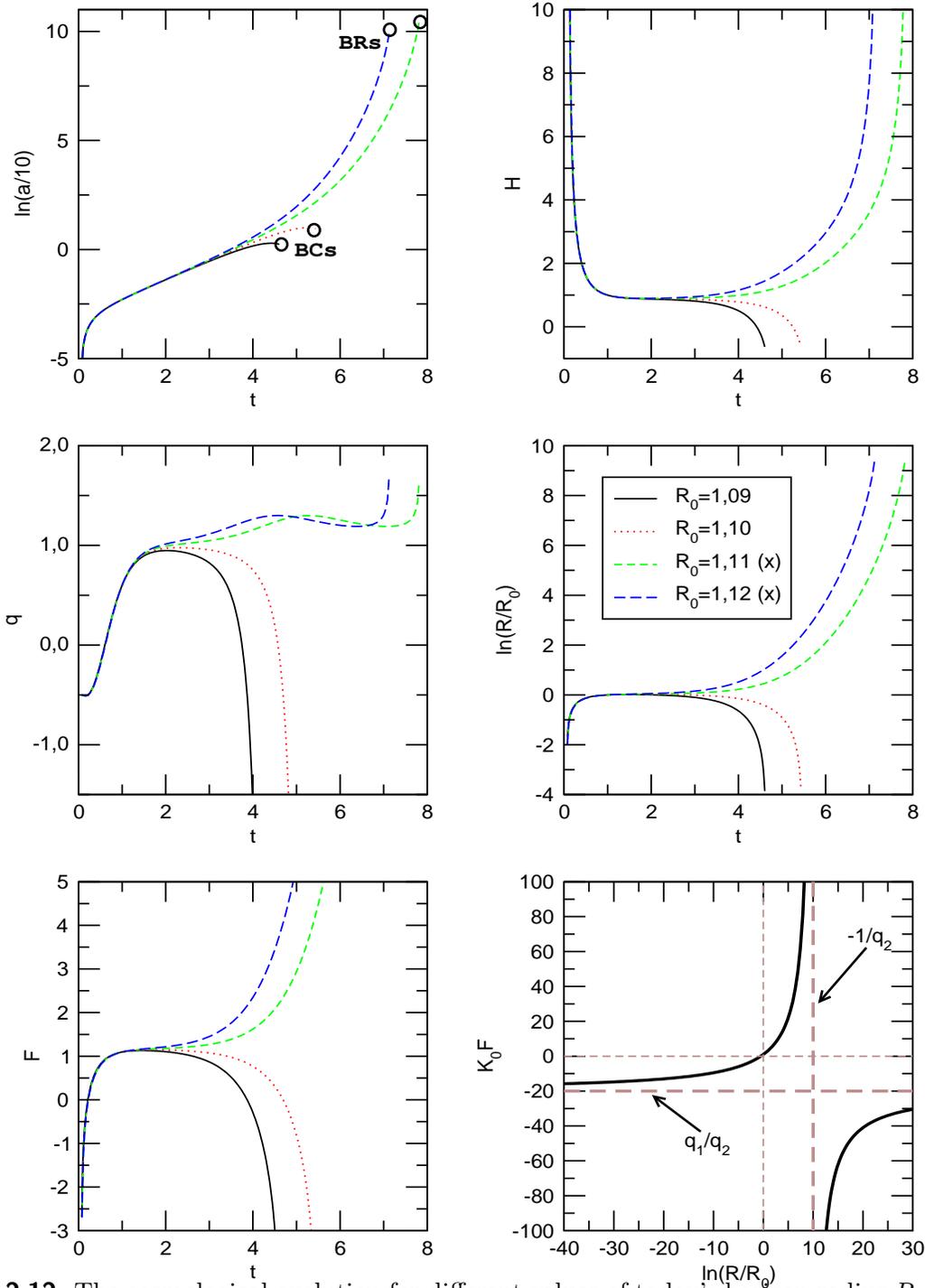}\par\end{centering}

\caption{\label{fig:RGE1-q1plu2-q2min01}The cosmological evolution for different
values of today's horizon radius~$R_{0}$ for the case~$q_{1}=+2$
and~$q_{2}=-0,1$. The solutions for~$R_{0}=1,09;\,1,10$ exhibit
a big crunch~(\texttt{BC}), whereas the initial conditions~$R_{0}=1,11;\,1,10$
lead to a big rip~(\texttt{BR}). $K_{0}F$ is bounded from below.
The numerical solutions marked by~(\texttt{x}) are not compatible
with the main equation~(\ref{eq:RGE1-K0FofR}), see Secs.~\ref{sec:RGE1-Late-time}
and~\ref{sec:RGE1-Ldot-Gdot} for further details. Nomenclature:
Scale factor~$a$, Hubble scale~$H=\frac{\dot{a}}{a}$, acceleration~$q=\frac{\ddot{a}a}{\dot{a}^{2}}$,
event horizon radius~$R$ and its initial value~$R_{0}$. For the
function~$K_{0}F$ see Eqs.~(\ref{eq:RGE1-K0FofR}) and~(\ref{eq:RGE1-K0def}),
for the mass parameters~$q_{1},q_{2}$ see Eqs.~(\ref{eq:RGE1-q1Def})
and~(\ref{eq:RGE1-q2Def}).}
\end{figure}

\begin{figure}[H]
\begin{centering}\includegraphics[clip,width=0.85\textwidth,height=19cm]{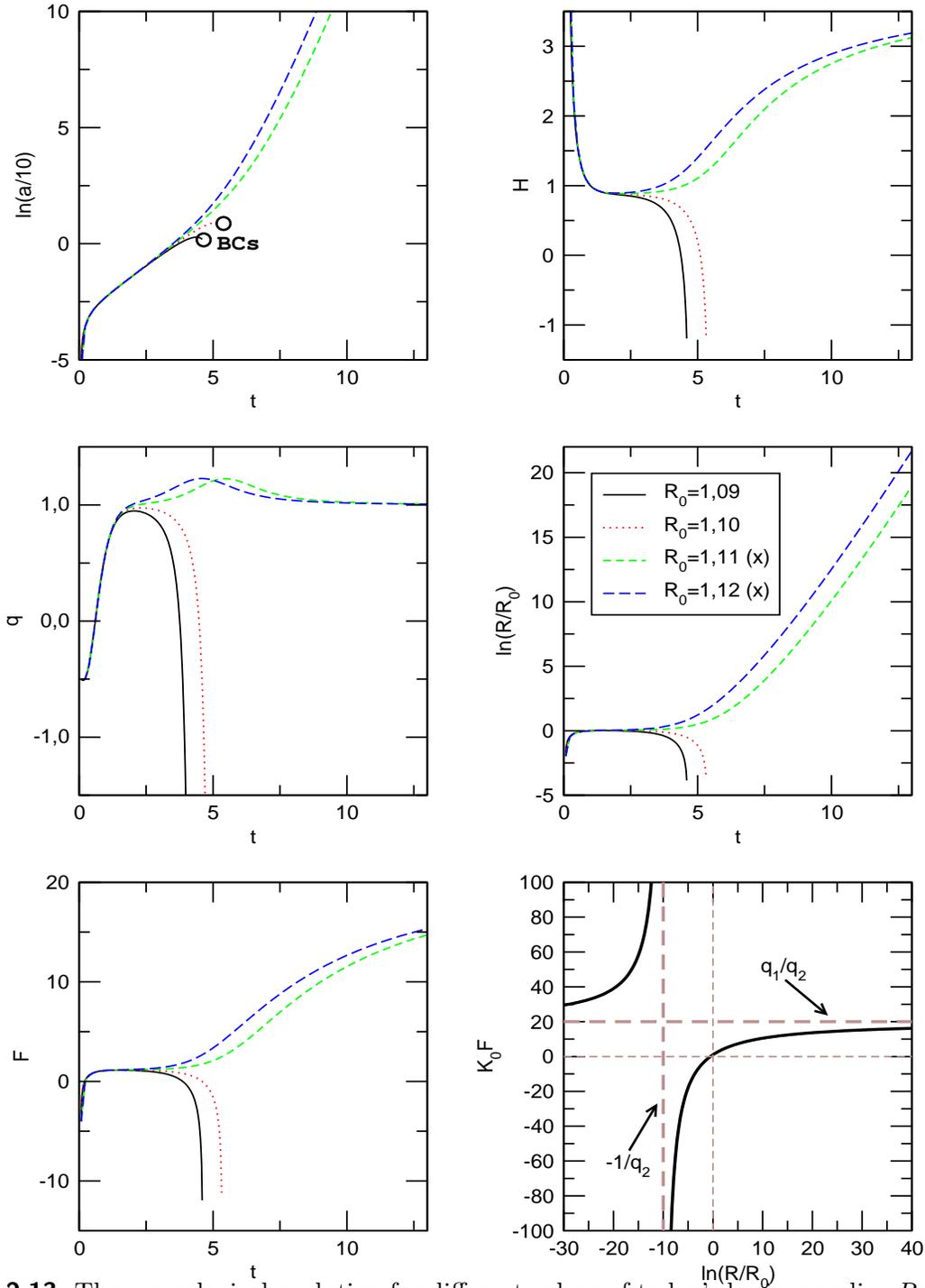}\par\end{centering}

\caption{\label{fig:RGE1-q1plu2-q2plu01}The cosmological evolution for different
values of today's horizon radius~$R_{0}$. Here, we choose~$q_{1}=+2$
and~$q_{2}=+0,1$. The solutions for~$R_{0}=1,09;\,1,10$ exhibit
a big crunch~(\texttt{BC}), where~$K_{0}F$ is unbounded from below.
For the initial conditions~$R_{0}=1,11;\,1,10$ the function~$F$
and the Hubble scale~$H$ approach a finite value, where the horizon
radius~$R$ diverges. The numerical solutions marked by~(\texttt{x})
are not compatible with the main equation~(\ref{eq:RGE1-K0FofR}),
see Secs.~\ref{sec:RGE1-Late-time} and~\ref{sec:RGE1-Ldot-Gdot}
for further details. Nomenclature: Scale factor~$a$, Hubble scale~$H=\frac{\dot{a}}{a}$,
acceleration~$q=\frac{\ddot{a}a}{\dot{a}^{2}}$, event horizon radius~$R$
and its initial value~$R_{0}$. For the function~$K_{0}F$ see Eqs.~(\ref{eq:RGE1-K0FofR})
and~(\ref{eq:RGE1-K0def}), for the mass parameters~$q_{1},q_{2}$
see Eqs.~(\ref{eq:RGE1-q1Def}) and~(\ref{eq:RGE1-q2Def}).}
\end{figure}

As an example we will discuss the combination of the RGEs~(\ref{eq:RGE1-RGE1})
for the CC and~(\ref{eq:RGE1-RGE1-G}) for~$G$ that were both derived
from the effective action~(\ref{eq:RGE1-Seff}). As renormalisation
scale~$\mu$ we use, according to Eq.~(\ref{eq:RGE1-Scale2}), the
inverse of the cosmological event horizon, $\mu=R^{-1}$. With these
preliminaries Eq.~(\ref{eq:RGE1-MainEqu}) reads\begin{equation}
K_{0}F=\frac{1+q_{1}\ln\frac{R}{R_{0}}}{1+q_{2}\ln\frac{R}{R_{0}}},\label{eq:RGE1-K0FofR}\end{equation}
leading to the ordinary differential equation~(\ref{eq:RGE1-ODE-a})
for the scale factor~$a(t)$. Unfortunately, finding explicit solutions
of this equation seems to be rather difficult because of the strongly
non-linear form of the equation. Therefore, we solve it numerically
to show the characteristic future cosmic evolution corresponding to
four different cases, which result from the parameter choices~$q_{1}=\pm2$
and~$q_{2}=\pm0,1$. These values for~$q_{1}$ mean that the relevant
mass scale~$m$ should be near~$\Lambda_{0}^{1/4}\sim10^{-3}\,\textrm{eV}$.
Actually, the only known particles with such a low mass are neutrinos.
This indicates that the influence of higher mass fields is suppressed,
or these fields have decoupled, respectively. Note that due to the
suppression by the Planck scale, the realistic value of~$q_{2}$
should be much lower than~$\pm0,1$. Here, we used a large value
for~$q_{2}$ just to explore the differences due to the sign of~$q_{2}$.
Concerning the differential equation, we fix the initial conditions
by using observational results, i.e., the relative vacuum energy density
is given by~$\Omega_{\Lambda0}=0,73$, no spatial curvature ($\Omega_{k}=0$)
and only dust (with an equation of state parameter~$Q=0,5$) and
the CC as relevant energy forms in the present-day universe. The acceleration
parameter~$q_{0}=\frac{\ddot{a}a}{\dot{a}^{2}}(t_{0})$ is determined
by Eq.~(\ref{eq:RGE1-Friedmann3}): $q=\Omega_{\Lambda}(1+Q)+Q(\Omega_{k}-1)$.
Today's value of the horizon radius~$R_{0}$ is unknown, so we have
to estimate it. Since it should be the largest physical length scale
and the universe seems to be almost de~Sitter-like, we assume the
horizon radius to be a bit larger than the inverse Hubble scale,~$R_{0}\gtrsim H_{0}^{-1}$.
Finally, all dimensionful quantities are expressed in terms of today's
Hubble scale~$H_{0}$ (Hubble units). Figures~\ref{fig:RGE1-q1min2-q2min01}--\ref{fig:RGE1-q1plu2-q2plu01}
show the numerical results for different values of the initial radius~$R_{0}$
of the event horizon. The graphs in each of the four figures illustrate
the scale factor~$a(t)$, the Hubble scale~$H(t)$, the acceleration~$q(t)$,
the event horizon radius~$R(t)$, and~$F(t)$ as functions of the
cosmic time~$t$, respectively. The last graph displays Eq.~(\ref{eq:RGE1-K0FofR}),
i.e.,~$K_{0}F$ as a function of the radius~$R/R_{0}$.

The first observation from the numerical solutions is, that for a
positive value of~$q_{1}$ the cosmic age decreases with respect
to the age~$t_{0}$ of the standard $\Lambda$CDM universe, whereas
for a negative~$q_{1}$ the age increases. In the latter case, we
observe only big rip solutions and de~Sitter final states. However,
for positive~$q_{1}$ a big crunch may also occur, but no stable
de~Sitter states exist. For positive values of~$q_{1}$ and~$q_{2}$
(see Fig.~\ref{fig:RGE1-q1plu2-q2plu01}), we have not observed any
big rip solutions. Hence, the final state may be either a big crunch
or a forever expanding universe, where the Hubble scale approaches
a finite positive value, but the event horizon radius~$R$ goes to
infinity. This is a contradiction, because an asymptotically constant
Hubble scale~$H>0$ implies a finite event horizon radius~$R\approx H^{-1}$
in the far future, which is not the case here. Obviously, this numerical
solution is not a solution of the original equation~(\ref{eq:RGE1-MainEqu}).
For~$q_{1}>0$ and~$q_{2}<0$ (see Fig.~\ref{fig:RGE1-q1plu2-q2min01})
the big rip events in the numerical solutions occur at a finite and
large value of the horizon radius~$R$. Again, this behaviour is
not compatible with the vanishing of the horizon radius at such an
event. Therefore, we can reject these numerical solutions, too. For
positive~$q_{1}$ we thus observe that if a solution exists it has
to be a big crunch. 

In summary, we conclude that the cosmological final states occurring
in the numerical solutions in this section do not differ from the
case with constant~$G$, discussed in Sec.~\ref{sub:RGE1-A2}. 

\newpage

\newpage

\newpage

\newpage

\newpage

\section{Summary}

Before we summarise the cosmological final states that we have found
in the previous sections, we have to mention that not all of the solutions
are likely to be realised even if the underlying assumptions were
correct. This is especially important for the extreme solutions that
exhibit future singularities. For these cases the strength of the
gravitational field becomes so large that other effects, e.g., higher
orders in the curvature scalar or unknown quantum gravity effects,
cannot be neglected anymore. Also matter sources like radiation would
become again dominant over dust matter in big crunch scenarios. Therefore,
it is possible that these cosmological final states are replaced by
other solutions. For instance, the avoidance of a big rip in phantom
cosmologies by such effects has been discussed in Refs.~\cite{BRip-QuCorr}.
On the other hand, in Ref.~\cite{BRip-QuCorr-2} one has an example,
where quantum effects were not able to prevent a big rip. 

Nevertheless, we stay in this analysis on the basic level and use
the results we found as an indicator for the (in)stability of the
cosmological fate. Indeed, we have found regular solutions in many
cases, which can be considered as realisable in nature. In this sense,
we have found de~Sitter solutions in the cases of Secs.~\ref{sub:RGE1-A1},
\ref{sub:RGE1-B1}, \ref{sub:RGE1-A2} and~\ref{sub:RGE1-B3}. Also
accelerating and decelerating power-law solutions for the scale factor
can be called regular, we have found them for all cases with the scaling
law~(\ref{eq:RGE1-RGE3}) in Secs.~\ref{sub:RGE1-C1}, \ref{sub:RGE1-C2}
and~\ref{sub:RGE1-C3}. Moreover, we have found in Sec.~\ref{sub:RGE1-A3}
a super-exponential expansion law of the type~$a(t)\sim\exp(ct^{2})$,
which however implies negative values of the matter energy density.
Future singularities of the big rip type are found for all cases with
the event horizon as renormalisation scale, see Secs.~\ref{sub:RGE1-A2},
\ref{sub:RGE1-B2} and~\ref{sub:RGE1-C2}. Big crunch solutions,
on the other hand, occur only for the scaling law~(\ref{eq:RGE1-RGE1})
as described in Secs.~\ref{sub:RGE1-A1}, \ref{sub:RGE1-A2} and
\ref{sub:RGE1-A3}. In Sec.~\ref{sec:RGE1-Ldot-Gdot} we have finally
analysed numerically the effect of a running Newton's constant~$G$,
but we did not observe any changes in the cosmological fates in comparison
to the cases with constant~$G$. To finish this chapter we give a
comprehensive overview of our results in Tab.~\ref{tab:RGE1-Final-States}.%
\begin{table}
\begin{centering}\begin{tabular}{|l||c|c|c|}
\hline 
&
$\Lambda=\Lambda_{0}(1-q_{1}\ln\frac{\mu}{\mu_{0}})$&
$\Lambda=\Lambda_{0}(L_{0}+L_{1}\frac{\mu^{2}}{\mu_{0}^{2}})$&
$\Lambda G=\Lambda_{0}G_{0}\frac{\mu^{2}}{\mu_{0}^{2}}$\tabularnewline
\hline
\hline 
$\mu=H$&
\texttt{dS}, \texttt{BC}&
\texttt{dS}&
\texttt{P}\tabularnewline
\hline 
$\mu=R^{-1}$&
\texttt{dS}, \texttt{BR}, \texttt{BC}&
\texttt{BR}&
\texttt{P}, \texttt{BR}\tabularnewline
\hline 
$\mu=T^{-1}$&
$\exp(\pm t^{2})$, \texttt{BC}&
\texttt{dS}&
\texttt{P}\tabularnewline
\hline
\end{tabular}\par\end{centering}

\caption{\label{tab:RGE1-Final-States}This table shows all types of cosmological
final states that we have found in Sec.~\ref{sec:RGE1-Late-time}
for the all combinations of scaling laws for the CC~$\Lambda$ (and~$G$
for the third law) from Sec.~\ref{sec:RGE1-Scaling-laws} and the
choices for the renormalisation scale~$\mu$ of Sec.~\ref{sec:RGE1-Choice-scale}.
Here, \texttt{dS} denotes de~Sitter solutions, \texttt{P} power-law
solutions, \texttt{BR} big rip and \texttt{BC} big crunch future singularities,
respectively. With~$\exp(\pm t^{2})$ we mean the late-time behaviour
of the scale factor, where~$a\propto\exp(-t^{2})$ finally leads
to a big crunch.}
\end{table}

\clearpage{\pagestyle{empty}\cleardoublepage}

\chapter{Vacuum Energy in Extra Dimensions\label{cha:DEED}}

\section{Introduction\label{sec:DEED-Intro}}

With this chapter we start to investigate the occurrence of dark energy
in models, where space-time has some non-trivial properties. One of
the most famous approaches for space-time modifications represents
the introduction of extra spatial dimensions. With the concept of
Kaluza-Klein~(KK) compactification~\cite{Kaluza:1921tu} one has
found a framework to unite gravity with electrodynamics. Decades later
EDs are still a popular way to obtain 4D theories from a simpler higher-dimensional
setup~\cite{Cremmer:1978ds}. In this approach, the 4D theory which
emerges after dimensional reduction is generally characterised by
a tower of KK modes \cite{Antoniadis:1990ew}. Also in string theories
EDs are a crucial ingredient~\cite{Antoniadis:1998ig} and are needed
for consistency reasons. Therefore EDs have become a common component
for modern model-building.

Another interesting aspect of compactified EDs is, that quantum fields
in such a non-trivial space-time give rise to the Casimir effect~\cite{Casimir:dh}.
In this scenario the bulk fields have to obey certain boundary conditions
thereby inducing a finite Casimir energy density, which depends on
the size and the topology of the EDs. The associated Casimir force
can be attractive and contract the compactified EDs to a size which
is sufficiently small so as to have escaped experimental detection
so far \cite{Appelquist:1982zs,CasimirAdS,Candelas:ae}. Upon integrating
out the EDs, the Casimir energies have additionally the interesting
property that they appear as an effective CC~$\Lambda$ or vacuum
energy in the 4D subspace. In the following chapters we will focus
mainly on this last point since the Casimir contributions to the vacuum
energy budget must be smaller than the observed value of the CC in
order to avoid fine-tuning. One can imagine that this requirement
leads to stringent bounds on the size of the ED. Later on in Chaps.~\ref{cha:DiED}
and~\ref{cha:VacD} we will investigate the Casimir effect for discretised
EDs and its applications in the context of deconstruction. To prepare
for this task we will first discuss the calculation the Casimir energy
density in the continuum. After that the results can be transferred
easily to the discretised case.

\section{The Casimir Effect\label{sec:DEED-Casimir}}

The Casimir effect is a notable exception from the normal ordering
procedure in quantum field theories. It occurs when quantum fields
have to obey certain boundary conditions, for instance the electric
component of the photon field, restricted between two parallel conducting
plates, has to vanish on the plates. This causes a geometry dependent
vacuum energy density inducing a force on the plates. Therefore, the
Casimir effect is a macroscopic quantum phenomenon, which is experimentally
well established~\cite{Sparnaay:1958wg}. For some recent reviews
of the effect and its applications see Refs.~\cite{Bordag:2001qi,Milton:2004ya}.
Global properties of a non-trivial space-time may also be probed by
the Casimir effect which is sensitive to the IR structure of a theory. 

Let us now consider the Casimir effect for a scalar field~$\phi$
in a 5D theory, where the ED is compactified on the circle~$\mathcal{S}^{1}$
with circumference~$R$. Hence, the whole manifold has the topology~$\mathcal{M}\times\mathcal{S}^{1}$,
where~$\mathcal{M}$ denotes Minkowski space. In order to find the
possible field theory configurations we have to specify the boundary
conditions for the scalar field on the circle. Motivated by periodicity
one obvious choice is given by\begin{equation}
\phi(y+R)=\phi(y),\label{eq:DEED-BC-untwist}\end{equation}
where~$y$ is the coordinate in the fifth dimension. This choice
implies a cylinder-like structure. But this not the only one. According
to Ref.~\cite{Isham:1978xxx} one can also form a structure with
the properties of a Möbius band, where $\phi$ obeys anti-periodic
boundary conditions\begin{equation}
\phi(y+mR)=(-1)^{m}\phi(y),\,\,\,\, m\in\mathbb{Z},\label{eq:DEED-BC-twist}\end{equation}
 In this case one must cycle twice through the circle~$\mathcal{S}^{1}$
to completely traverse the Möbius band. In the latter case, the field~$\phi$
is called a twisted field, whereas the fields with periodic boundary
conditions are called \emph{}untwisted fields. Locally, both cases
have the same product structure, but globally they differ significantly.
Since they yield inequivalent degrees of freedom of the field~$\phi$,
both must be considered in the Casimir effect.

Turning to the calculation of the Casimir energy, we call~$q$ the
momentum corresponding to the position~$y$ in the extra-dimensional
space. Since the 5D manifold is flat, it is sensible to use for the
scalar field a plane wave ansatz given by \begin{equation}
\phi(t,\vec{x},y)=J\cdot\exp(i\omega t-i\vec{p}\vec{x}-iqy),\label{eq:DEED-PlaneWave}\end{equation}
where~$J$ is a normalisation factor,~$\omega$ the energy and~$\vec{p}$
the 3-momentum corresponding to the time~$t$ and respectively the
3-coordinates~$\vec{x}$. As discussed above, the untwisted field
configuration is fixed by periodic boundary conditions,\[
\phi(y+R)=\phi(y)\,\,\,\,\Longrightarrow\,\,\,\, e^{-iqR}=1,\]
 implying a discrete momentum spectrum,\begin{equation}
q=2\pi\frac{n}{R},\,\,\,\, n\in\mathbb{Z}.\label{eq:DEED-qSpectUntwist}\end{equation}
For twisted fields we have to use anti-periodic boundary conditions\[
\phi(y+mR)=(-1)^{m}\phi(y),\,\,\,\, m\in\mathbb{Z}\,\,\,\,\Longrightarrow\,\,\,\, e^{-iqR}=-1,\]
which yield the discrete momentum spectrum\begin{equation}
q=2\pi\frac{(n-\frac{1}{2})}{R},\,\,\,\, n\in\mathbb{Z}.\label{eq:DEED-qSpectTwist}\end{equation}
Since this is the only difference which is relevant for the following
calculations, we will work with untwisted fields and replace~$n$
by~$n-1/2$ when needed. 

To normalise the field modes in Eq.~(\ref{eq:DEED-PlaneWave}), we
define the following scalar product for two modes $\phi_{1,2}(\vec{p},n)$
by\[
(\phi_{1},\phi_{2}):=i\int\textrm{d}^{3}x\,\int\textrm{d}y\,\left(\phi_{1}^{\dag}(\partial_{t}\phi_{2})-(\partial_{t}\phi_{1})^{\dag}\phi_{2}\right),\]
so that the normalisation factor~$J$ can be fixed by demanding the
orthonormality relation\begin{equation}
\left(\phi(\vec{p},n),\phi(\vec{p}^{\,\prime},n^{\prime})\right)=-V_{3}^{-1}\delta^{(3)}(\vec{p}-\vec{p}^{\,\prime})\delta_{nn^{\prime}},\label{eq:DEED-Orthonorm}\end{equation}
where~$V_{3}$ is an arbitrary $3$-volume factor, which leaves the
scalar product dimensionless. With the ansatz in Eq.~(\ref{eq:DEED-PlaneWave}),
we find\begin{equation}
J^{\dag}J=\frac{1}{2\omega(2\pi)^{3}V_{3}R},\label{eq:DEED-NormScalar}\end{equation}
where we have applied the relations\[
\int\textrm{d}^{3}x\cdot\exp(i\vec{x}(\vec{p}-\vec{p}^{\,\prime}))=(2\pi)^{3}\delta^{(3)}(\vec{p}-\vec{p}^{\,\prime})\,\,\,\,\textrm{and}\,\,\,\,\int_{0}^{R}\textrm{d}y\,\exp(2\pi iy\frac{n-n^{\prime}}{R})=R\delta_{nn^{\prime}}.\]

The equation of motion for the real 5D scalar field~$\phi$ with
a bulk mass~$M_{\textrm{s}}$ is given by the Klein-Gordon equation\begin{equation}
\left[\frac{\partial^{2}}{\partial t^{2}}-\nabla^{2}-\partial_{5}^{\dag}\partial_{5}+M_{\textrm{s}}^{2}\right]\phi=0,\label{eq:DEED-KG-Glg}\end{equation}
which determines the energy~$\omega$ of a field mode with the momenta~$\vec{p}$
and~$q$:\[
\omega^{2}=\vec{p}^{\,2}+m^{2}.\]
Here we have introduced the squared effective 4D mass\[
m^{2}:=q^{2}+M_{\textrm{s}}^{2}.\]
Furthermore, we need the 5D energy-momentum tensor~$T_{AB}$ of the
real scalar field~$\phi$ to calculate the energy density and pressure
of the field. It has the form~\cite{QFTonCurve-1}\begin{equation}
T_{AB}=(\phi_{,A})^{\dag}(\phi_{,B})-\frac{1}{2}g_{AB}g^{CD}(\phi_{,C})^{\dag}(\phi_{,D})+\frac{1}{2}g_{AB}M_{\textrm{s}}^{2}\phi^{\dag}\phi,\label{eq:DEED-Scalar5D-EPTensor}\end{equation}
where~$A,B,C,\dots$ are 5D coordinate indices. Here,~$A=0$ is
the time-like index,~$A=1,2,3$ are spatial indices, corresponding
to the uncompactified $3$-space, and~$A=5$ characterises the extra
spatial dimension. The 5D energy density~$\rho_{5}$ is then given
by the $00$-component of~$T_{AB}$:\begin{equation}
\rho_{5}=T_{00}=\frac{1}{2}(\partial_{0}\phi)^{\dag}(\partial_{0}\phi)+\frac{1}{2}(\nabla\phi)^{\dag}(\nabla\phi)+\frac{1}{2}(\partial_{5}\phi)^{\dag}(\partial_{5}\phi)+\frac{1}{2}M_{\textrm{s}}^{2}\phi^{\dag}\phi.\label{eq:DEED-rho5-class}\end{equation}
 By averaging over all directions of the isotropic $3$-space, we
obtain the pressure~$p_{5}$ of the scalar field~$\phi$: \begin{equation}
p_{5}=\frac{1}{3}\sum_{i=1}^{3}T_{ii}=\frac{1}{3}\left[(\nabla\phi)^{\dag}(\nabla\phi)+\frac{1}{2}(\partial_{0}\phi)^{\dag}(\partial_{0}\phi)-\frac{1}{2}(\nabla\phi)^{\dag}(\nabla\phi)-\frac{1}{2}(\partial_{5}\phi)^{\dag}(\partial_{5}\phi)-\frac{1}{2}M_{\textrm{s}}^{2}\phi^{\dag}\phi\right].\label{eq:DEED-p5-class}\end{equation}

Let us now perform the canonical quantisation of the field~$\phi$
by introducing the field operator\begin{equation}
\hat{\phi}(t,\vec{x},y)=\sqrt{V_{3}}\int\textrm{d}^{3}p\sum_{n=1}^{N}\left(\phi(\vec{p},n)a_{\vec{p},n}+\phi^{\dag}(\vec{p},n)a_{\vec{p},n}^{\dag}\right),\label{eq:DEED-Skalar-Feldop}\end{equation}
where~$a_{p,n}$ and~$a_{p,n}^{\dag}$ obey bosonic commutator relations:\[
[a_{\vec{p},n},a_{\vec{p}^{\,\prime},n^{\prime}}^{\dag}]=V_{3}^{-1}\delta(\vec{p}-\vec{p}^{\,\prime})\delta_{nn^{\prime}}\,,\,\,\,\,[a,a]=[a^{\dag},a^{\dag}]=0.\]

Now, the $00$-component $\hat{T}_{00}$ and the averaged $ii$-components
$\frac{1}{3}\sum_{i=1}^{3}\hat{T}_{ii}$ of the energy-momentum operator
$\hat{T}_{AB}$ follow from substituting the field operator $\hat{\phi}$
in Eq.~(\ref{eq:DEED-Skalar-Feldop}) into Eqs.~(\ref{eq:DEED-rho5-class})
and~(\ref{eq:DEED-p5-class}). Here, it is useful to consider the
relations\begin{eqnarray}
(\partial_{0}\hat{\phi})^{\dag}(\partial_{0}\hat{\phi}) & = & V_{3}^{2}\int\textrm{d}^{3}p\int d^{3}p^{\prime}\sum_{n,n^{\prime}=-\infty}^{\infty}\left[\partial_{0}\phi(\vec{p},n)a_{\vec{p},n}\cdot\partial_{0}\phi(\vec{p}^{\,\prime},n^{\prime})^{\dag}a_{\vec{p}^{\,\prime},n^{\prime}}^{\dag}+\cdots\right]\nonumber \\
 & = & V_{3}\int\textrm{d}^{3}p\sum_{n=-\infty}^{\infty}\, A^{\dag}A\cdot\omega^{2}+\cdots\nonumber \\
(\partial_{b}\hat{\phi})^{\dag}(\partial_{b}\hat{\phi}) & = & V_{3}\int\textrm{d}^{3}p\sum_{n=-\infty}^{\infty}\, A^{\dag}A\cdot p_{b}^{2}+\cdots\,\,\,\,\forall\, b=1,2,3\nonumber \\
(\partial_{5}\hat{\phi})^{\dag}(\partial_{5}\hat{\phi}) & = & V_{3}\int\textrm{d}^{3}p\sum_{n=-\infty}^{\infty}\, A^{\dag}A\cdot q^{2}+\cdots,\label{eq:DEED-ScalarEPTparts}\end{eqnarray}
where the ellipses ($\cdots$) denote the terms which vanish in the
VEVs~$\left\langle 0\right|\hat{T}_{AB}\left|0\right\rangle $ due
to~$\left\langle 0\right|a^{\dag}=a\left|0\right\rangle =0$. When
we insert the terms from Eqs.~(\ref{eq:DEED-ScalarEPTparts}) into
Eqs.~(\ref{eq:DEED-rho5-class}) and~(\ref{eq:DEED-p5-class}),
one obtains, after taking the VEVs of $\hat{T}_{00}$ and $\hat{T}_{ii}$,
the energy density~$\rho_{5}$ and the pressure~$p_{5}$ of the
quantised field $\phi$,\begin{eqnarray*}
\rho_{5}=\left\langle 0\right|\hat{T}_{00}\left|0\right\rangle  & = & V_{3}\int\textrm{d}^{3}p\sum_{n=-\infty}^{\infty}\, J^{\dag}J\cdot\left[\frac{1}{2}\omega^{2}+\frac{1}{2}\vec{p}^{\,2}+\frac{1}{2}q^{2}+\frac{1}{2}M_{\textrm{s}}^{\textrm{2}}\right]\\
 & = & V_{3}\int\textrm{d}^{3}p\sum_{n=-\infty}^{\infty}\, J^{\dag}J\cdot\omega^{2},\end{eqnarray*}
\begin{eqnarray*}
p_{5}=\frac{1}{3}\sum_{i=1}^{3}\left\langle 0\right|\hat{T}_{ii}\left|0\right\rangle  & = & V_{3}\int\textrm{d}^{3}p\sum_{n=-\infty}^{\infty}\, J^{\dag}J\cdot\left[\frac{1}{3}\vec{p}^{\,2}+\frac{1}{2}\omega^{2}-\frac{1}{2}\vec{p}^{\,2}-\frac{1}{2}q^{2}-\frac{1}{2}M_{\textrm{s}}^{2}\right]\\
 & = & V_{3}\int\textrm{d}^{3}p\sum_{n=-\infty}^{\infty}\, J^{\dag}J\cdot\frac{\vec{p}^{\,2}}{3\omega},\end{eqnarray*}
where we have used the energy-momentum relation~$\omega^{2}=\vec{p}^{\,2}+q^{2}+M_{\textrm{s}}^{2}$.
With the normalisation factor~$J$ from Eq.~(\ref{eq:DEED-NormScalar})
we finally obtain the energy density~$\rho_{5}$ and the pressure~$p_{5}$
of the quantised 5D field $\phi$,\begin{eqnarray}
\rho_{5}=\left\langle 0\right|\hat{T}_{00}\left|0\right\rangle  & = & \frac{1}{2(2\pi)^{3}R}\int\textrm{d}^{3}p\sum_{n=-\infty}^{\infty}\,\omega,\label{eq:DEED-rho5-unregul}\\
p_{5}=\frac{1}{3}\sum_{i=1}^{3}\left\langle 0\right|\hat{T}_{ii}\left|0\right\rangle  & = & \frac{1}{2(2\pi)^{3}R}\int\textrm{d}^{3}p\sum_{n=-\infty}^{\infty}\,\frac{\vec{p}^{\,2}}{3\omega}.\label{eq:DEED-p5-unregul}\end{eqnarray}

The momentum integral~$\int\textrm{d}^{3}p=4\pi\int_{0}^{\infty}\textrm{d}p\cdot p^{2}$
in these equations and thus~$\rho_{5}$ and~$p_{5}$ are divergent.
We therefore have to apply a regularisation procedure to obtain meaningful,
finite expressions. Consider first Eq.~(\ref{eq:DEED-rho5-unregul}).
By introducing an exponential suppression factor~$e^{-(dp)^{2}}$
as regulator function for~$d>0$, we find\begin{eqnarray}
\int\textrm{d}p\cdot p^{2}\omega & \longrightarrow & \int_{0}^{\infty}\textrm{d}p\cdot p^{2}\omega e^{-(dp)^{2}}\nonumber \\
 & = & \frac{1}{4}m^{2}d^{-2}e^{\frac{1}{2}(dm)^{2}}K_{1}\left(\kl{\frac{1}{2}}d^{2}m^{2}\right)\label{eq:DEED-rho-regul}\\
 & = & \frac{1}{2}d^{-4}+\frac{1}{4}m^{2}d^{-2}+\frac{1}{8}m^{4}\left(\frac{1}{4}+\frac{1}{2}\gamma-\ln2+\ln(dm)\right)+\mathcal{O}(d^{6}m^{6}),\nonumber \end{eqnarray}
where~$\omega=\sqrt{\vec{p}^{\,2}+m^{2}}$ and~$K_{n}(x)$ denotes
the modified Bessel function of the second kind of the order~$n$
and~$\gamma=0,577\dots$ is the Euler-Mascheroni constant. The limit~$d\rightarrow0$
removes the regulator and recovers the divergence. Before taking this
limit, a renormalisation has to be carried out to remove the potentially
divergent terms. Alternatively to the exponential regulator function
in Eq.~(\ref{eq:DEED-rho-regul}), one can also apply dimensional
regularisation by moving to~$n$ space-time dimensions. To be specific,
\begin{eqnarray*}
\int\textrm{d}^{3}p\cdot\omega & \longrightarrow & \mu^{(4-n)}S(n-1)\cdot\int_{0}^{\infty}\textrm{d}p\cdot p^{(n-2)}\cdot\omega\\
 & = & -\frac{1}{2}m^{4}\left(\frac{\mu}{m}\right)^{(4-n)}\pi^{(n/2-1)}\Gamma\left(-\frac{n}{2}\right)\\
 & \stackrel{n\rightarrow4}{=} & (4\pi)\frac{1}{8}m^{4}\left[(n-4)^{-1}+\frac{1}{2}\gamma+\ln\left(\frac{m}{\mu}\right)+\mathcal{O}(n-4)\right]_{n\rightarrow4},\end{eqnarray*}
where the renormalisation scale~$\mu$ has been introduced to keep
the mass dimension of the whole term constant, and~$S(n)$ is the
surface area of an $n$-ball. The regularisation of the divergent
integral in Eq.~(\ref{eq:DEED-p5-unregul}) with the exponential
suppression factor with~$d>0$ goes along the same lines as above:\begin{eqnarray}
\int\textrm{d}p\cdot p^{2}\frac{p^{2}}{\omega} & \longrightarrow & \int_{0}^{\infty}\textrm{d}p\cdot p^{2}\frac{p^{2}}{\omega}e^{-(dp)^{2}}\label{eq:DEED-p-regul}\\
 & = & \frac{1}{4}m^{2}d^{-2}e^{\frac{1}{2}(dm)^{2}}\left(d^{2}m^{2}K_{0}\left(\kl{\frac{1}{2}}d^{2}m^{2}\right)+(1-d^{2}m^{2})K_{1}\left(\kl{\frac{1}{2}}d^{2}m^{2}\right)\right)\nonumber \\
 & = & \frac{1}{2}d^{-4}-\frac{1}{4}m^{2}d^{-2}-\frac{3}{8}m^{4}\left(\frac{7}{12}+\frac{\gamma}{2}-\ln2+\ln(dm)\right)+\mathcal{O}(d^{6}m^{6}).\nonumber \end{eqnarray}

In the next step we have to remove the potential divergences in the
regularised expressions. For a curved background space-time~\cite{QFTonCurve-1},
one would therefore decompose~$\rho$ into a divergent and a finite
term, so that the former one has the form of a cosmological term in
Einsteins´s equations. Then the divergences would be absorbed to yield
renormalised coupling constants (which is the origin of some RGEs
for~$\Lambda$ and~$G$ in Chap.~\ref{cha:RGE1-TimeDepCC}), and
the finite remainder is called the renormalised energy density or,
in our case, the Casimir energy density. Here, such a general treatment
is not necessary because the divergence also arises in flat space-time,
like our~$\mathcal{M}\times\mathcal{S}^{1}$-manifold, but there
are neither cosmological terms nor Einstein´s equations. In order
to get rid of the divergence, one simply subtracts the corresponding
part of the energy density of the same field in a Minkowski-like space-time
with the same dimensions, \emph{}i.e.,~$\mathcal{M}\times\mathbb{R}^{1}$
in our case. In this case there is no IR cutoff and one has to integrate
over the 5-momentum instead of summing over it. This kind of renormalisation
works because the 5D Minkowski space suffers from the same divergence
as the~$\mathcal{M}\times\mathcal{S}^{1}$ space-time but exhibits
no Casimir effect. For the energy density~$\rho_{5}$ from Eq.~(\ref{eq:DEED-rho5-unregul}),
where the $p$-integral has been regularised by using Eq.~(\ref{eq:DEED-rho-regul}),
the renormalisation subtraction can be written as\begin{eqnarray}
\rho_{5,\textrm{renorm}} & \propto & \sum_{n=-\infty}^{\infty}\, f(n)-\int_{-\infty}^{\infty}\textrm{d}n\, f(n)\label{eq:DEED-Ren-Subtraction}\end{eqnarray}
with\[
f(n):=\frac{1}{2}d^{-4}+\frac{1}{4}m^{2}d^{-2}+\frac{1}{8}m^{4}\left(\frac{1}{4}+\frac{1}{2}\gamma-\ln2+\ln(dm)\right)+\mathcal{O}(d^{6}m^{6}).\]
 Above we have applied the substitution~$q=2\pi n/R$ to rewrite
the integral over~$q$:\[
\int\frac{\textrm{d}q}{2\pi}\, f(q)=\frac{1}{R}\int\textrm{d}n\, f(2\pi\frac{n}{R}).\]
Notice that when keeping in Eqs.~(\ref{eq:DEED-rho-regul}) and~(\ref{eq:DEED-p-regul})
only terms proportional to~$m^{4}\ln m$, we obtain the equation
of state~$p=-\rho$ of the CC. As we will see next, all other terms
including the divergences will vanish due to the subtraction in Eq.~(\ref{eq:DEED-Ren-Subtraction})
and the subsequent regularisation removal with~$d\rightarrow0$.
The finite result of the subtraction can be calculated by using the
Abel-Plana formulas~\cite{Saharian:2000xx} given by\begin{eqnarray}
\sum_{n=0}^{\infty}f(n)-\int_{0}^{\infty}\textrm{d}n\cdot f(n) & = & \frac{1}{2}f(0)+i\int_{0}^{\infty}\textrm{d}n\cdot\frac{f(+in)-f(-in)}{\exp(2\pi n)-1},\label{eq:DEED-AbelPlana1}\\
\sum_{n=0}^{\infty}f(n+\kl{\frac{1}{2}})-\int_{0}^{\infty}\textrm{d}n\cdot f(n) & = & -i\int_{0}^{\infty}\textrm{d}n\cdot\frac{f(+in)-f(-in)}{\exp(2\pi n)+1}.\label{eq:DEED-AbelPlana2}\end{eqnarray}
With~$m=\sqrt{k^{2}n^{2}+M_{\textrm{s}}^{2}}$ and~$k:=2\pi/R$
we immediately see that terms like~$m^{0}$, $m^{2}$, $m^{4}$ in
the function~$f(n)$ are cancelled on the right-hand side. And the
term~$\frac{1}{2}f(0)$ can be dropped as well since we consider
the range~$n=-\infty\dots\infty$. Only terms of the form ~$m^{4}\ln m$
survive, which we write for untwisted fields in the form\[
A:=\sum_{n=-\infty}^{\infty}m^{4}(n)\ln[m(n)]-\int_{-\infty}^{\infty}\textrm{d}n\cdot m^{4}(n)\ln[m(n)].\]
By applying the first Abel-Plana formula~(\ref{eq:DEED-AbelPlana1})
we therefore obtain\[
A=2i\int_{0}^{\infty}\textrm{d}n\,\,\frac{m^{4}(+in)\ln[m(+in)]-m^{4}(-in)\ln[m(-in)]}{\exp(2\pi n)-1}.\]
Assuming~$k,n,M\geq0$, we have to consider two cases for the root~$m(\pm in)$
in the last equation:\[
\left[k^{2}(\pm in)^{2}+M^{2}\right]^{\frac{1}{2}}=\left\{ \begin{array}{ll}
(M^{2}-k^{2}n^{2})^{\frac{1}{2}} & \,\,\,\textrm{for}\,\,\, M>kn,\\
\pm i(k^{2}n^{2}-M^{2})^{\frac{1}{2}} & \,\,\,\textrm{for}\,\,\, M<kn.\end{array}\right.\]
For~$a>0$ we can write the logarithm as $\ln(\pm i\cdot a)=\pm i\pi/2+\ln a$,
and with~$x:=M/k$ we obtain the result\begin{equation}
A(M)=-2\pi k^{4}\int_{x}^{\infty}\textrm{d}n\frac{(n^{2}-x^{2})^{2}}{\exp(2\pi n)-1},\label{eq:DEED-CasAofM}\end{equation}
which has, in the massless case~($x=0$), the value \[
A=-2\pi k^{4}\frac{3}{4\pi^{5}}\cdot\zeta(5)=-8R^{-4}3\zeta(5).\]

For twisted fields we have to use the second Abel-Plana formula~(\ref{eq:DEED-AbelPlana2}).
Analogously, we write\[
B(M)=-2i\int_{0}^{\infty}\textrm{d}n\frac{m^{4}(+in)\ln[m(+in)]-m^{4}(-in)\ln[m(-in)]}{\exp(2\pi n)+1},\]
using~$m(n)=\sqrt{k^{2}n^{2}+M^{2}}$, and not~$m(n)=\sqrt{k^{2}(n+\frac{1}{2})^{2}+M^{2}}$.
Thus, the result becomes\begin{equation}
B(M)=+2\pi k^{4}\int_{x}^{\infty}\textrm{d}n\frac{(n^{2}-x^{2})^{2}}{\exp(2\pi n)+1},\label{eq:DEED-CasBofM}\end{equation}
which has for massless fields~($x=0$) the value \[
B=+2\pi k^{4}\frac{45}{64\pi^{5}}\cdot\zeta(5)=\frac{15}{2}R^{-4}3\zeta(5).\]
For large masses~($x\gg1$), approximate expressions for~$A(M)$
and~$B(M)$ can be given by neglecting the~$1$ in the denominator
of the Eqs.~(\ref{eq:DEED-CasAofM}) and~(\ref{eq:DEED-CasBofM}):\begin{equation}
\int_{x}^{\infty}\textrm{d}n\frac{(n^{2}-x^{2})^{2}}{\exp(2\pi n)\pm1}\stackrel{x\gg1}{\sim}\int_{x}^{\infty}\textrm{d}n\frac{(n^{2}-x^{2})^{2}}{\exp(2\pi n)}=\frac{4\pi^{2}x^{2}+6\pi x+3}{4\pi^{5}}e^{-2\pi x}.\label{eq:DEED-mass-suppression}\end{equation}

Combining the prefactors in Eq.~(\ref{eq:DEED-rho5-unregul}) and
the finite results~$A(M)$, $B(M)$ from the renormalisation in Eq.~(\ref{eq:DEED-Ren-Subtraction})
we have found the finite 5D energy density. To obtain the effective
4D Casimir energy~$\rho_{4}$ density we just have to integrate over
the fifth dimension~$\int_{0}^{R}\textrm{d}y$, which simply yields
a factor~$R$. Finally, we find for untwisted scalar fields\begin{eqnarray}
\rho_{4} & = & \frac{1}{8(2\pi)^{2}}\cdot A(M)\,\,\stackrel{M=0}{=}\,\,-1\cdot(2\pi)^{-2}R^{-4}3\zeta(5),\label{eq:DEED-rho4-Muntwist}\end{eqnarray}
whereas twisted scalar fields yield\begin{eqnarray}
\rho_{4} & = & \frac{1}{8(2\pi)^{2}}\cdot B(M)\,\,\stackrel{M=0}{=}\,\,+\frac{15}{16}\cdot(2\pi)^{-2}R^{-4}3\zeta(5).\label{eq:DEED-rho4-Mtwist}\end{eqnarray}
Obviously, untwisted and twisted fields provide energy densities~$\rho_{4}$
of different sign. Note that the values for~$\rho_{4}$ in Eqs.~(\ref{eq:DEED-rho4-Muntwist})
and~(\ref{eq:DEED-rho4-Mtwist}) agree with the results in Refs.~\cite{Candelas:ae,Kantowski:ct}
for~$M=0$. 

As explained above, the corresponding effective 4D pressure~$p_{4}$
is just the negative of~$\rho_{4}$. Therefore the Casimir energy
density from the ED represents a finite contribution to the 4D CC.
From these results we see that~$\rho_{4}$ scales like the inverse
fourth power of the size~$R$ of the fifth dimension. For small EDs
this could cause problems with the observed tiny value of the CC.
Apart from fine-tuning the field content, there is another possibility
to solve this problem. From expression~(\ref{eq:DEED-mass-suppression})
we obtain an approximately exponential suppression of the Casimir
energy by bulk field masses~$M=xk$. We will make use of this behaviour
in Secs.~\ref{sec:DiED-Massive} and~\ref{sec:VacD-VacEng}, where
the Casimir energy becomes small for large bulk masses even when the
ED is small.

Finally, we mention that from statistical arguments and counting degrees
of freedom one can immediately conclude that the Casimir energy densities
for Dirac fermions are just~$(-4)$ times the values of real scalars.
We will use this later in Sec.~\ref{sec:DiED-Casimir-Fermion}.

\chapter{Discretised Extra Dimensions\label{cha:DiED}}

\section{Introduction}

In the Chap.~\ref{cha:DEED} we have discussed dark energy in the
form of Casimir energy in the context of a continuous ED. We have
found there that the Casimir effect yields a contribution the effective
4D CC. In this chapter we go one step further and consider discretised
EDs. As already mentioned in Chap.~\ref{cha:Introduction} a discrete
space-time structure might not only serve as an UV regulator in theories
of quantum gravity but it is also a crucial ingredient of deconstruction.
In this framework the phenomenology of discretised higher dimensions
is exactly reproduced by a deconstruction model working in a 4D continuous
space-time. This concept helps to circumvent some disadvantages occurring
in extra-dimensional theories.

In the following sections we will derive the Casimir energy density
for scalar and fermion fields in a discretised ED. Since we we are
working effectively on a transverse lattice we expect to encounter
typical lattice effects like fermion doubling. Furthermore, the scaling
of the Casmir energy density with the number of lattice sites and
the suppression by bulk masses will be discussed in detail. In Chap.~\ref{cha:VacD}
we will apply these results to a deconstruction model and propose
a prescription to calculate the absolute value of the vacuum energy
of 4D quantum fields in the deconstruction framework.

\section{Casimir Effect for a Scalar Field\label{sec:DiED-Casimir-Scalar}}

In this section, we consider a scalar quantum field~$\phi$ in a
space-time with the topology~$\mathcal{M}\times\mathcal{S}_{\textrm{lat}}^{1}$,
where~$\mathcal{M}$ is the continuous Minkowski space and~$\mathcal{S}_{\textrm{lat}}^{1}$
denotes the discrete fifth dimension compactified on the circle. Taking
the discrete nature of the fifth dimension into account, the discretisation
of the circle~$\mathcal{S}^{1}$ also forces the coordinate~$y$
in the fifth dimension to be discrete. Assuming~$N$ lattice sites
with a universal lattice spacing~$a$, the circumference of the fifth
dimension is given by~$R=Na$, and the position~$y$ of each site
can be described by a coordinate index~$j$,\begin{equation}
y=a\cdot j,\,\,\,\, j=1,\dots,N.\label{eq:DiED-S-discreteY}\end{equation}
From the standard definition for a derivative in the continuum,\[
\frac{\partial\phi}{\partial y}(y)=\lim_{a\rightarrow0}\frac{\phi(y+a)-\phi(y)}{a},\]
 follows the discrete forward and backward difference operators~$\partial_{5}\phi$
and~$(\partial_{5}\phi)^{\dag}$:\begin{eqnarray*}
\frac{\partial\phi}{\partial y}\longrightarrow(\partial_{5}\phi) & := & \frac{\phi(j+1)-\phi(j)}{a},\\
\left(\frac{\partial\phi}{\partial y}\right)^{\dag}\longrightarrow(\partial_{5}\phi)^{\dag} & := & \frac{\phi(j)-\phi(j-1)}{a}.\end{eqnarray*}
By inserting the ansatz~(\ref{eq:DEED-PlaneWave}) for~$\phi$ we
find\[
\partial_{5}\phi=a^{-1}(e^{-iqa}-1)\phi,\,\,\,\,(\partial_{5}\phi)^{\dag}=a^{-1}(1-e^{iqa})\phi,\]
and therefore\[
\partial_{5}^{\dag}\partial_{5}\phi=-2a^{-2}(1-\cos qa)\phi.\]
Taking into account the discrete derivatives~$\partial_{5}$ and~$\partial_{5}^{\dag}$,
the continuum Klein-Gordon equation~(\ref{eq:DEED-KG-Glg}) for a
real 5D scalar field~$\phi$ with bulk mass~$M_{\textrm{s}}$ becomes\[
\left[\frac{\partial^{2}}{\partial t^{2}}-\nabla^{2}-\partial_{5}^{\dag}\partial_{5}+M_{\textrm{s}}^{2}\right]\phi=0.\]
Thus the the energy~$\omega$ of a field mode with the momenta~$\vec{p}$
and~$q$ reads\begin{equation}
\omega^{2}=\vec{p}^{\,2}+m^{2},\,\,\,\, m^{2}:=2a^{-2}(1-\cos qa)+M_{\textrm{s}}^{2},\label{eq:DiED-Scalar-Spectrum}\end{equation}
where the values~$q$ depend on the boundary conditions as in Eqs.~(\ref{eq:DEED-qSpectUntwist})
and~(\ref{eq:DEED-qSpectTwist}). However, here~$n$ has to be in
the range~$1\dots N$ since the lattice introduces an UV cutoff.
Values of~$n$ outside the range~$1\dots N$ are mapped back into
this range due to the $2\pi$-periodicity in the discrete derivatives.
Therefore, the momentum spectrum is finite and given by\begin{equation}
q=2\pi\frac{n}{R}=\frac{2\pi}{a}\cdot\frac{n}{N},\,\,\,\, n=1\dots N\label{eq:DiED-qSpectUntwist}\end{equation}
for untwisted fields. For twisted fields we respectively find\begin{equation}
q=2\pi\frac{(n-\frac{1}{2})}{R}=\frac{2\pi}{a}\frac{(n-\frac{1}{2})}{N},\,\,\,\, n=1\dots N.\label{eq:DiED-qSpectTwist}\end{equation}

We can now easily transfer many of the continuum results from Sec.~\ref{sec:DEED-Casimir}
to the discretised case by replacing the infinite sum~$\sum_{n=-\infty}^{\infty}$
by the finite sum~$\sum_{n=1}^{N}$. For instance, the scalar product
for the field modes now reads\[
(\phi_{1},\phi_{2}):=i\int\textrm{d}^{3}x\, a\sum_{j=1}^{N}\,\left(\phi_{1}^{\dag}(\partial_{t}\phi_{2})-(\partial_{t}\phi_{1})^{\dag}\phi_{2}\right),\]
which yields via Eq.~(\ref{eq:DEED-Orthonorm}) the normalisation
factor~$J$ from Eq.~(\ref{eq:DEED-NormScalar}) by using the relation\[
\sum_{j=1}^{N}\exp\left(2\pi i\frac{n-n^{\prime}}{N}j\right)=N\delta_{nn^{\prime}}.\]
The quantisation of the scalar field and the determination of the
corresponding energy density~$\rho_{5}$ and pressure~$p_{5}$ goes
along the same lines as in the continuum case:\begin{eqnarray}
\rho_{5}=\left\langle 0\right|\hat{T}_{00}\left|0\right\rangle  & = & \frac{1}{2(2\pi)^{3}R}\int\textrm{d}^{3}p\sum_{n=1}^{N}\,\omega,\label{eq:DiED-S-rho5-unregul}\\
p_{5}=\frac{1}{3}\sum_{i=1}^{3}\left\langle 0\right|\hat{T}_{ii}\left|0\right\rangle  & = & \frac{1}{2(2\pi)^{3}R}\int\textrm{d}^{3}p\sum_{n=1}^{N}\,\frac{\vec{p}^{\,2}}{3\omega}.\label{eq:DiED-S-p5-unregul}\end{eqnarray}
Also, the regularisation of the momentum integral~$\int\textrm{d}^{3}p$
can be performed exactly like in the continuum case. Here, we also
consider the method with the exponential regulator function~$e^{-(dp)^{2}}$
with~$d>0$. However, in the renormalisation procedure we cannot
make use of the Abel-Plana formulas since they only treat infinite
sums. In order to obtain the finite Casimir energy density, it is
necessary to compare the discrete mode sums belonging to the momenta
in~$\mathcal{S}_{\textrm{lat}}^{1}$ with the energy density and
pressure of a field in a space-time with a non-compactified but still
discretised ED. Regarding Eqs.~(\ref{eq:DiED-S-rho5-unregul}) and~(\ref{eq:DiED-S-p5-unregul}),
the mode sum with respect to the fifth momentum coordinate~$q=2\pi n/(aN)$
is of the type\[
\sum_{n=1}^{N}f(n/N),\]
where~$f(n/N)$ summarises the terms\[
\frac{1}{2}d^{-4}+\frac{1}{4}m^{2}d^{-2}+\frac{1}{8}m^{4}\left(\frac{1}{4}+\frac{1}{2}\gamma-\ln2+\ln(d)\right)+\mathcal{O}(d^{6}m^{6}),\]
following from the last line in Eq.~(\ref{eq:DEED-rho-regul}). From
this sum, the mode integral corresponding to a non-compactified~$\mathbb{R}^{1}$-dimension
can be obtained by cutting out a section of length~$R$ of an~$\mathbb{R}^{1}$-dimension.
This means, that we take the limit of an infinite number~$M$ of
lattice sites,~$M\rightarrow\infty$, while keeping the spacing~$a$
constant: \[
\left.\frac{R}{Ma}\sum_{n=1}^{M}f\left(\frac{n}{M}\right)\right|_{M\rightarrow\infty}=\frac{R}{a}\left[\sum_{n=1}^{M}\frac{\Delta n}{M}f\left(\frac{n}{M}\right)\right]_{M\rightarrow\infty}\stackrel{s:=n/M}{=}N\cdot\int_{0}^{1}\textrm{d}s\cdot f(s),\]
where~$Ma$ becomes the infinite {}``length'' of~$\mathbb{R}^{1}$
and~$f(s)$ is the same function as in the~$\mathcal{S}_{\textrm{lat}}^{1}$
mode sum. In the last equation, we have substituted~$s:=n/M$ and
inserted~$\Delta n=1$ so that~$\textrm{d}s=\Delta n/M$ for~$M\rightarrow\infty$.
Both the sum and the integral are finite since the lattice introduces
an UV cutoff. Then the renormalisation is performed by subtracting
the integral from the sum, \begin{equation}
\sum_{n=1}^{N}f\left(\frac{n}{N}\right)-N\cdot\int_{0}^{1}\textrm{d}s\cdot f(s)=\sum_{n=1}^{N}m^{4}\ln m-N\cdot\int_{0}^{1}\textrm{d}s\cdot m^{4}\ln m,\label{eq:DiED-Sum1NMinusInt01}\end{equation}
where only~$m^{4}\ln m$ survives since all other terms like~$m^{0}$,~$m^{2}$,~$m^{4}$
either vanish when the regularisation is removed for $d\rightarrow0$
or are completely subtracted due to the following identities:\begin{eqnarray}
\sum_{n=1}^{N}(1-\cos2\pi\kl{\frac{n}{N}}+\kl{\frac{1}{2}}a^{2}M_{\textrm{s}}^{2}) & = & N\cdot\int_{0}^{1}\textrm{d}s\cdot(1-\cos2\pi s+\kl{\frac{1}{2}}a^{2}M_{\textrm{s}}^{2})\\
 & = & N(1+\kl{\frac{1}{2}}a^{2}M_{\textrm{s}}^{2}),\end{eqnarray}
 \begin{eqnarray}
\sum_{n=1}^{N}(1-\cos2\pi\kl{\frac{n}{N}}+\kl{\frac{1}{2}}a^{2}M_{\textrm{s}}^{2})^{2} & = & N\cdot\int_{0}^{1}\textrm{d}s\cdot(1-\cos2\pi s+\kl{\frac{1}{2}}a^{2}M_{\textrm{s}}^{2})^{2}\\
 & = & N(\kl{\frac{3}{2}}+a^{2}M_{\textrm{s}}^{2}+(\kl{\frac{1}{2}}a^{2}M_{\textrm{s}}^{2})^{2}).\end{eqnarray}
 This is also the case for twisted fields, where~$n$ is replaced
by~$n-\frac{1}{2}$. Note that in the discretised case we obtain
the CC equation of state~$p=-\rho$, too. Finally, we find the renormalised
Casimir energy density to be\begin{eqnarray}
\rho_{5} & = & \frac{1}{2(2\pi)^{3}R}\cdot\frac{4\pi}{8}\left[\sum_{n=1}^{N}m^{4}\ln m-N\cdot\int_{0}^{1}\textrm{d}s\cdot m^{4}\ln m\right]\nonumber \\
 & = & +R^{-5}S_{1}(N)=-p_{5},\,\,\,\,\textrm{(untwisted)}\label{eq:DiED-S-rho5-ren}\end{eqnarray}
where~$m^{2}=2a^{-2}(1-\cos qa)+M_{\textrm{s}}^{2}$ and where we
have introduced the function\begin{eqnarray}
S_{1}(N) & := & \frac{1}{4(2\pi)^{2}}N^{4}\cdot\left[\sum_{n=1}^{N}\left(1-\cos2\pi\frac{n}{N}+\frac{1}{2}a^{2}M_{\textrm{s}}^{2}\right)^{2}\ln\left(1-\cos2\pi\frac{n}{N}+\frac{1}{2}a^{2}M_{\textrm{s}}^{2}\right)\right.\nonumber \\
 & - & \left.N\cdot\int_{0}^{1}\textrm{d}s\cdot\left(1-\cos2\pi s+\frac{1}{2}a^{2}M_{\textrm{s}}^{2}\right)^{2}\ln\left(1-\cos2\pi s+\frac{1}{2}a^{2}M_{\textrm{s}}^{2}\right)\right].\label{eq:DiED-CasimirFuncS1}\end{eqnarray}
In the limit~$N\rightarrow\infty$ and~$M_{\textrm{s}}=0$, the
function~$S_{1}(N)$ converges to the value of a continuous fifth
dimension:\[
\lim_{N\rightarrow\infty}S_{1}(N)=-\frac{1}{4(2\pi)^{2}}3\zeta(5)\cdot4.\]
By integrating out the fifth dimension, we obtain the 4D energy density
\begin{equation}
\rho_{4}=\int_{0}^{R}\textrm{d}r\cdot\rho_{5}=R\cdot\rho_{5}=-\frac{3\zeta(5)}{(2\pi)^{2}R^{4}}=\frac{1}{R^{4}}\cdot(-0,0787970\dots).\,\,\,\,(\textrm{untwisted})\label{eq:DiED-S-rho4-untwist}\end{equation}

In the case of a twisted scalar field everything is like above, but
the energy density reads~\begin{eqnarray*}
\rho_{5} & = & +R^{-5}S_{2}(N)\\
 & = & -p_{5},\,\,\,\,\textrm{(twisted)}\end{eqnarray*}
where~$S_{2}(N)$ is the function~$S_{1}(N)$ with~$n$ replaced
by~$n-\frac{1}{2}$. For massless fields~($M_{\textrm{s}}=0$) we
obtain in the continuum limit\[
\lim_{N\rightarrow\infty}S_{2}(N)=+\frac{1}{4(2\pi)^{2}}3\zeta(5)\cdot(4-\kl{\frac{1}{4}}),\]
and after integrating out the fifth dimension the 4D energy density
reads\begin{equation}
\rho_{4}=R\cdot\rho_{5}=+\frac{15}{16}\cdot\frac{3\zeta(5)}{(2\pi)^{2}R^{4}}=\frac{1}{R^{4}}\cdot(+0,0738722\dots).\,\,\,\,\textrm{(twisted)}\label{eq:DiED-S-rho4-twist}\end{equation}
In the continuum limit the lattice results for~$\rho_{4}$ found
here are consistent with the continuum Casimir energy densities from
Eqs.~(\ref{eq:DEED-rho4-Muntwist}) and~(\ref{eq:DEED-rho4-Mtwist}).

\section{Casimir Effect for a Dirac Fermion\label{sec:DiED-Casimir-Fermion}}

In analogy with the treatment of scalar fields in previous section,
we will now calculate the Casimir energy density of Dirac fermions.
Therefore, a plane wave Ansatz for Dirac spinor fields~$\Psi$ in
the~$\mathcal{M}\times\mathcal{S}^{1}$ manifold is a convenient
choice, too: \begin{equation}
\Psi=\psi\exp(-i\omega t+i\vec{p}\vec{x}+iqy).\label{eq:DiED-Fermion-PlaneWave}\end{equation}
The boundary conditions, associated with the compactified~$\mathcal{S}^{1}$-dimension,
provide the discrete momentum spectra. For twisted and untwisted fields
we have as before~$q=2\pi(n-\frac{1}{2})/(aN)$ and~$q=2\pi n/(aN)$,
respectively. Like in Eq.~(\ref{eq:DiED-S-discreteY}) the coordinate~$y$
corresponding to the fifth dimension is discrete,~$y=a\cdot j$,
where~$j=1,\dots,N$, and implies an upper bound for the momentum~$q$.

Unlike the Klein-Gordon equation for scalars fields, the Dirac equation
is linear in the derivatives, and therefore we need a symmetric derivative
operator for the discrete~$y$-coordinate:\[
\partial_{5}\Psi(j):=\frac{1}{2a}\left(\Psi(j+1)-\Psi(j-1)\right).\]
With the Ansatz~(\ref{eq:DiED-Fermion-PlaneWave}) we obtain\begin{eqnarray*}
\partial_{5}\Psi(j) & = & \frac{1}{2a}\psi\cdot\exp(-i\omega t+i\vec{p}\vec{x})\left[\exp(iqa(j+1))-\exp(iqa(j-1))\right]\\
 & = & \Psi(j)\cdot\left(+\frac{i}{a}\sin(qa)\right),\end{eqnarray*}
and together with the 5D Dirac equation%
\footnote{A fifth Dirac matrix $\gamma^{5}:=i\gamma_{\textrm{4D}}^{5}$ has
to be introduced, where $\gamma_{\textrm{4D}}^{5}$ is the usual $\gamma^{5}$
matrix of the 4D Dirac theory~\cite{Pilaftsis:1999jk}.%
} for a Dirac field with mass~$M_{\textrm{f}}$, \[
(i\gamma^{A}\partial_{A}-M_{\textrm{f}})\Psi=0,\,\,\,\, A=0,\dots,3,5,\]
 the energy-momentum relation is determined to be \begin{equation}
\omega^{2}=\vec{p}^{\,2}+m^{2},\,\,\,\, m^{2}:=a^{-2}\sin^{2}qa+M_{\textrm{f}}^{2}.\label{eq:DiED-Fermion-EP-Rel}\end{equation}
The energy-momentum tensor~$T_{AB}$ for the Dirac field~$\Psi$
has the form~\cite{QFTonCurve-1}\[
T_{AB}=\frac{1}{4}i[\overline{\Psi}\gamma_{A}\partial_{B}\Psi+\overline{\Psi}\gamma_{B}\partial_{A}\Psi-\overline{(\partial_{A}\Psi)}\gamma_{B}\Psi-\overline{(\partial_{B}\Psi)}\gamma_{A}\Psi],\]
and the usual canonical quantisation procedure parallels that for
scalar fields up to replacing the bosonic commutator relations by
the fermionic anti-commutator relations, which give an overall minus
sign in the result. The Dirac fermion also has four times the degrees
of freedoms of a real scalar, describing particles and anti-particles
with two spin states each. In total, the energy density~$\rho_{5}$
and pressure~$p_{5}$ of a quantised Dirac field differ from the
scalar results of Eqs.~(\ref{eq:DiED-S-rho5-unregul}) and~(\ref{eq:DiED-S-p5-unregul})
only by a factor of~$(-4)$ and in the modified energy-momentum relation
of Eq.~(\ref{eq:DiED-Fermion-EP-Rel})\emph{,} i.e.\emph{,}\begin{eqnarray}
\rho_{5} & = & -\frac{2}{(2\pi)^{3}R}\sum_{n=1}^{N}\int\textrm{d}^{3}p\cdot\omega,\label{eq:DiED-F-rho5-unregul}\\
p_{5} & = & -\frac{2}{(2\pi)^{3}R}\sum_{n=1}^{N}\int\textrm{d}^{3}p\cdot\frac{\vec{p}^{\,2}}{\omega},\label{eq:DiED-F-p5-unregul}\end{eqnarray}
where~$\omega^{2}=\vec{p}^{\,2}+a^{-2}\sin^{2}qa+M_{\textrm{f}}^{2}$.
By replacing back the sum~$\sum_{n=1}^{N}$ to the continuum version~$\sum_{n=-\infty}^{\infty}$
and additionally using the continuum energy momentum relation~$\omega^{2}=\vec{p}^{\,2}+q^{2}+M_{\textrm{f}}^{2}$,
we now explicitly see by comparison with Eqs.~(\ref{eq:DEED-rho5-unregul})
and~(\ref{eq:DEED-p5-unregul}) that the corresponding fermionic
Casimir energy density would indeed be $(-4)$ times the scalar values.
In the following, however, we will observe that on the lattice this
is not true anymore due to fermion doublers.

\begin{figure}[t]
\begin{centering}\includegraphics[clip,width=0.95\textwidth,keepaspectratio]{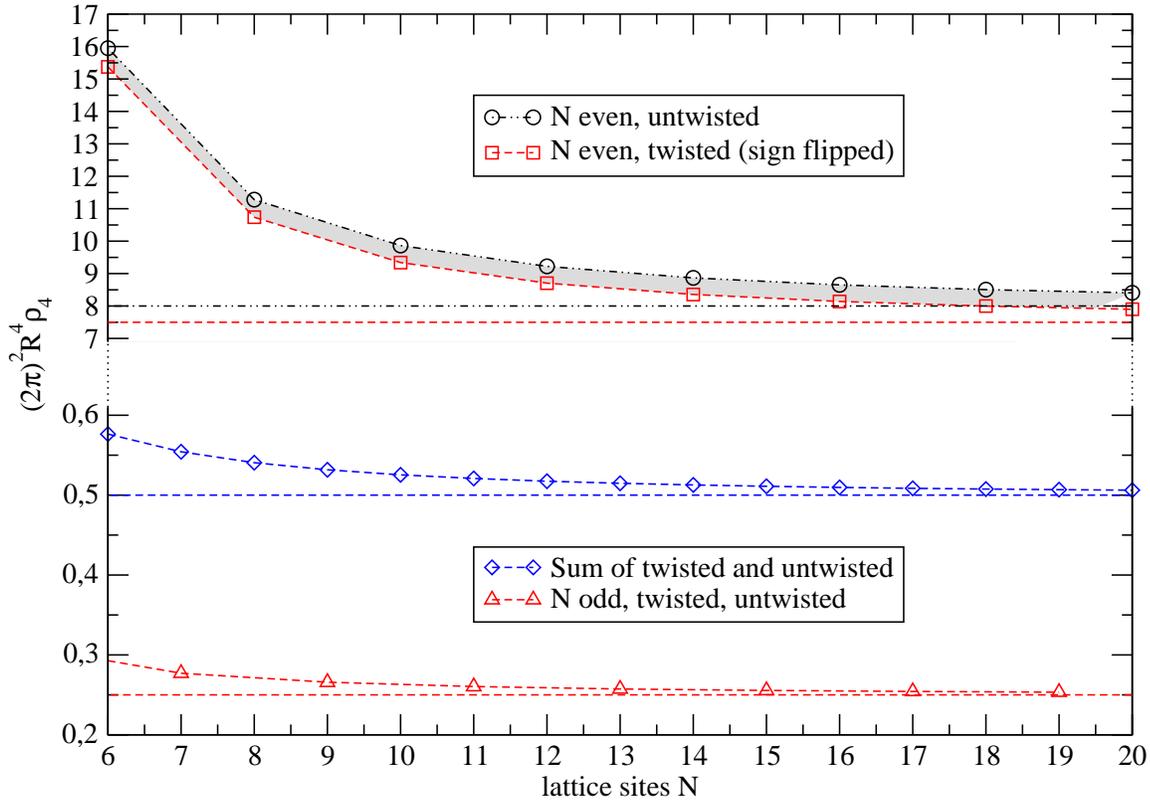}\par\end{centering}

\caption{\label{fig:DiED-OddEvenArtifact}Illustration of the odd-even artefact
for fermion fields. The cases of odd and even~$N$ are plotted separately,
and the circumference~$R$ of the fifth dimension is kept constant.
For a better representation, we flipped the sign in the results for
twisted fields and even~$N$. The horizontal lines correspond to
the continuum values given in Table~\ref{tab:DiED-rho4-Ninfty}.}
\end{figure}

From here on, the regularisation and renormalisation procedures are
identical to the scalar case in Sec.~\ref{sec:DiED-Casimir-Scalar}.
This also implies that the equation of state of the fermionic vacuum
energy is that of a cosmological constant,~$p=-\rho$. Thus, it is
sufficient to give the renormalised energy density in five dimensions
\begin{eqnarray}
\rho_{5} & = & -\frac{2}{(2\pi)^{3}R}4\pi\frac{1}{8}\left[\sum_{n=1}^{N}m^{4}\ln m-N\cdot\int_{0}^{1}\textrm{d}s\cdot m^{4}\ln m\right]\nonumber \\
 & = & +R^{-5}F_{1}(N),\label{eq:DiED-F-rho5-ren}\end{eqnarray}
where the function~$F_{1}(N)$ in the last equation is defined as\begin{eqnarray}
F_{1}(N) & := & -\frac{1}{4(2\pi)^{2}}N^{4}\cdot\left[\sum_{n=1}^{N}\left(\sin^{2}2\pi\frac{n}{N}+a^{2}M_{\textrm{f}}^{2}\right)^{2}\ln\left(\sin^{2}2\pi\frac{n}{N}+a^{2}M_{\textrm{f}}^{2}\right)\right.\nonumber \\
 & - & \left.N\cdot\int_{0}^{1}\textrm{d}s\cdot\left(\sin^{2}2\pi\frac{n}{N}+a^{2}M_{\textrm{f}}^{2}\right)^{2}\ln\left(\sin^{2}2\pi\frac{n}{N}+a^{2}M_{\textrm{f}}^{2}\right)\right],\label{eq:DiED-CasimirFuncF1}\end{eqnarray}
with~$R=Na$. For the twisted Dirac field we have \[
\rho_{5}=R^{-5}F_{2}(N),\]
where~$F_{2}(N)$ is the function~$F_{1}(N)$ with~$n$ replaced
by~$n-\frac{1}{2}$. Unlike the functions~$S_{1,2}(N)$ for the
scalar fields, the functions~$F_{1,2}(N)$ for the fermionic fields
have two limit points each, which depend on whether the number of
lattice sites~$N$ is even or odd. For massless fermions~($M_{\textrm{f}}=0$)
and even~$N$ we obtain\begin{eqnarray*}
\lim_{N\rightarrow\infty}F_{1}(N) & = & -\frac{1}{4(2\pi)^{2}}\cdot3\zeta(5)\cdot(-32),\,\,\,\,(\textrm{untwisted})\\
\lim_{N\rightarrow\infty}F_{2}(N) & = & -\frac{1}{4(2\pi)^{2}}\cdot3\zeta(5)\cdot(+30).\,\,\,\,(\textrm{twisted})\end{eqnarray*}
After integrating out the fifth dimension, the 4D Casimir energy densities
read\begin{eqnarray}
\rho_{4} & = & \frac{32\cdot3\zeta(5)}{4(2\pi)^{2}R^{4}}=\frac{1}{R^{4}}\cdot(+0,630376\dots),\,\,\,\,(\textrm{untwisted})\label{eq:DiED-F-rho4-untwist}\\
\rho_{4} & = & \frac{-30\cdot3\zeta(5)}{4(2\pi)^{2}R^{4}}=\frac{1}{R^{4}}\cdot(-0,590978\dots).\,\,\,\,(\textrm{twisted})\label{eq:DiED-F-rho4-twist}\end{eqnarray}
In the case of odd~$N$, both functions have the same limit\[
\lim_{N\rightarrow\infty}F_{1}(N)=\lim_{N\rightarrow\infty}F_{2}(N)=\frac{1}{4(2\pi)^{2}}\cdot3\zeta(5),\]
\[
\rho_{4}=\frac{3\zeta(5)}{4(2\pi)^{2}R^{4}}=\frac{1}{R^{4}}\cdot(+0,019699\dots).\]
Obviously, this behaviour is an effect of the lattice, in Ref.~\cite{Hill:2002me}
it is called an odd-even artefact. We also notice, that the~$N\rightarrow\infty$
limit of the sum of twisted and untwisted results does not depend
on whether~$N$ is even or odd. Therefore, it is reasonable to consider
only this sum as a physical quantity. For finite~$N$, this odd-even
artefact is illustrated in Fig.~\ref{fig:DiED-OddEvenArtifact}.
Note again, that the continuum results for~$\rho_{4}$ in Eqs.~(\ref{eq:DiED-F-rho4-untwist})
and~(\ref{eq:DiED-F-rho4-twist}) are identical with the values in
Ref.~\cite{Candelas:ae}.

\section{Massless Fields\label{sec:DiED-Massless}}

The calculations of Sec.~\ref{sec:DiED-Casimir-Scalar} show that
the Casimir effect for a real scalar field in the transverse lattice
space-time~$\mathcal{M}\times\mathcal{S}_{\textrm{lat}}^{1}$ induces
a negative vacuum energy density~$\rho$ and therefore a negative
contribution to the effective 4D CC. On the other hand, the fermionic
Dirac field of Sec.~\ref{sec:DiED-Casimir-Fermion} yields a positive
contribution to the CC.

\begin{figure}
\begin{centering}\includegraphics[clip,width=1\textwidth,keepaspectratio]{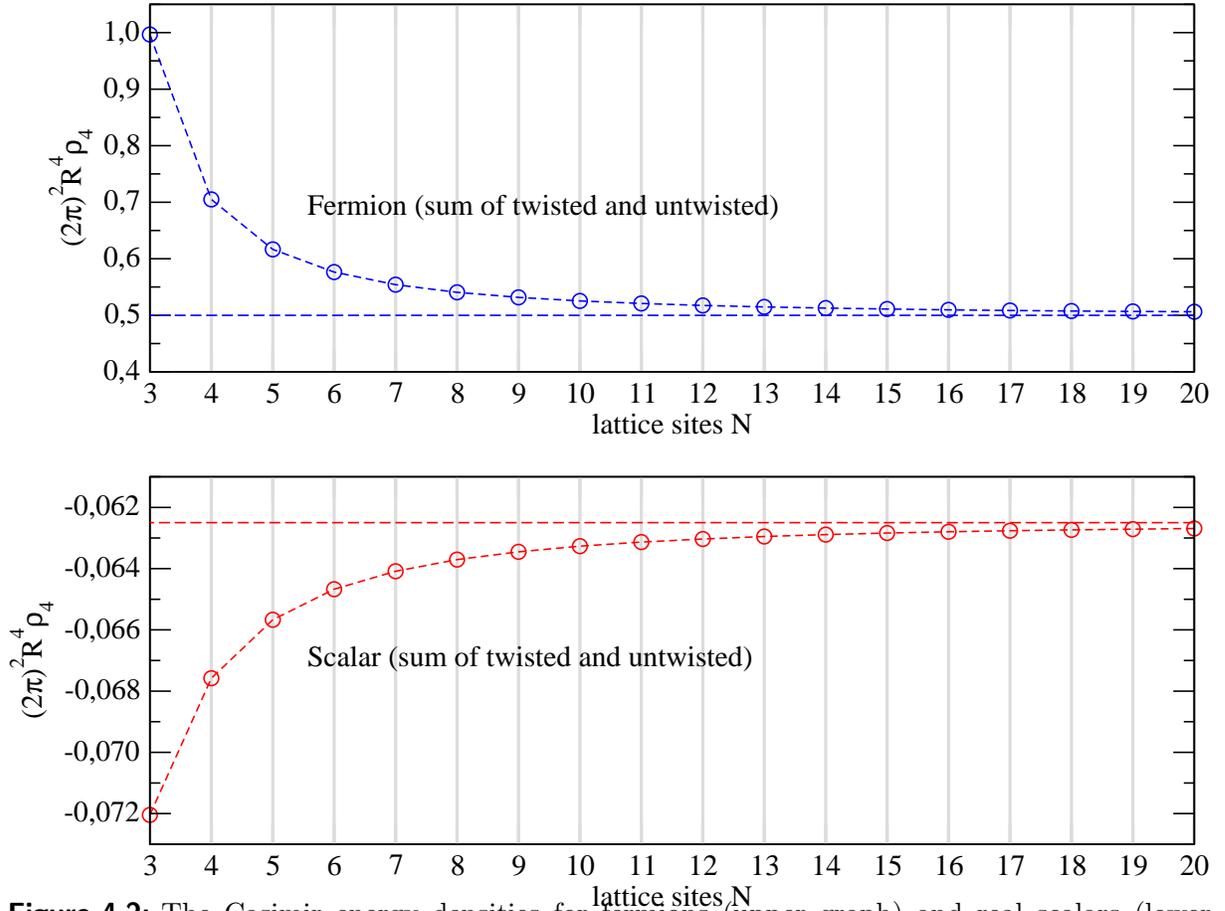}\par\end{centering}

\caption{\label{fig:DiED-SF-rho4ofN}The Casimir energy densities for fermions
(upper graph) and real scalars (lower graph). In this plot,~$R$
is fixed so that the lattice spacing decreases for increasing~$N$.
The dashed horizontal lines denote the continuum limit~$N\rightarrow\infty$,
which has for scalars the value~$-\frac{1}{16}$ and for fermions~$+\frac{1}{2}$.}
\end{figure}

We have already concluded that only the sum of twisted and untwisted
fields can be regarded as a physical quantity, and we note that its
sign is independent of~$N$. Moreover, for a constant circumference~$R$
and small~$N$, the Casimir energy density~$\rho_{4}(N)$ in the
transverse lattice setup has already the same order of magnitude as
the energy density~$\rho_{4}(N\rightarrow\infty)$ in the continuum
limit. Specifically, for~$N\gtrsim10$ the continuum result is approximated
at the few percent level. Even for a number of lattice sites which
is as small as~$N=3$, the results differ at most by a factor of~$2$,
which is clearly shown in Fig.~\ref{fig:DiED-SF-rho4ofN}. 

In the limit~$N\rightarrow\infty$~(Tab.~\ref{tab:DiED-rho4-Ninfty}),
the results for real scalars are the same as in the non-lattice calculation~\cite{Candelas:ae,Kantowski:ct},
but for the fermions there is an extra factor of~$2$ in the energy
density of our lattice calculation because of the fermion doubling
phenomenon in lattice theory. In a calculation for continuous dimensions,
one usually expects, from counting degrees of freedom, that the energy
density for Dirac fermions is~$(-4)$~times the value of real scalars. 

Up to now, we have investigated the Casimir effect for Dirac fermions
and real scalars having twisted and untwisted field configurations.
When passing to a complex scalar field which transforms under a~$U(1)$
gauge group there exist only trivial (untwisted) structures and therefore
the charged scalar obeys only periodic boundary conditions. For fermions,
on the other hand, the appearance of twisted field modes is related
to the double covering map~$SL(2,\mathbb{C})\rightarrow SO(3,1)$
which gives rise to inequivalent spin connections~\cite{Isham:1978xxx}.
Consequently, even in presence of a simply connected gauge group like~$U(1)$
we still have also the anti-periodic boundary condition for the fermions.

\begin{table}
\begin{centering}\begin{tabular}{|c|c|c||c|}
\hline 
$\rho_{4}R^{4}$&
untwisted&
twisted&
sum\tabularnewline
\hline
\hline 
real scalar&
$-1\cdot(2\pi)^{-2}3\zeta(5)$&
$+\frac{15}{16}\cdot(2\pi)^{-2}3\zeta(5)$&
$-\frac{1}{16}\cdot(2\pi)^{-2}3\zeta(5)$\tabularnewline
\hline 
fermion, $N$ even&
$+8\cdot(2\pi)^{-2}3\zeta(5)$&
$-\frac{15}{2}\cdot(2\pi)^{-2}3\zeta(5)$&
$+\frac{1}{2}\cdot(2\pi)^{-2}3\zeta(5)$\tabularnewline
\hline
fermion, $N$ odd&
$+\frac{1}{4}\cdot(2\pi)^{-2}3\zeta(5)$&
$+\frac{1}{4}\cdot(2\pi)^{-2}3\zeta(5)$&
$+\frac{1}{2}\cdot(2\pi)^{-2}3\zeta(5)$\tabularnewline
\hline
\end{tabular}\par\end{centering}

\caption{\label{tab:DiED-rho4-Ninfty}The Casimir energy density~$\rho_{4}$
multiplied by~$R^{4}$ for real massless scalars and Dirac fermions
($M_{\textrm{s,f}}=0$) in the limit of an infinite number of lattice
sites ($N\rightarrow\infty$). }
\end{table}

\section{Exponential Suppression by Massive Fields\label{sec:DiED-Massive}}

\begin{figure}
\begin{centering}\includegraphics[clip,width=0.95\textwidth,keepaspectratio]{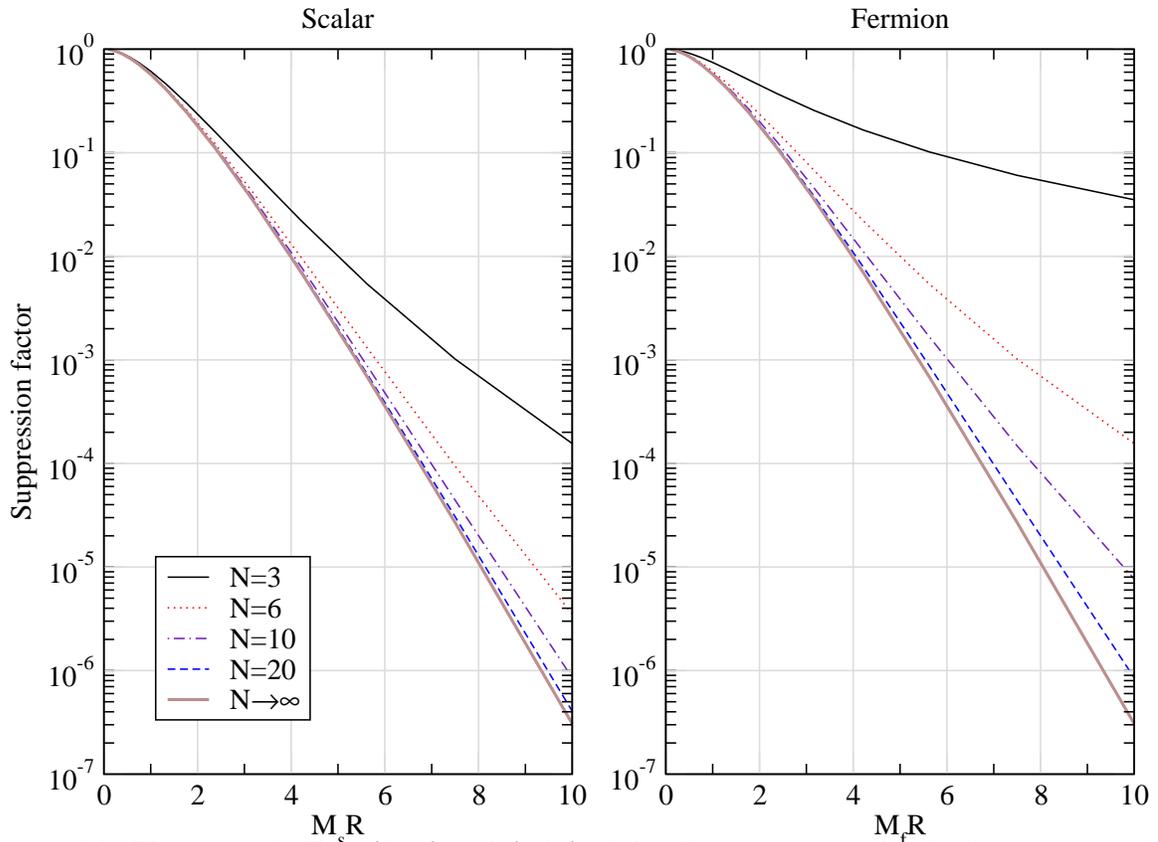}\par\end{centering}

\caption{\label{fig:DiED-BulkMassRatios}The ratios in Eqs.~(\ref{eq:DiED-S-BulkMassRatio})
and~(\ref{eq:DiED-F-BulkMassRatio}) of the Casimir energy density
between massive and massless fields. The values for~$N\rightarrow\infty$
are taken from the analytic formulas in Eqs.~(\ref{eq:DEED-rho4-Muntwist})
and~(\ref{eq:DEED-rho4-Mtwist}).}
\end{figure}

So far, we have given results only in the case of vanishing bulk mass~($M_{\textrm{s,f}}=0$).
For massive 5D fields we observe an approximately exponential suppression
of the Casimir energy. This behaviour becomes obvious in the analytical
calculation for a continuous ED, which is given in Sec.~\ref{sec:DEED-Casimir}.
But it is also achieved for the discretised case of this chapter,
where in the limit of an infinite number of lattice sites~($N\rightarrow\infty$)
we approach the values of the analytical formulas~(\ref{eq:DEED-rho4-Muntwist})
and~(\ref{eq:DEED-rho4-Mtwist}). To investigate the suppression
behaviour depending on the mass~$M_{\textrm{s},\textrm{f}}$ and
the number~$N$, we examine the ratio between the energy density
of fields with mass~$M_{\textrm{s,f}}$ and that of massless fields.
For scalar fields this ratio is defined by\begin{equation}
\frac{S_{1}(M_{\textrm{s}}R)+S_{2}(M_{\textrm{s}}R)}{S_{1}(0)+S_{2}(0)},\label{eq:DiED-S-BulkMassRatio}\end{equation}
where the functions~$S_{1,2}(N)$ are taken from Eq.~(\ref{eq:DiED-CasimirFuncS1})
of Sec.~\ref{sec:DiED-Casimir-Scalar}. Analogously, using the functions~$F_{1,2}(N)$
from Eq.~(\ref{eq:DiED-CasimirFuncF1}) of Sec.~\ref{sec:DiED-Casimir-Fermion},
the ratio for fermionic fields reads \begin{equation}
\frac{F_{1}(M_{\textrm{f}}R)+F_{2}(M_{\textrm{f}}R)}{F_{1}(0)+F_{2}(0)}.\label{eq:DiED-F-BulkMassRatio}\end{equation}
Both ratios are plotted in Fig.~\ref{fig:DiED-BulkMassRatios} for
a range of values of~$N$ and~$M_{\textrm{s},\textrm{f}}R$. The
suppression by a bulk mass is most minimal for small~$N$. In the
case of~$N=3$ lattice sites, the corresponding ratios are given
in Table~\ref{tab:DiED-BulkMassRatioN3}.

\begin{table}[t]
\begin{centering}\begin{tabular}{|c|c|c|c|c|c|c|}
\hline 
$M_{\textrm{s,f}}R$&
$1$&
$10$&
$100$&
$1000$&
$10^{4}$&
$10^{5}$\tabularnewline
\hline
\hline 
Scalar&
$0,61$&
$1,5\cdot10^{-4}$&
$2,9\cdot10^{-12}$&
$3,0\cdot10^{-20}$&
$3,0\cdot10^{-28}$&
$3,0\cdot10^{-36}$\tabularnewline
\hline 
Fermion&
$0,74$&
$3,5\cdot10^{-2}$&
$3,7\cdot10^{-4}$&
$3,8\cdot10^{-6}$&
$3,7\cdot10^{-8}$&
$3,7\cdot10^{-10}$\tabularnewline
\hline
\end{tabular}\par\end{centering}

\caption{\label{tab:DiED-BulkMassRatioN3}The exponential suppression factors
in Eqs.~(\ref{eq:DiED-S-BulkMassRatio}) and~(\ref{eq:DiED-F-BulkMassRatio})
of the Casimir energy densities for $N=3$ lattice sites, where $M_{\textrm{s},\textrm{f}}$
denotes the bulk masses of the quantum fields.}
\end{table}

\clearpage{\pagestyle{empty}\cleardoublepage}

\chapter{Vacuum Energy in Deconstruction\label{cha:VacD}}

In this chapter we will discuss the vacuum energy of 4D quantum fields
occurring in a deconstruction scenario. It is well known that the
absolute value of the zero-point energy density~$\int\textrm{d}^{3}p\cdot\omega(p)$
of 4D quantum fields is formally divergent and completely unconstrained
from a theory point of view. This means that without any further information
we do not know how to assign a sensible and finite vacuum energy value
to these fields. In Sec.~\ref{sec:RGE1-Quantum-CC} we already mentioned
this problem and investigated the change of the vacuum energy density
with respect to a renormalisation scale, but the absolute value still
remained undetermined there. 

Here, however, we propose a prescription how to treat this problem
by using the correspondence of a discretised fifth dimension and deconstruction.
For a 5D field obeying certain boundary conditions we are able to
derive the part of its vacuum energy that depends on the boundary
via the Casimir effect. Since in deconstruction the 5D field is described
by many 4D fields it seems plausible to identify the vacuum energies
of all the 4D fields with the Casimir energy density of the 5D field.
Going this way we have found a definite and sensible prescription
to determine the zero-point energies of quantum fields in the deconstruction
framework. Due to the lack of alternatives to handle this problem,
our idea represents at least a well motivated ansatz to obtain a reasonable
result, in contrast to the naive cutoff method given in Eq.~(\ref{eq:RGE1-Cutoff}). 

In the next section we will briefly introduce a deconstruction model
that describes a discretised fifth dimension and give the values of
masses and VEVs of the fields that appear in this framework. After
that, we will use the results of the previous chapter to determine
the vacuum energy of the 4D fields by using the prescription given
above.

\section{Deconstruction Model\label{sec:VacD-DecModel}}

In continuous EDs the maximum number of KK modes is usually restricted
by an UV cutoff which reflects the fact that non-Abelian gauge theories
in higher dimensions are non-renormalisable. Although this leads below
the cutoff to a renormalisable effective 4D theory, the full higher-dimensional
gauge-invariance is in general lost. To circumvent these problems
the deconstruction framework has been proposed~\cite{Arkani-Hamed:2001ca,Hill:2000mu}
representing manifestly gauge-invariant and renormalisable 4D gauge
theories, which reproduce higher-dimensional physics in their IR limit.
These theories use the transverse lattice technique~\cite{Bardeen:1976tm}
as a gauge-invariant regulator to describe the EDs and yield viable
UV completions extra-dimensional theories~\cite{Arkani-Hamed:2001nc}.

\begin{figure}
\begin{centering}\includegraphics[clip,keepaspectratio]{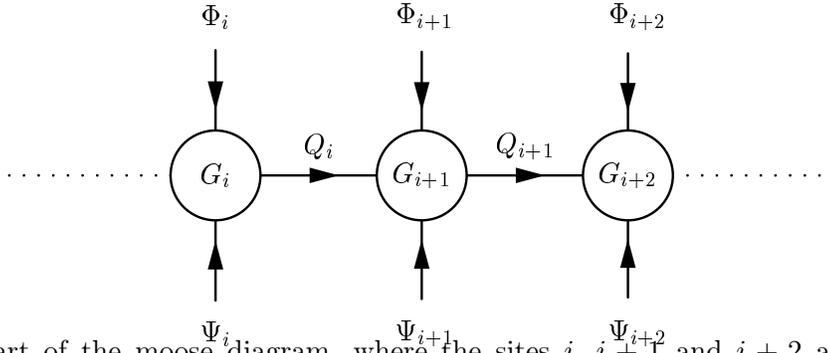}\par\end{centering}

\caption{\label{fig:VacD-Moose-Diagramme}Part of the moose diagram, where
the sites~$i$,~$i+1$ and~$i+2$ are shown for a deconstructed
extra dimension compactified on~$\mathcal{S}^{1}$. Each circle~$G_{i}$
corresponds to the $U(1)_{i}$ gauge group and each arrow pointing
towards (outwards) a circle represents a field with negative (positive)
charge under this group~\cite{Bauer:2003mh}.}
\end{figure}

Here, we consider a specific 5D deconstruction framework which was
introduced in Ref.~\cite{Bauer:2003mh} and which represents a periodic
model for a deconstructed 5D~$U(1)$ gauge theory compactified on
the circle~$\mathcal{S}^{1}$. The model is defined by an~$U(1)^{N}=\Pi_{i=1}^{N}U(1)_{i}$
product gauge group with~$N$ scalar link fields~$Q_{i}$ $(i=1,\ldots,N)$,
which carry the~$U(1)$-charges~$(q,-q)$ under the neighbouring
groups~$U(1)_{i}\times U(1)_{i+1}$. The identification~$i+N=i$
establishes the periodicity of the lattice corresponding to untwisted
quantum fields in the language of Sec.~\ref{sec:DEED-Casimir}, whereas
twisted quantum fields are described by an anti-periodic lattice with
the condition~$i+N=-i$. On the $i^{\textrm{th}}$ lattice site,
we put one Dirac fermion~$\Psi_{i}$ and one scalar~$\Phi_{i}$
which carry both the charge~$-q$ of the group $U(1)_{i}$. The fermions~$\Psi_{i}$
are SM-singlets and correspond to a right-handed bulk neutrino in
the ADD scheme~\cite{Arkani-Hamed:1998rs,Arkani-Hamed:1998vp}. As
an illustration of this setup the corresponding {}``moose''~\cite{geor86}
(or {}``quiver''~\cite{douglas:1996xx}) diagram is shown in Fig.~\ref{fig:VacD-Moose-Diagramme}.%
 Let us split up the Lagrangian of the model into several parts,\[
\mathcal{L}=\mathcal{L}_{\textrm{kin}}[\Phi_{i},Q_{i}]+\mathcal{L}_{\textrm{kin}}[A_{i}^{\mu}]+\mathcal{L}_{\textrm{kin}}[\Psi_{i}]+\mathcal{L}_{\textrm{mass}}[\Psi_{i},Q_{i}]-V,\]
where~$\mathcal{L}_{\textrm{kin}}[A_{i}^{\mu}]$ and~$\mathcal{L}_{\textrm{kin}}[\Psi_{i}]$
are the standard kinetic terms for the gauge bosons~$A_{i}^{\mu}$
and the fermions~$\Psi_{i}$ given by\begin{eqnarray*}
\mathcal{L}_{\textrm{kin}}[A_{i}^{\mu}] & = & -\frac{1}{4}\sum_{n=1}^{N}(\partial_{\mu}A_{n\nu}-\partial_{\nu}A_{n\mu})^{2},\\
\mathcal{L}_{\textrm{kin}}[\Psi_{i}] & = & \sum_{n=1}^{N}\overline{\Psi}_{n}\gamma_{\mu}(\partial^{\mu}-ig_{n}A_{n}^{\mu})\Psi_{n}.\end{eqnarray*}
Moreover, the kinetic terms~$\mathcal{L}_{\textrm{kin}}[\Phi_{i},Q_{i}]$
for the scalars~$\Phi_{i}$ and~$Q_{i}$, which will later provide
the gauge boson masses, can be written as\begin{equation}
\mathcal{L}_{\textrm{kin}}[\Phi_{i},Q_{i}]=\sum_{i=1}^{N}\left|(\partial_{\mu}+ig_{i}A_{i\mu})\Phi_{i}\right|^{2}+\left|(\partial_{\mu}+ig_{i}A_{i\mu}-ig_{i+1}A_{(i+1)\mu})Q_{i}\right|^{2},\label{eq:VacD-Lkin-Phi-Q}\end{equation}
where~$g_{i}$ are the gauge couplings corresponding to the gauge
group~$U(1)_{i}$. Finally, the mass and mixing terms involving the
fermions~$\Psi_{i}$ and the link fields~$Q_{i}$ are combined into\begin{equation}
\mathcal{L}_{\textrm{mass}}[\Psi_{i},Q_{i}]=\frac{u}{2}\cdot\sum_{n=1}^{N}\overline{\Psi}_{n\textrm{L}}\left[\frac{Q_{n}^{\dag}}{u}\Psi_{(n+1)\textrm{R}}-\frac{Q_{n-1}}{u}\Psi_{(n-1)\textrm{R}}\right]+\textrm{h.c.},\label{eq:VacD-Psi-Massterms}\end{equation}
where~$\Psi_{\textrm{L,R}}:=\frac{1}{2}(1\mp\gamma^{5})$ denote
the left- and right-handed components of the Dirac fermion~$\Psi$.
In the last line we have already introduced the universal real VEV~$\langle Q_{i}\rangle=u$
of the link fields~$Q_{i}$. At this point let us denote the VEV
of the scalars~$\Phi_{i}$ by~$\langle\Phi_{i}\rangle=v$, which
is also real and universal for all~$i=1\dots N$. These properties
of the VEV structure follow from the renormalisable potential~$V$,
which is not shown here since it not needed for the determination
of vacuum energy in this setup. The potential has been discussed and
minimised in great detail in Ref.~\cite{Bauer:2003mh}, where it
leads via a type-II seesaw mechanism~\cite{typeII} to the following
values for the VEVs:\[
\langle Q_{i}\rangle=u\sim10^{-2}\,\textrm{eV},\,\,\,\langle\Phi_{i}\rangle=v\sim10^{2}\,\textrm{GeV}.\]

Once~$Q_{i}$ and~$\Phi_{i}$ have acquired their VEVs the gauge
bosons~$A_{i\mu}$ become massive via the Higgs mechanism. Assuming
universal gauge couplings~$g_{i}=g$ in the kinetic terms~(\ref{eq:VacD-Lkin-Phi-Q})
of the scalars one finds the mass terms for the gauge bosons\begin{equation}
g^{2}\sum_{i=1}^{N}\left[v^{2}A_{i\mu}A_{i}^{\mu}+u^{2}(A_{i\mu}-A_{(i+1)\mu})^{2}\right].\label{eq:VacD-A-Massterms}\end{equation}
After diagonalisation, the mass eigenvalues~$M_{n}$ of the gauge
bosons read\begin{equation}
M_{n}^{2}=g^{2}v^{2}+2g^{2}u^{2}\left(1-\cos2\pi\frac{n}{N}\right),\,\,\,\, n=1,\dots,N,\label{eq:VacD-A-MassSpect}\end{equation}
which can be interpreted for~$n\ll N$ as a linear KK spectrum~$\sim n/R=gu\cdot n/N$
with a mass scale~$u\sim10^{-2}\,\textrm{eV}$. The scalar fields~$\Phi_{i}$
provide in addition a constant bulk (or kink) mass of the order~$v\sim10^{2}\,\textrm{GeV}$.
At this stage we can already observe the correspondence with the discretised
boson spectrum from Eq.~(\ref{eq:DiED-Scalar-Spectrum}), by identifying
the VEV~$u$ with the inverse lattice spacing~$a^{-1}$, and respectively
the VEV~$v$ with the bulk mass~$M_{\textrm{s}}$.

Furthermore, we find in~$\mathcal{L}_{\textrm{mass}}[\Psi_{i},Q_{i}]$
from Eq.~(\ref{eq:VacD-Psi-Massterms}) terms of the type~$Q_{i}^{\dag}\overline{\Psi}_{i\textrm{L}}\Psi_{(i+1)\textrm{R}}+\textrm{h.c.}$
which yield fermion masses of order~$u$ when~$Q_{i}$ has taken
on its VEV. In the next section we will see that the~$\Psi_{i}$
mass spectrum corresponds to the fermionic mass spectrum found in
Eq.~(\ref{eq:DiED-Fermion-EP-Rel}), where the fermionic Casimir
effect was calculated. Also in this case we will be able to identify~$u$
with the inverse lattice spacing~$a^{-1}$ in the ED. 

It is also possible to interpret the set of scalars~$\Phi_{i}$ as
one 5D massive scalar on a transverse lattice, since in the potential~$V$
there exist a bulk mass term and terms of the type~$\sum_{i=1}^{N}M_{\textrm{b}}^{2}\left|\Phi_{i}-Q_{i}\Phi_{i+1}/u\right|^{2}$
that can be interpreted as the discretised version of the continuum
KK mass term~$\int_{0}^{R}\textrm{d}y\,(\partial_{5}\Phi)^{2}$.
The resulting KK mass spectrum corresponds to the one for the gauge
bosons in Eq.~(\ref{eq:VacD-A-MassSpect}), but with an inverse lattice
spacing of the order~$M_{\textrm{b}}$ and a constant kink mass~$M_{\Phi}\sim10^{2}\,\textrm{GeV}$.

Let us close this section with a summary of the relevant scales that
follow from this model. For the gauge fields~$A_{i\mu}$ and the
fermions~$\Psi_{i}$ we find the inverse lattice spacing to be of
the order~$u\sim10^{-2}\,\textrm{eV}$ with an additional bulk gauge
mass term of the order~$v\sim10^{2}\,\textrm{GeV}$. The fermions
do not have a bulk mass term. Finally, the scalars~$\Phi_{i}$ with
a bulk mass~$M_{\Phi}\sim10^{2}\,\textrm{GeV}$ yield a lattice spacing
given by~$M_{\textrm{b}}\sim(10^{4}\dots10^{5})\,\textrm{eV}$.

\section{Vacuum Energy in Deconstruction\label{sec:VacD-VacEng}}

In this section we show that by applying the correspondence between
gauge theories in geometric and in deconstructed higher dimensions,
it is possible to transfer the methods for calculating finite Casimir
energy densities in higher dimensions to the 4D deconstruction setup.
One therefore obtains an unambiguous and well-defined prescription
to determine finite vacuum energies of 4D quantum fields which have
a higher-dimensional correspondence. We will demonstrate this procedure
explicitly with the deconstruction model given in the previous section,
which finally yields a 4D vacuum energy density that is comparable
with the observed value~$\rho_{\textrm{obs}}\sim10^{-47}\,\textrm{GeV}^{4}$. 

Let us start with the energy density~$\rho$ of the 4D quantum fields
in deconstruction with~$N$ KK modes which is schematically given
by\begin{equation}
\rho\propto\sum_{n=1}^{N}\int\textrm{d}^{3}p\cdot\sqrt{\vec{p}^{\,2}+M_{n}^{2}},\label{eq:VacD-4d-VacEng}\end{equation}
 where the masses~$M_{n}$ depend on~$N$,~$R$, and the spin of
the fields. Without any knowledge of the fifth dimension, it would
not be clear how to put these UV divergent expressions into a sensible
(finite) form. However, since we now interpret the KK tower in terms
of an underlying higher-dimensional theory with certain boundary conditions,
the 5D Casimir effect provides a well-known procedure to handle these
UV divergences in four dimensions and yields a finite result. 

We can make the correspondence even more definite by comparing the
zero-point modes in Eq.~(\ref{eq:VacD-4d-VacEng}), which follow
from the deconstruction mass spectra, with the energy momentum relations~(\ref{eq:DiED-Scalar-Spectrum})
and~(\ref{eq:DiED-Fermion-EP-Rel}) appearing in the Casmir calculations.
In the previous section we have already found this relation explicitly
by confronting the gauge boson/scalar masses from Eq.~(\ref{eq:VacD-A-MassSpect}),\[
M_{n}^{2}=g^{2}v^{2}+2g^{2}u^{2}\left(1-\cos2\pi\frac{n}{N}\right),\]
with Eq.~(\ref{eq:DiED-Scalar-Spectrum}),\[
m^{2}=2a^{-2}(1-\cos2\pi\frac{n}{N})+M_{\textrm{s}}^{2}.\]
Here, we directly observe that the inverse lattice spacing~$a^{-1}$
corresponds to~$gu$, and respectively the bulk mass~$M_{\textrm{s}}$
to~$gv$. For simplicity, we treat all bosons like real scalars since
they differ only by their number of degrees of freedom, and for our
purposes even this small factor~$2$ or~$3$ can be neglected. In
this case the overall prefactor in Eq.~(\ref{eq:VacD-4d-VacEng})
is exactly given by that of Eq.~(\ref{eq:DiED-S-rho5-unregul}) after
integrating out the ED. This shows that the formal correspondence
is exact for the bosons. 

Now, we show this correspondence also for the fermions~$\Psi_{i}$.
Let us first mention that in this case the prefactor in~Eq.~(\ref{eq:VacD-4d-VacEng})
also exactly matches that of Eq.~(\ref{eq:DiED-F-rho5-unregul})
after integrating out the ED. To determine the mass spectrum~$M_{n}$
for the Dirac spinor~$\Psi_{n}=(\Psi_{n\textrm{L}},\Psi_{n\textrm{R}})^{\textrm{T}}$
we start with Eq.~(\ref{eq:VacD-Psi-Massterms}), where the link
fields~$Q_{i}$ have acquired their VEVs~$u$,\[
\mathcal{L}_{\textrm{mass}}[\Psi_{i},Q_{i}]\rightarrow\frac{u}{2}\cdot\sum_{n=1}^{N}\overline{\Psi}_{n\textrm{L}}\left[\Psi_{(n+1)\textrm{R}}-\Psi_{(n-1)\textrm{R}}\right]+\textrm{h.c.}.\]
 This sum contains the boundary terms~$-\frac{u}{2}\overline{\Psi}_{1\textrm{L}}\Psi_{0\textrm{R}}$
and~$\frac{u}{2}\overline{\Psi}_{N\textrm{L}}\Psi_{(N+1)\textrm{R}}$,
and~$\Psi_{0\textrm{R}}$ and~$\Psi_{(N+1)\textrm{R}}$ are defined
by\[
\Psi_{mN+1}=T^{m}\cdot\Psi_{1}\,\,\,\,\textrm{and}\,\,\,\,\Psi_{0}=T^{m}\cdot\Psi_{mN},\,\,\,\, m\in\mathbb{Z},\]
where we distinguish between untwisted~($T=+1$) and twisted~($T=-1$)
fermionic fields. Note that this is the discretised version of the
continuum boundary condition\[
\Psi(y+mR)=T^{m}\cdot\Psi(y),\,\,\,\, m\in\mathbb{Z}.\]
 Now, the Lagrangian in matrix form reads\[
\mathcal{L}_{\textrm{mass}}[\Psi_{i}]=\sum_{n,k=1}^{N}\overline{\Psi}_{n\textrm{L}}M_{nk}\Psi_{k\textrm{R}}+\textrm{h.c.},\]
where the mass matrix~$M_{nk}$ and its square~$M^{2}=MM^{\dag}=M^{\dag}M$
are explicitly given by\[
M=\frac{u}{2}\left[\begin{array}{rrrcr}
0 & +1 &  &  & -T\\
-1 & 0 & +1\\
\\ & \ddots & \ddots & \ddots\\
\\ &  & -1 & 0 & +1\\
+T &  &  & -1 & 0\end{array}\right],\,\,\,\, M^{2}=\frac{u^{2}}{4}\left[\begin{array}{rrrrrrr}
2 & 0 & -1 &  &  & -T & 0\\
0 & 2 & 0 & -1 &  &  & -T\\
-1 & 0 & 2 & 0 & -1\\
 & \ddots & \ddots & \ddots & \ddots & \ddots\\
 &  & -1 & 0 & 2 & 0 & -1\\
-T &  &  & -1 & 0 & 2 & 0\\
0 & -T &  &  & -1 & 0 & 2\end{array}\right].\]
The squared masses~$m_{n}^{2}$ of the fermions are found to be the
eigenvalues of~$M^{2}$. Thus, the mass spectrum for untwisted fields
reads\begin{equation}
m_{n}^{2}=u^{2}\sin^{2}2\pi\frac{n}{N},\,\,\,\, n=1,\dots,N,\label{eq:VacD-F-spect-untwist}\end{equation}
and for the twisted fields we obtain\begin{equation}
m_{n}^{2}=u^{2}\sin^{2}2\pi\frac{n-\frac{1}{2}}{N},\,\,\,\, n=1,\dots,N,\label{eq:VacD-F-spect-twist}\end{equation}
which is consistent with the spectra found in Ref.~\cite{Hill:2002me}.
Note that only for odd~$N$, both spectra become identical, which
has already been discussed in the context of the odd-even artefact
in Sec.~\ref{sec:DiED-Massless}. With these mass spectra the Casimir
energy of the fermions is~$(-8)$ times the Casimir energy of a real
scalar, whereas in a non-lattice calculation the energies differ only
by a factor of~$(-4)$ (see Sec.~\ref{sec:DiED-Massless}). This
comes again from the well known phenomenon of fermion doubling in
lattice theory. 

Note that the fermionic spectrum given in Eq.~(\ref{eq:VacD-F-spect-untwist})
(and respectively Eq.~(\ref{eq:VacD-F-spect-twist}) for twisted
fields) is exactly that of Eq.~(\ref{eq:DiED-Fermion-EP-Rel}),\[
m^{2}=a^{-2}\sin^{2}2\pi\frac{n}{N}+M_{\textrm{f}}^{2},\]
 where the inverse lattice spacing~$a^{-1}$ is identified with the
VEV~$u$ and vanishing bulk mass~$M_{\textrm{f}}$. 

We therefore conclude that with the prescription motivated in this
section, we are able to assign the finite Casimir energy density as
calculated in Secs.~\ref{sec:DiED-Casimir-Scalar} and~\ref{sec:DiED-Casimir-Fermion}
to the vacuum energy densities of 4D quantum fields in deconstruction.
Before we discuss the phenomenological consequences of this result,
let us first treat the problem of fermion doubling. 

One way to handle the fermion doubling is to add a Wilson term\emph{~}\cite{Wilson:1974sk}
to~$\mathcal{L}_{\textrm{mass}}[\Psi_{i},Q_{i}]$. This leads to
a modified fermion mass Lagrangian,\begin{equation}
\mathcal{L}_{\textrm{mass}}=u\cdot\sum_{n=1}^{N}\left[\overline{\Psi}_{n\textrm{L}}\left(\frac{Q_{n}^{\dagger}}{u}\Psi_{(n+1)\textrm{R}}-\Psi_{n\textrm{R}}\right)-\overline{\Psi}_{n\textrm{R}}\left(\Psi_{n\textrm{L}}-\frac{Q_{n-1}}{u}\Psi_{(n-1)\textrm{L}}\right)\right],\label{eq:VacD-WilsonMassLagr}\end{equation}
which yields with~$\left\langle Q_{n}\right\rangle =u$ the mass
matrix~$M$. The squared masses~$m_{n}^{2}$ of the fermions are
now given by the eigenvalues of~$M^{2}$, which is exactly the mass
matrix for the bosons:\[
M=u\cdot\left[\begin{array}{rrrrr}
-1 & +1 &  &  & 0\\
 & -1 & +1\\
 &  & \ddots & \ddots\\
 &  &  & -1 & +1\\
+T &  &  &  & -1\end{array}\right],\,\,\,\, M^{2}=u^{2}\cdot\left[\begin{array}{rrrrr}
2 & -1 &  &  & -T\\
-1 & 2 & -1\\
 & \ddots & \ddots & \ddots\\
 &  & -1 & 2 & -1\\
-T &  &  & -1 & 2\end{array}\right].\]
For the untwisted fields we get\begin{equation}
m_{n}^{2}=2u^{-2}\left(1-\cos2\pi\frac{n}{N}\right),\,\,\,\, n=1\dots N,\label{eq:VacD-MassSpecScalarWilsFerm}\end{equation}
and for the twisted ones~$n$ is replaced by~$n-\frac{1}{2}$. These
mass spectra are identical with the mass spectra of real scalars and
yields the usual (continuum) factor~$(-4)$ in the vacuum energy
density between the two field species. Looking at the above mass spectra,
we remark that the spectrum~(\ref{eq:VacD-MassSpecScalarWilsFerm})
for scalars and Wilson-modified fermions does not contain a zero mode
in the case of twisted fields.

Let us now discuss the overall value of vacuum energy density~$\rho$
originating from the deconstruction model of Sec.~\ref{sec:VacD-DecModel}.
According to the discussion in Sec.~\ref{sec:DiED-Massless}, the
gauge bosons and scalars give a negative contribution to the CC. Without
bulk masses its magnitude would of the order~$R^{-4}$, where the
size~$R$ of the ED is given by~$R=Na$. For a small number~$N$
of lattice sites this would induce a large negative contribution~$|\rho|\gg\rho_{\textrm{obs}}$
to the CC due to the small lattice spacing~$a^{-1}=M_{\textrm{b}}\sim10^{4\dots5}\,\textrm{eV}$
for the scalars~$\Phi_{i}$. Fortunately, the scalars and also the
gauge bosons are equipped with large bulk masses, which sufficiently
suppress the Casimir energy thereby avoiding serious problems with
observations. In contrast to this, the fermionic fields with KK masses
of the order of the small VEV~$u\sim10^{-2}\,\textrm{eV}$ induce
a positive contribution~$(u/N)^{4}\sim(10^{-3}\,\textrm{eV})^{4}$
to the CC which is of the observed order of magnitude already for
a small number~$N=\mathcal{O}(1)$ of lattice sites. In Table~\ref{tab:VacD-OffsetMassSuppInModel}
we give the relevant quantities for the vacuum energy suppression
of the fields. Finally, it should be noted that we have determined
only the vacuum energy contributions of quantum fields in the deconstruction
setup which have a higher-dimensional correspondence. Other sources
of vacuum energy could still lead to a CC that is in contrast with
its observed value.%
\begin{table}
\begin{centering}\begin{tabular}{|c|c|c|c|}
\hline 
field&
inverse lattice spacing $a^{-1}$&
bulk mass $M$&
$MR$\tabularnewline
\hline
\hline 
scalars $\Phi_{i}$&
$M_{\textrm{b}}\sim10^{4}-10^{5}\,\textrm{eV}$&
$m\sim10^{2}\,\textrm{GeV}$&
$N\cdot10^{6}$\tabularnewline
\hline 
gauge bosons $A_{i}$&
$u\sim10^{-2}\,\textrm{eV}$&
$v\sim10^{2}\,\textrm{GeV}$&
$N\cdot10^{13}$\tabularnewline
\hline 
fermions $\Psi_{i}$&
$u\sim10^{-2}\,\textrm{eV}$&
$0$&
$0$\tabularnewline
\hline
\end{tabular}\par\end{centering}

\caption{\label{tab:VacD-OffsetMassSuppInModel}The 4D fields in the deconstruction
scenario of Sec.~\ref{sec:VacD-DecModel}, which can be interpreted
as KK modes of a 5D field on a transverse lattice. The bulk mass is
denoted by~$M$, and~$R=Na$ is the circumference of~$\mathcal{S}_{\textrm{lat}}^{1}$
with~$N$ sites and a lattice spacing~$a$. For large values of~$MR$
the Casimir effect will be highly suppressed according to Fig.~\ref{fig:DiED-BulkMassRatios}
and Table~\ref{tab:DiED-BulkMassRatioN3}. Only the fermions contribute
significantly to the CC.}
\end{table}
 Further discussions about dark energy in the context of discretised
space-times and theory spaces can be in Refs.~\cite{Arkani-Hamed:2001ed,DE-Discrete-1,DE-Discrete-2}.

\clearpage{\pagestyle{empty}\cleardoublepage}

\chapter{Dark Energy and Discretised Gravity\label{cha:DiGr}}

Up to now, we have ignored gravity in the ED since in the usual approach
of deconstruction gravity is assumed to have completely decoupled
from the matter fields. In a gravitational ED, however, the gravity
field has a discrete structure, too. By transferring the methods of
deconstruction to this new setup one obtains an effective theory of
a discretised gravitational ED \cite{Arkani-Hamed:2002sp,Arkani-Hamed:2003vb}.
One has found that the gravitons that arise in such a framework exhibit
a non-trivial feature in the form of a strong coupling behaviour that
might occur significantly below the usual scale, where the effective
theory breaks down. In addition, this new strong coupling scale depends
on the size of the ED thereby leading to a so-called UV/IR-connection.
In this chapter we will investigate the vacuum energy arising from
the Casimir effect of massive quantum fields, where the strong coupling
scale sets an upper limit for the bulk field masses. To avoid Casimir
energy densities above the observed value of the CC large bulk field
masses are needed for small EDs, but here they cannot be arbitrarily
large due to the strong coupling. Because of this feature and the
UV/IR-connection we are able to derive a lower bound on the size of
the ED~\cite{Bauer:2005hb}.

In the next section we will briefly introduce the effective theory
of one discrete gravitational ED and the strong coupling scale arising
in this model. Afterwards we will discuss the Casimir energy in this
setup and derive the bounds on the fifth dimension.

\section{Gravitational Extra Dimensions\label{sec:DiGr}}

Since we will perform the discretisation of a 6D gravitational model
in great detail in Chap.~\ref{cha:Disk} we keep the discussion of
the 5D case relatively short. In the following we therefore present
mainly results from the literature. For the 5D case of this chapter,
an effective theory for a single discrete ED has been recently proposed
\cite{Arkani-Hamed:2002sp,Arkani-Hamed:2003vb} by implementing gravity
in theory space similar to the concept of deconstruction. In this
model gravity remains continuous in the bulk. 

\begin{figure}
\begin{centering}\includegraphics[clip,keepaspectratio]{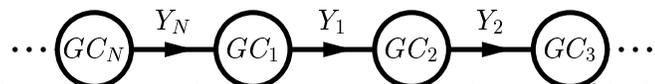}\par\end{centering}

\caption{\label{fig:DiGr-moose}Part of the gravitational theory space for
a discrete fifth dimension compactified on the circle~$\mathcal{S}^{1}$.
Each site corresponds to one general coordinate invariance~$GC_{i}$
($i=1,2,\ldots,N$), where two neighbouring sites~$i$ and~$i+1$
are connected by one link field~$Y_{i}$ and we identify~$i+N=i$.}
\end{figure}

Let us consider the theory space as given by the moose diagram shown
in Fig.~\ref{fig:DiGr-moose}.%
 Each of the sites shown as circles corresponds to one general coordinate
invariance (GC) symmetry $GC_{i}$ with~$i=1\dots N$ and can be
identified with the 4D metric $g_{\mu\nu}^{i}$ for this site. Every
site~$i$ is connected to its neighbouring site~$i+1$ by the link
field~$Y_{i\mu}$, which transforms as a vector under the two neighbouring
GCs. This is shown as an arrow connecting the GCs in the diagram.
Moreover, the theory space and thus the ED is compactified on a circle
by the identification $i+N=i$. Finally, we have on each site the
usual 4D Einstein--Hilbert action as described by\begin{eqnarray}
S_{\textrm{site}}^{g} & = & \sum_{i=1}^{N}M_{4}^{2}\int\textrm{d}^{4}x\,\sqrt{|g^{i}|}R_{4\textrm{D}}(g^{i}),\label{eq:DiGr-Site-action}\end{eqnarray}
 where $R_{4\textrm{D}}(g^{i})$ is the Ricci scalar on the site $i$,
and~$M_{4}$ denotes the universal Planck scale on the sites, which
is related to the continuous Planck scale~$M_{\textrm{Pl}}=1/\sqrt{8\pi G}$
by~$M_{4}^{2}=M_{\textrm{Pl}}^{2}/N$ with~$G$ as the 4D Newton's
constant. From Eq.~(\ref{eq:DiGr-Site-action}) we observe, that
the action~$S_{\textrm{site}}^{g}$ is invariant under the product
group~$\Pi_{i=1}^{N}GC_{i}$, which is explicitly broken to the diagonal
GC by the gravitational interactions~$S_{\textrm{link}}^{g}$ between
the sites. In our minimal discretisation with only nearest neighbour
interactions, the action~$S_{\textrm{link}}^{g}$ has the form a
Fierz-Pauli~\cite{Fierz:1939ix} mass term%
\footnote{The Fierz-Pauli form for graviton mass terms ensures the absence of
ghosts in the spectrum. For a recent discussion of ghosts in massive
gravity, see Ref.~\cite{Creminelli:2005qk}. For other aspects in
this context see, e.g., Refs.~\cite{Boulanger:2000rq,Damour:2002wu}.%
}\begin{eqnarray}
S_{\textrm{link}}^{g} & = & \sum_{i=1}^{N}M_{4}^{2}\int\textrm{d}^{4}x\,\sqrt{|g^{i}|}m^{2}(g_{\mu\nu}^{i}-g_{\mu\nu}^{i+1})(g_{\alpha\beta}^{i}-g_{\alpha\beta}^{i+1})(g^{i\mu\nu}g^{i\mu\nu}-g^{i\mu\alpha}g^{i\nu\beta}),\label{eq:DiGr-Link-action}\end{eqnarray}
 where the mass~$m$ corresponds to the inverse lattice spacing~$a^{-1}$.
The size of the fifth dimension is therefore given by~$R=N/m$ and
the 5D Planck scale by~$M_{5}=(M_{\textrm{Pl}}^{2}/R)^{1/3}$, which
defines the usual UV cutoff of the 5D theory. Let us now expand in
the weak field limit the metrics~$g_{\mu\nu}^{i}$ around flat space,~$g_{\mu\nu}^{i}=\eta_{\mu\nu}+h_{\mu\nu}^{i}$,
where~$\eta_{\mu\nu}$ is the Minkowski metric. This leads to Fierz-Pauli
graviton mass terms given by \begin{equation}
S_{ij}^{\textrm{FP}}=\int\textrm{d}^{4}x\, M^{2}m^{2}(2\delta_{i,j}-\delta_{i,j+1}-\delta_{i,j-1})(h_{\mu\nu}^{i}h^{\mu\nu,j}-h_{\mu}^{\mu,i}h_{\nu}^{\nu,j}).\label{eq:DiGr-FP-massterms}\end{equation}
Note that this is the same mass matrix as for the gauge bosons and
scalars in Sec.~\ref{sec:VacD-DecModel} implying the graviton mass
spectrum \begin{equation}
m_{n}^{2}=2m^{2}(1-\cos(2\pi\frac{n}{N}))=4m^{2}\sin^{2}(\frac{\pi n}{N}),\,\,\,\, n=1,2,\ldots,N.\label{eq:DiGr-graviton-spectrum}\end{equation}
 This spectrum describes one diagonal zero-mode graviton which corresponds
to the unbroken GC and a phonon-like spectrum of massive gravitons
that approximates for~$n\ll N$ a linear KK tower. At this level,
the phenomenology of the model appears to be very similar to that
of a deconstructed gauge theory. An important qualitative difference
to deconstruction, however, reveals itself in the peculiar strong
coupling effects of the theory.

In Ref.~\cite{Arkani-Hamed:2002sp} it was shown that the strong
coupling behaviour is most conveniently discussed by making use of
the Callan-Coleman-Wess-Zumino formalism for effective field theories~\cite{Coleman:1969sm}.
Following this lead, the product symmetry group $\Pi_{i=1}^{N}GC_{i}$,
which is broken by~$S_{\textrm{link}}^{g}$, can be formally restored
in~$S_{\textrm{link}}^{g}$ by including Goldstone bosons. Therefore,
one expands each link field around~$x^{\mu}$ as~$Y_{i}^{\mu}=x^{\mu}+\pi_{i}^{\mu}$,
where the Goldstone bosons~$\pi_{i}^{\mu}$ transform non--linearly
under~$GC_{i}$ and~$GC_{i+1}$. The massive Goldstone vector bosons,
which have three degrees of freedom, are eaten by the massless gravitons
with two degrees of freedom, to generate the five polarisations of
the massive gravitons with the spectrum given by Eq.~(\ref{eq:DiGr-graviton-spectrum}).
Including the Goldstone bosons, the link action therefore acquires
the form\begin{eqnarray}
S_{\textrm{link}}^{g} & = & \sum_{i=1}^{N}M_{4}^{2}\int\textrm{d}^{4}x\,\sqrt{|g^{i}|}m^{2}(g^{i\mu\nu}g^{i\mu\nu}-g^{i\mu\alpha}g^{i\nu\beta})\label{eq:DiGr-Link-GB-action}\\
 & \times & (g_{\mu\nu}^{i}-\partial_{\mu}Y_{i}^{\gamma}\partial_{\nu}Y_{i}^{\delta}g_{\gamma\delta}^{i+1})(g_{\alpha\beta}^{i}-\partial_{\alpha}Y_{i}^{\rho}\partial_{\beta}Y_{i}^{\sigma}g_{\rho\sigma}^{i+1}).\end{eqnarray}
Let us further decompose the Goldstone bosons~$\pi_{i}^{\mu}$ into
its transverse~$A_{i}^{\mu}$ and longitudinal~$\phi_{i}$ components
as~$\pi_{i}^{\mu}=A_{i}^{\mu}+\partial_{\mu}\phi_{i}$. By using
this form in Eq.~(\ref{eq:DiGr-Link-GB-action}) one finally obtains
the interaction terms for the~$\phi_{i}$ that are responsible for
the strong coupling, they have the form\begin{equation}
S_{\textrm{link}}^{g}=\dots(\partial^{2}\phi)(\partial^{2}\phi)(\partial^{2}\phi).\label{eq:DiGr-d6phi3}\end{equation}
Actually, the interactions of the lowest lying scalar longitudinal
component~$\phi$ of the Goldstone bosons lead to scattering amplitudes
that quickly grow with the energy~$E$ of the scalars~$\phi$ and
lead to unitarity violation once the strong coupling scale is reached.
For the present model it was found that the amplitude~$\mathcal{A}(\phi\phi\rightarrow\phi\phi)$
for~$\phi-\phi$ scattering is of the order~$\mathcal{A}\sim E^{10}/\Lambda_{4}^{10}$,
where \begin{equation}
\Lambda_{4}:=\left(\frac{M_{\textrm{Pl}}}{R^{3}}\right)^{1/4}\label{eq:DiGr-Lambda4}\end{equation}
 is the strong coupling scale of the theory, that is set by the triple
vertex of~$\phi$ as described by Eq.~(\ref{eq:DiGr-d6phi3}). From
Eq.~(\ref{eq:DiGr-Lambda4}), it is seen that the UV cutoff scale~$\Lambda_{4}$
of the effective theory depends on the IR scale~$R$ of the compactified
ED. This phenomenon has been called UV/IR connection~\cite{Arkani-Hamed:2003vb}
since in a sensible effective theory for massive gravitons the lattice
spacing~$a=m^{-1}$ must always be larger than the minimal lattice
spacing defined by~$a_{\textrm{min}}\sim\Lambda_{4}^{-1}$. This
implies that the theory does not possess a naive continuum limit.
In other words, for a given radius~$R$, the effective theory is
characterised by a highest possible number of lattice sites~$N_{\textrm{max}}=R\Lambda_{4}$,
which limits how fine grained the lattice can be made.

In addition to the triple derivative coupling of~$\phi$, the Goldstone
boson action contains other types of vertices, each of which can be
associated with a characteristic strong coupling scale for that interaction~\cite{Schwartz:2003vj}.
As two such typical examples, we will consider the scales \begin{equation}
\Lambda_{3}=\left(\frac{M_{\textrm{Pl}}}{R^{2}}\right)^{1/3}\,\,\,\,\textrm{and}\,\,\,\,\Lambda_{5}=\left(\frac{M_{\textrm{Pl}}}{R^{4}}\right)^{1/5},\label{eq:DiGr-Lambda35}\end{equation}
 which we will later compare with~$\Lambda_{4}$. It is important
to note that here the existence of the strong coupling is qualitatively
different from the UV cutoff in deconstructed gauge theories. The
strong coupling scale in deconstruction associated with the non--linear
sigma model approximation is of the order the inverse lattice spacing.
Therefore deconstruction may provide, unlike the effective theory
of massive gravitons discussed here, an UV completion of higher-dimensional
gauge theories.

\section{Bounds on the Size of the Extra Dimension\label{sec:DiGr-Bounds-R}}

Let us now investigate the Casimir energies of matter fields propagating
in the discrete fifth dimension introduced in the previous section.
For this purpose, we treat the gravitational theory space as a flat
background for quantum fields propagating in the space-time with a
discretised fifth dimension. Since these Casimir energy densities
contribute to the CC, we require that they lie below the observed
value~$\rho_{\textrm{obs}}\sim10^{-47}\,\textrm{GeV}^{4}$, associated
with the accelerated expansion of the universe. In Sec.~\ref{sec:DiED-Casimir-Scalar}
we have found that for massless bulk fields the 4D Casimir energy
density~$\rho$ scales with the size~$R$ of the ED as $|\rho|\sim R^{-4}$,
which would lead to a lower bound%
\footnote{A scenario for obtaining the observed CC from a 5D Casimir effect
of massless bulk matter fields with a sub--mm extra dimension has
been proposed, e.g., in Ref.~\cite{Milton:2001np}. Current Cavendish--type
experiments, however, put already very stringent upper bounds of the
order $R\lesssim0,1\:\textrm{mm}$ on the possible size~$R$ of extra
dimensions~\cite{Hoyle:2004cw}.%
}~$R\gtrsim(10^{-3}\,\textrm{eV})^{-1}\sim0,1\,\textrm{mm}$. A much
smaller size~$R$ becomes possible, if the bulk fields have nonzero
masses~$M$, which implies according to Sec.~\ref{sec:DiED-Massive}
an exponentially suppression for~$M\gg R^{-1}$. In the discrete
gravitational EDs, this suppression is only limited by the strong
coupling scale~$\Lambda$ of the theory, since in a sensible effective
field theory~$M$ should be smaller than the UV cutoff~$\Lambda$.
By virtue of the UV/IR connection, however, the cutoff~$\Lambda$
depends on~$R$ and can be much lower than the usual 4D Planck scale~$M_{\textrm{Pl}}\sim10^{19}\,\textrm{GeV}$.
As a consequence, we expect from the Casimir effect a smallest possible
value or lower limit on the size~$R$, when~$M$ can at most be
as large as the strong coupling scale~$\Lambda$.

In the ED, the boundary conditions for the quantum fields can be periodic
or anti--periodic corresponding to untwisted and twisted fields, respectively.
We have found in Secs.~\ref{sec:DiED-Casimir-Scalar} and~\ref{sec:DiED-Casimir-Fermion}
that the Casimir energy densities of these field configurations differ
by a small factor and have opposite sign (ignoring the odd-even artefact).
Following Eq.~(\ref{eq:DiED-S-rho5-ren}), the 4D Casimir energy
density of a single untwisted real scalar field in the discretised
fifth dimension can be written as\begin{equation}
\rho_{\textrm{untwisted}}=\frac{1}{2(2\pi)^{3}}\cdot\frac{4\pi}{8}\left[\sum_{n=1}^{N}m_{n}^{4}\ln m_{n}-N\cdot\int_{0}^{1}\textrm{d}s\cdot m_{s}^{4}\ln m_{s}\right],\label{eq:DiGr-rho-untwist-lat}\end{equation}
 where the ED has been integrated out. The mass spectrum~$m_{n}$
is given by Eq.~(\ref{eq:DiED-Scalar-Spectrum}) and reads~$m_{n}^{2}=4m^{2}\sin^{2}(\pi n/N)+M^{2}$.
Furthermore, the variable~$s$ is treated in the integral as a continuous
parameter which replaces~$n/N$ in the sine function. As long as
the number of lattice sites is~$N\gtrsim\mathcal{O}(10)$, the Casimir
energy density on the transverse lattice in Eq.~(\ref{eq:DiGr-rho-untwist-lat})
differs less than~$\lesssim1\%$ from the value in the naive continuum
limit $N\rightarrow\infty$, which is demonstrated in Fig.~\ref{fig:DiED-BulkMassRatios}.
In the remainder of this section, we will therefore employ the expressions
for the Casimir energy densities of quantum fields in the continuum
theory as discussed in Sec.~\ref{sec:DEED-Casimir}. In this approximation,
the vacuum energy density of a real (un)twisted scalar field reads
\begin{equation}
\rho_{\textrm{(un)twisted}}=\frac{\pm1}{8(2\pi)^{2}}\frac{(2\pi)^{5}}{R^{4}}\int_{x}^{\infty}\textrm{d}n\frac{(n^{2}-x^{2})^{2}}{\exp(2\pi n)\pm1},\label{eq:DiGr-rho-cont}\end{equation}
 where the plus and minus signs belong to twisted and untwisted fields,
respectively, and~$x=MR/(2\pi)$, in which~$M$ denotes the bulk
mass of the scalar field. The integral in Eq.~(\ref{eq:DiGr-rho-cont})
can be performed exactly after neglecting the term~$\pm1$ in the
denominator. Hence, both densities differ only in an overall sign:\begin{equation}
\rho_{\textrm{(un)twisted}}=\pm\frac{(MR)^{2}+3MR+3}{(2\pi)^{2}R^{4}}e^{-MR}.\label{eq:DiGr-rho-approx}\end{equation}
 When taking the sum of contributions for twisted and untwisted fields,
the integrals must be added before carrying out the approximation,
which gives \begin{equation}
\rho_{\textrm{sum}}=-\frac{4(MR)^{2}+6MR+3}{16(2\pi)^{2}R^{4}}e^{-2MR}.\label{eq:DiGr-rho-sum}\end{equation}
 The corresponding energy densities of Dirac fermions are obtained
by simply multiplying the scalar densities~$\rho_{\textrm{(un)twisted}}$
by~$-4$, where we assume that the fermion doubling has been taken
care of, e.g., in the way discussed in Sec.~\ref{sec:VacD-VacEng}.
Note that the applied approximation works fine even in the limit of
vanishing bulk masses~$M\rightarrow0$. The basic feature expressed
in Eqs.~(\ref{eq:DiGr-rho-approx}) and (\ref{eq:DiGr-rho-sum})
is that for large bulk masses~$M\gg R^{-1}$, the energy density
of massive matter fields becomes exponentially suppressed, which compensates
for the large factor~$R^{-4}$ when~$R$ is comparatively small.

Now, we are in a position to calculate the Casimir energy densities
with the bulk masses~$M$ set equal to the strong coupling scales~$\Lambda_{3}$,~$\Lambda_{4}$
and~$\Lambda_{5}$ given in Eqs.~(\ref{eq:DiGr-Lambda4}) and~(\ref{eq:DiGr-Lambda35}).
The effective field theory description suggests that these are the
largest possible values that~$M$ can take in the gravitational theory
space. If the UV cutoff~$\Lambda$ is much larger than~$\sim R^{-1}$,
the expressions in Eqs.~(\ref{eq:DiGr-rho-approx}) and (\ref{eq:DiGr-rho-sum})
are dominated by the exponential damping factors, such that the Casimir
energy densities are most strongly suppressed when~$M$ becomes of
the order the strong coupling scale~$\Lambda$, with $\Lambda=\Lambda_{3},\Lambda_{4},\Lambda_{5}$.
Moreover, from Fig.~\ref{fig:DiED-BulkMassRatios} and Sec.~\ref{sec:DiED-Massive}
we know that this suppression is most effective, when the number of
lattice sites~$N$ is maximised. Therefore we choose the inverse
lattice spacing~$m=N/R$ to be also of the order~$\Lambda$.

\begin{figure}
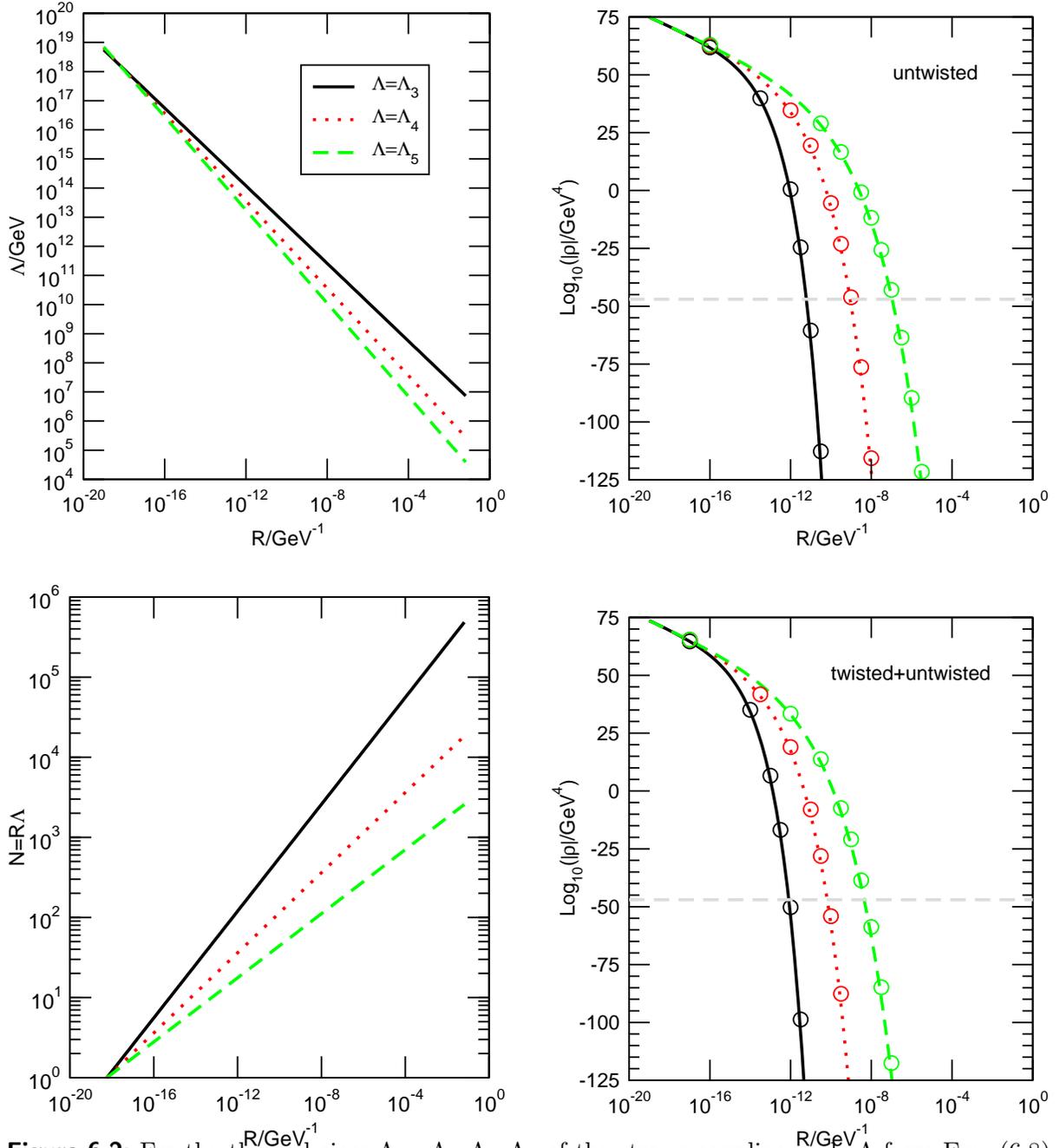

\begin{raggedleft}\includegraphics[clip,width=0.48\columnwidth,keepaspectratio]{UV-Casimir-Lambda-2}\hfill{}\includegraphics[clip,width=0.48\columnwidth,keepaspectratio]{UV-Casimir-rho-untwisted-mitGitter-2}\\
\vspace{0.5cm}\includegraphics[clip,width=0.48\columnwidth,keepaspectratio]{UV-Casimir-N-2}\hfill{}\includegraphics[clip,width=0.48\columnwidth,keepaspectratio]{UV-Casimir-rho-sum-mitGitter-2}\par\end{raggedleft}

\caption{\label{fig:DiGr-rho-Lambda-N}For the three choices~$\Lambda=\Lambda_{3},\Lambda_{4},\Lambda_{5}$
of the strong coupling scale~$\Lambda$ from Eqs.~(\ref{eq:DiGr-Lambda4})
and~(\ref{eq:DiGr-Lambda35}), we have plotted the values of~$\Lambda$,
the Casimir energy densities~$\rho$, and the corresponding number~$N=R\Lambda$
of lattice sites as functions of the size~$R$ of the fifth dimension.
The energy densities~$\rho$ are given for the untwisted scalar field,
see Eq.~(\ref{eq:DiGr-rho-approx}), and for the sum of one untwisted
and one twisted scalar field as given by Eq.~(\ref{eq:DiGr-rho-sum}).
Note, that~$\rho$ is negative in both cases, and the bulk masses
of the fields have their maximal values, given by~$\Lambda$, according
to Sec.~\ref{sec:DiGr-Bounds-R}. In the plots of~$\rho$, the horizontal
dashed line marks the observed value~$\rho_{\textrm{obs}}\sim10^{-47}\,\textrm{GeV}^{4}$
of the DE density and the circles represent exact lattice values from
Eq.~(\ref{eq:DiGr-rho-untwist-lat}).}
\end{figure}

Now the lower limit~$R_{\textrm{min}}$ on the size~$R$ of the
ED emerges from requiring that the Casimir energy densities remain
below the observed value~$\rho_{\textrm{obs}}\sim10^{-47}\:\textrm{GeV}^{4}$
of the dark energy density. The results for an untwisted scalar field
and the sum of twisted and untwisted fields are plotted in Fig.~\ref{fig:DiGr-rho-Lambda-N}.
Since the smallest value~$R_{\textrm{min}}$ that~$R$ can take
is due to the UV/IR connection a function of $\Lambda$, we have considered~$R_{\textrm{min}}(\Lambda)$
for all three scales~$\Lambda=\Lambda_{3},\Lambda_{4},\Lambda_{5}$.
These values together with the corresponding maximum number of lattice
sites~$N=R_{\textrm{min}}\cdot\Lambda(R_{\textrm{min}})$, where~$\Lambda(R_{\textrm{min}})$
is the strong coupling scale associated with~$R_{\textrm{min}}$,
are summarised in Tab.~\ref{tab:DiGr-Rmin}. Note that we can apply
here the relations from the continuum theory, since the number~$N$
of lattice sites is of the order~$10^{2}$. Furthermore, the lattice
calculation leads to energy densities (drawn as circles in Fig.~\ref{fig:DiGr-rho-Lambda-N}),
that agree very well with the values in the continuum theory. For~$R_{\textrm{min}}$
the values of the continuum and lattice formulas differ by about~$15\%$,
which is negligible, since the strong coupling scales~$\Lambda_{3,4,5}$
from Eqs.~(\ref{eq:DiGr-Lambda4}) and~(\ref{eq:DiGr-Lambda35})
are order of magnitude estimates. For instance, the lattice calculation
for an untwisted scalar field and~$\Lambda=\Lambda_{3}$ gives $R_{\textrm{min}}=6.8\cdot10^{-12}\,\textrm{GeV}^{-1}$,
whereas the continuum approximation yields~$R_{\textrm{min}}=6.1\cdot10^{-12}\,\textrm{GeV}^{-1}$. 

Let us now discuss the results. For the sum of a twisted and an untwisted
field, we observe that the Casimir energy density of massive bulk
fields exhibits a stronger suppression due to the different signs
of both components. From Fig.~\ref{fig:DiGr-rho-Lambda-N}, we read
off that the minimal radius~$R_{\textrm{min}}$ of the discrete gravitational
extra dimension lies in the range \begin{equation}
(10^{12}\,\textrm{GeV})^{-1}\lesssim R_{\textrm{min}}\lesssim(10^{7}\,\textrm{GeV})^{-1},\label{eq:DiGr-Rmin-range}\end{equation}
 where we typically find~$\Lambda(R_{\textrm{min}})\sim10^{2}\times R_{\textrm{min}}^{-1}$.
For a radius~$R$ which is much smaller than the range given in Eq.~(\ref{eq:DiGr-Rmin-range}),
the Casimir energy densities of the bulk matter fields would significantly
exceed~$\rho_{\textrm{obs}}$ and thus run into conflict with observation.
Of course, there may be other possible sources of dark energy which
might be responsible for the accelerated expansion of the universe,
but it seems unlikely that they could exactly cancel the potentially
large contributions from the Casimir effect in EDs. %

\begin{table}
\begin{centering}\begin{tabular}{|c|c|c|c|}
\hline 
untwisted&
 $R_{\textrm{min}}$&
 $\Lambda(R_{\textrm{min}})$&
 $N=R_{\textrm{min}}\cdot\Lambda(R_{\textrm{min}})$\tabularnewline
\hline
$\Lambda_{3}$&
 $6,1\cdot10^{-12}\,\textrm{GeV}^{-1}$&
 $3,6\cdot10^{13}\,\textrm{GeV}$&
 $219$\tabularnewline
\hline
$\Lambda_{4}$&
 $9,0\cdot10^{-10}\,\textrm{GeV}^{-1}$&
 $2,2\cdot10^{11}\,\textrm{GeV}$&
 $198$\tabularnewline
\hline
$\Lambda_{5}$&
 $1,1\cdot10^{-7}\,\textrm{GeV}^{-1}$&
 $1,7\cdot10^{9}\,\textrm{GeV}$&
 $179$ \tabularnewline
\hline
\end{tabular}\par\end{centering}

\begin{centering}\begin{tabular}{|c|c|c|c|}
\hline 
sum&
 $R_{\textrm{min}}$&
 $\Lambda(R_{\textrm{min}})$&
 $N=R_{\textrm{min}}\cdot\Lambda(R_{\textrm{min}})$\tabularnewline
\hline
$\Lambda_{3}$&
 $8,2\cdot10^{-13}\,\textrm{GeV}^{-1}$&
 $1,4\cdot10^{14}\,\textrm{GeV}$&
 $112$\tabularnewline
\hline
$\Lambda_{4}$&
 $6,6\cdot10^{-11}\,\textrm{GeV}^{-1}$&
 $1,6\cdot10^{12}\,\textrm{GeV}$&
 $103$\tabularnewline
\hline
$\Lambda_{5}$&
 $4,4\cdot10^{-9}\,\textrm{GeV}^{-1}$&
 $2,1\cdot10^{10}\,\textrm{GeV}$&
 $95$ \tabularnewline
\hline
\end{tabular}\par\end{centering}

\caption{\label{tab:DiGr-Rmin}The lower bounds~$R_{\textrm{min}}$ on the
size~$R$ of the ED for an untwisted real scalar field and the sum
of a twisted and an untwisted scalar. Additionally, the values of
the strong coupling scale~$\Lambda$ and the number of lattice sites~$N$
are given when~$R$ is equal to~$R_{\textrm{min}}$. For the scale~$\Lambda$,
we considered each of the three choices~$\Lambda=\Lambda_{3},\Lambda_{4},\Lambda_{5}$
from Eqs.~(\ref{eq:DiGr-Lambda4}) and~(\ref{eq:DiGr-Lambda35}).
The lower bound~$R_{\textrm{min}}$ emerges from the requirement
that the absolute Casimir energy density lies below the observed value~$\rho_{\textrm{obs}}$
of the DE density, when the bulk field mass~$M$ takes the largest
possible value~$M\sim\Lambda$.}
\end{table}

\clearpage{\pagestyle{empty}\cleardoublepage}

\chapter{Discretised Curved Disk\label{cha:Disk}}

\begin{figure}[t]
\begin{centering}\includegraphics[clip,width=0.55\columnwidth,keepaspectratio]{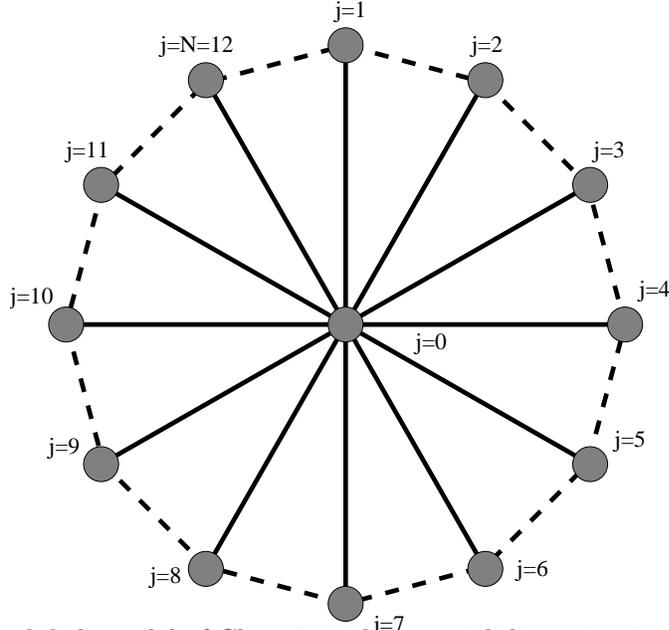}\par\end{centering}

\caption{The curved disk model of Chap.~\ref{cha:Disk} with a special discretisation,
where~$N$ lattice sites are located on the boundary and only one
site~($j=0$) sits in the centre. In this example we chose~$N=12$.
In the effective theory of massive 4D gravitons the solid lines in
radial direction correspond to the mass scale~$m_{\star}$ from Eq.~(\ref{eq:Disk-mstar-m}),
whereas~$m$ characterises the interactions in angular direction,
which are shown as dashed lines.\label{fig:Disk-Simple-Disk}}
\end{figure}

This chapter is devoted to a special 6D model, where as before the
4D subspace is a flat Minkowski space-time. Both higher dimensions,
however, form the discrete version of a disk with constant curvature.
In the case of positive curvature, for example, the disk would be
part of a 2-sphere. Moreover, we apply a very special discretisation,
where on the disk boundary we place~$N$ lattice sites and only one
site in the centre of the disk (Fig.~\ref{fig:Disk-Simple-Disk}).%
 For gauge theories and gravitational EDs similar setups have been
investigated before in Refs.~\cite{Bauer:2003mh,Hallgren:2004mw},
but here we include the possibility that the disk has a constant curvature,
too~\cite{Bauer:2006xxx}. This more general case allows a flexible
hierarchy in the mass structures of the effective 4D fields, which
can be controlled entirely by the geometry parameters of the disk.
Another motivation for the setup of this chapter was found in the
context of multi-throat geometries, where the possibility of hiding
large EDs was discussed~\cite{Kim:2005aa}.

In the following we will first discuss the 6D model in the continuum
and show that it is a solution of Einstein's equations. Afterwards
we explicitly derive the discretisation of the EDs and determine the
mass spectrum for the effective 4D gravitons and the value for the
effective 4D Planck mass. All these scales will be found to depend
directly on the parameters of the curved disk. The explicit calculations
also help to understand the 5D case of Sec.~\ref{sec:DiGr}, especially
the origin of the Fierz-Pauli graviton mass terms in Eq.~(\ref{eq:DiGr-FP-massterms})
and the link action~(\ref{eq:DiGr-Link-action}). Finally, we show
how to implement fermions into this scenario, where we will find that
the corresponding mass spectrum for the 4D fermions is directly related
to the graviton spectrum. With these results we finally show a simple
application to generate small fermion masses.

\section{Curved Disk Geometry\label{sec:Disk}}

Let us consider an effective field theory for massive gravitons, which
arise after compactification of two gravitational EDs that have been
discretised. We start out with a continuum theory, where 6D general
relativity is compactified to four dimensions on a two-dimensional
disk of constant curvature. In this space, the 6D coordinates are
denoted by~$x^{M}$ with capital Latin letters $M=0,1,2,3,5,6$,
while Greek letters appear in the usual 4D coordinates~$x^{\mu}$
as $\mu=0,1,2,3$. The position of a point on the disk is described
by the radial coordinate~$r:=x^{5}\in[0,L]$ and the polar coordinate~$\varphi:=x^{6}\in[0,2\pi]$.
The 6D Minkowski metric is given by $\eta_{MN}=\textrm{diag}(1,-1,\dots,-1)$.
In the following, we consider a metric~$g_{MN}$ that is defined
by the line element \begin{equation}
\textrm{d}s^{2}=g_{\mu\nu}(x^{M})\textrm{d}x^{\mu}\textrm{d}x^{\nu}-\frac{1}{1-er^{2}}\textrm{d}r^{2}-r^{2}\textrm{d}\varphi^{2},\label{eq:Disk-6D-metric}\end{equation}
 where~$1/\sqrt{|e|}$ is the curvature radius of the disk and~$g_{\mu\nu}(x^{M})$
the metric of the 4D subspace. Note that the parameter~$e$ controls
the curvature of the disk, for~$e>0$ the disk is spherically curved,
and~$e<0$ leads to a hyperbolic disk%
\footnote{Compact hyperbolic extra dimensions have been discussed, e.g., in
Refs.~\cite{Kaloper:2000jb,Neupane:2003cs}.%
}, respectively.~$e=0$ corresponds to a flat disk. From Eq.~(\ref{eq:Disk-6D-metric}),
we read off $g_{55}=-(1-er^{2})^{-1}$ and~$g_{66}=-r^{2}$. We denote
partial and covariant derivatives by commas and semicolons, respectively.
Using~$g_{55,5}=-2er/(1-er^{2})^{2}$ and~$g_{66,5}=-2r$ we find
for the non-zero Christoffel symbols \begin{eqnarray}
\Gamma_{\mu\nu}^{\sigma} & = & \stackrel{4D}{\Gamma_{\mu\nu}^{\sigma}},\quad\Gamma_{\nu5}^{\mu}\:=\:\frac{1}{2}g^{\mu\rho}g_{\nu\rho,5},\quad\Gamma_{\nu6}^{\mu}\:=\:\frac{1}{2}g^{\mu\rho}g_{\nu\rho,6},\quad\Gamma_{\mu\nu}^{5}\:=\:-\frac{1}{2}g^{55}g_{\mu\nu,5},\nonumber \\
\Gamma_{\mu\nu}^{6} & = & -\frac{1}{2}g^{66}g_{\mu\nu,6},\quad\Gamma_{55}^{5}\:=\:\frac{1}{2}g^{55}g_{55,5}\:=\:\frac{er}{1-er^{2}},\nonumber \\
\Gamma_{66}^{5} & = & -\frac{1}{2}g^{55}g_{66,5}\:=\:-r(1-er^{2}),\quad\Gamma_{56}^{6}\:=\:\frac{1}{2}g^{66}g_{66,5}=\frac{1}{r},\label{eq:Disk-Christoffels}\end{eqnarray}
where~$\stackrel{4D}{\Gamma_{\mu\nu}^{\sigma}}:=\frac{1}{2}g^{\sigma\rho}(g_{\mu\rho,\nu}+g_{\rho\nu,\mu}-g_{\mu\nu,\rho})$.
With our conventions the Riemann (curvature) tensor~$R_{\,\,\, MND}^{A}$,
the Ricci tensor~$R_{MN}$ and the Ricci scalar~$R$ are defined
by\[
R_{\,\,\, MND}^{A}:=\Gamma_{MD,N}^{A}-\Gamma_{MN,D}^{A}+\Gamma_{BN}^{A}\Gamma_{MD}^{B}-\Gamma_{MN}^{B}\Gamma_{BD}^{A},\]
\[
R_{MN}:=R_{\,\, MNA}^{A}=\Gamma_{MA,N}^{A}-\Gamma_{MN,A}^{A}+\Gamma_{BN}^{A}\Gamma_{MA}^{B}-\Gamma_{MN}^{B}\Gamma_{BA}^{A},\]
\[
R:=R_{MN}g^{MN}.\]

The 6D Einstein-Hilbert action%
\footnote{In~$n$ space-time dimensions the Einstein-Hilbert action with Planck
scale~$M_{n}$ reads\[
S=\int\textrm{d}^{n}x\sqrt{|g|}M_{n}^{(n-2)}(R-2\lambda)+S_{\textrm{matter}},\]
yielding Einstein's equations in the form\[
R_{MN}-\frac{1}{2}g_{MN}R+\lambda g_{MN}=-\frac{1}{2M_{n}^{(n-2)}}T_{MN}.\]
} including a CC~$\lambda$ and some matter action~$S_{\textrm{matter}}$
reads \begin{equation}
S=M_{6}^{4}\int\textrm{d}^{6}x\,\sqrt{|g|}(R-2\lambda)+S_{\textrm{matter}},\label{eq:Disk-6D-EH-matter}\end{equation}
 where~$g=\det g_{MN}$ and~$M_{6}$ denotes the 6D Planck scale.
Note that in this chapter we denote the CC by the quantity~$\lambda$,
which has mass dimension two and is related to the vacuum energy density~$\Lambda$
via the Planck mass. The requirement that the variation~$\delta S$
with respect to the metric vanishes, leads to Einstein's equations\[
R_{MN}-\frac{1}{2}g_{MN}R+\lambda g_{MN}=-\frac{1}{2M_{6}^{4}}T_{MN},\]
and the energy-momentum tensor~$T_{MN}$ follows from\[
\int\textrm{d}^{6}x\sqrt{|g|}T_{MN}=2\frac{\delta S_{\textrm{matter}}}{\delta g^{MN}}.\]
Now we are interested in the Einstein-Hilbert action corresponding
to the specific 6D metric given by Eq.~(\ref{eq:Disk-6D-metric}).
Under the 6D integral we therefore calculate~$\sqrt{|g|}R=\sqrt{|g|}(g^{\mu\nu}R_{\mu\nu}+g^{55}R_{55}+g^{66}R_{66})$
in the following. The first term reads\begin{eqnarray*}
\sqrt{|g|}g^{\mu\nu}R_{\mu\nu} & = & \sqrt{|g|}g^{\mu\nu}\big[\Gamma_{\mu\alpha,\nu}^{\alpha}-\Gamma_{\mu\nu,\alpha}^{\alpha}+\Gamma_{\beta\nu}^{\alpha}\Gamma_{\mu\alpha}^{\beta}-\Gamma_{\mu\nu}^{\beta}\Gamma_{\beta\alpha}^{\alpha}\\
 & - & \Gamma_{\mu\nu,5}^{5}+\Gamma_{\beta\nu}^{5}\Gamma_{\mu5}^{\beta}+\Gamma_{5\nu}^{\alpha}\Gamma_{\mu\alpha}^{5}-\Gamma_{\mu\nu}^{5}\Gamma_{5A}^{A}\\
 & - & \Gamma_{\mu\nu,6}^{6}+\Gamma_{\beta\nu}^{6}\Gamma_{\mu6}^{\beta}+\Gamma_{6\nu}^{\alpha}\Gamma_{\mu\alpha}^{6}-\Gamma_{\mu\nu}^{6}\Gamma_{6A}^{A}\big].\end{eqnarray*}
The derivative of the Christoffel symbol with respect to~$r$ (and
similarly with respect to~$\varphi$) can be written as \[
-\sqrt{|g|}g^{\mu\nu}\Gamma_{\mu\nu,5}^{5}=-\left[\sqrt{|g|}g^{\mu\nu}\Gamma_{\mu\nu}^{5}\right]_{,5}+\sqrt{|g|}g^{\mu\nu}\Gamma_{A5}^{A}\Gamma_{\mu\nu}^{5}-2\sqrt{|g|}g^{\mu\alpha}\Gamma_{\alpha5}^{\nu}\Gamma_{\mu\nu}^{5},\]
where on the right-hand side we have used~$(\sqrt{|g|})_{,A}=\sqrt{|g|}\Gamma_{BA}^{B}$
in the second term and~$g_{\,\,\,\,\,,5}^{\mu\nu}=-g^{\mu\alpha}g^{\nu\beta}g_{\alpha\beta,5}=-2g^{\mu\alpha}\Gamma_{\alpha5}^{\nu}$
in the third term. Thus we find\[
\sqrt{|g|}g^{\mu\nu}R_{\mu\nu}=\sqrt{|g|}R_{4\textrm{D}}-\left[\sqrt{|g|}g^{\mu\nu}\Gamma_{\mu\nu}^{5}\right]_{,5}-\left[\sqrt{|g|}g^{\mu\nu}\Gamma_{\mu\nu}^{6}\right]_{,6}\]
with~$R_{4\textrm{D}}$ denoting the usual 4D curvature scalar. Similarly
we find for the other terms\begin{eqnarray*}
\sqrt{|g|}g^{55}R_{55} & = & \sqrt{|g|}g^{55}\left[\Gamma_{5A,5}^{A}-\Gamma_{55,A}^{A}+\Gamma_{B5}^{A}\Gamma_{5A}^{B}-\Gamma_{55}^{B}\Gamma_{BA}^{A}\right]\\
 & = & \left[\sqrt{|g|}g^{55}\Gamma_{A5}^{A}\right]_{,5}-\left[\sqrt{|g|}g^{55}\Gamma_{55}^{A}\right]_{,A}\\
 & + & \sqrt{|g|}g^{55}\big[2\Gamma_{55}^{5}\Gamma_{A5}^{A}-\Gamma_{B5}^{B}\Gamma_{A5}^{A}+\Gamma_{BA}^{B}\Gamma_{55}^{A}\\
 &  & -2\Gamma_{55}^{5}\Gamma_{55}^{5}+\Gamma_{B5}^{A}\Gamma_{A5}^{B}-\Gamma_{55}^{B}\Gamma_{BA}^{A}\big].\\
 & = & \left[\sqrt{|g|}g^{55}\Gamma_{A5}^{A}\right]_{,5}-\left[\sqrt{|g|}g^{55}\Gamma_{55}^{A}\right]_{,A}\\
 & + & \sqrt{|g|}g^{55}[\Gamma_{\beta5}^{\alpha}\Gamma_{\alpha5}^{\beta}-\Gamma_{\alpha5}^{\alpha}\Gamma_{\beta5}^{\beta}-2\Gamma_{\alpha5}^{\alpha}\Gamma_{65}^{6}]\end{eqnarray*}
and\begin{eqnarray*}
\sqrt{|g|}g^{66}R_{66} & = & \sqrt{|g|}g^{66}\left[\Gamma_{6A,6}^{A}-\Gamma_{66,A}^{A}+\Gamma_{B6}^{A}\Gamma_{6A}^{B}-\Gamma_{66}^{B}\Gamma_{BA}^{A}\right]\\
 & = & \left[\sqrt{|g|}g^{66}\Gamma_{A6}^{A}\right]_{,6}-\left[\sqrt{|g|}g^{66}\Gamma_{66}^{A}\right]_{,A}\\
 & + & \sqrt{|g|}g^{66}\big[-2\Gamma_{56}^{6}\Gamma_{66}^{5}-\Gamma_{B6}^{B}\Gamma_{A6}^{A}+\Gamma_{BA}^{B}\Gamma_{66}^{A}\\
 &  & +\Gamma_{B6}^{A}\Gamma_{A6}^{B}-\Gamma_{66}^{B}\Gamma_{BA}^{A}\big].\\
 & = & \left[\sqrt{|g|}g^{66}\Gamma_{A6}^{A}\right]_{,6}-\left[\sqrt{|g|}g^{66}\Gamma_{66}^{A}\right]_{,A}\\
 & + & \sqrt{|g|}g^{66}[\Gamma_{\beta6}^{\alpha}\Gamma_{\alpha6}^{\beta}-\Gamma_{\alpha6}^{\alpha}\Gamma_{\beta6}^{\beta}].\end{eqnarray*}
Within these terms we collect all total derivatives that read altogether\[
\left[\sqrt{|g|}(-g^{\mu\nu}\Gamma_{\mu\nu}^{6}+g^{66}\Gamma_{\alpha6}^{\alpha})\right]_{,6}+\left[\sqrt{|g|}(-g^{\mu\nu}\Gamma_{\mu\nu}^{5}+g^{55}\Gamma_{\alpha5}^{\alpha}-g^{66}\Gamma_{66}^{5}+g^{55}\Gamma_{65}^{6})\right]_{,5},\]
where we now write the last two terms as\begin{equation}
+2\sqrt{|g|}g^{55}\Gamma_{\alpha5}^{\alpha}\Gamma_{65}^{6}-2\sqrt{|g|}g^{55}\Gamma_{55}^{5}\Gamma_{65}^{6}.\label{eq:Disk-extra-CC-terms}\end{equation}
Here, we have used the relations~$g^{66}\Gamma_{66}^{5}=-g^{55}\Gamma_{65}^{6}$
and~$\Gamma_{65,5}^{6}=-(\Gamma_{65}^{6})^{2}$ following from Eqs.~(\ref{eq:Disk-Christoffels}).
Therefore, the first term in Eq.~(\ref{eq:Disk-extra-CC-terms})
cancels one term in~$g^{55}R_{55}$ given above, and the last terms
yields the cosmological term~$\sqrt{|g|}(2e)$.

Finally, we are able to divide the 6D action~(\ref{eq:Disk-6D-EH-matter})
without matter into three parts,\begin{equation}
S=S_{4\textrm{D}}+S_{\textrm{surface}}+S_{\textrm{mass}},\label{eq:Disk-6D-action-parts}\end{equation}
where the (modified) Einstein-Hilbert action~$S_{4\textrm{D}}$ for
the 4D subspace, the remaining surface terms~$S_{\textrm{surface}}$
and some terms~$S_{\textrm{mass}}$, that will later become graviton
mass terms, are respectively given by\begin{eqnarray}
S_{4\textrm{D}} & := & M_{6}^{4}\int\textrm{d}^{6}x\sqrt{|g|}\,(R_{4\textrm{D}}+2e),\label{eq:Disk-S-4D}\\
S_{\textrm{surface}} & := & M_{6}^{4}\int\textrm{d}^{6}x\,\left(2\left[\sqrt{|g|}g^{55}\Gamma_{\alpha5}^{\alpha}\right]_{,5}+2\left[\sqrt{|g|}g^{66}\Gamma_{\alpha6}^{\alpha}\right]_{,6}\right)\label{eq:Disk-S-surface}\\
S_{\textrm{mass}} & := & M_{6}^{4}\int\textrm{d}^{6}x\sqrt{|g|}\, g^{55}[\Gamma_{\beta5}^{\alpha}\Gamma_{\alpha5}^{\beta}-\Gamma_{\alpha5}^{\alpha}\Gamma_{\beta5}^{\beta}].\nonumber \\
 & + & M_{6}^{4}\int\textrm{d}^{6}x\sqrt{|g|}\, g^{66}[\Gamma_{\beta6}^{\alpha}\Gamma_{\alpha6}^{\beta}-\Gamma_{\alpha6}^{\alpha}\Gamma_{\beta6}^{\beta}].\label{eq:Disk-S-mass}\end{eqnarray}
Let us first consider~$S_{\textrm{surface}}$, which reads in our
setup\begin{eqnarray*}
S_{\textrm{surface}} & = & M_{6}^{4}\int\textrm{d}^{4}x\int_{0}^{2\pi}\textrm{d}\varphi\int_{0}^{L}\textrm{d}r\,\left(\left[\sqrt{|g|}g^{55}g^{\mu\nu}g_{\mu\nu,5}\right]_{,5}+\left[\sqrt{|g|}g^{66}g^{\mu\nu}g_{\mu\nu,6}\right]_{,6}\right).\end{eqnarray*}
The $\varphi$-integral over the second term vanishes if we apply
periodic boundary conditions, which is plausible for a disk, and the
remaining term\begin{equation}
S_{\textrm{surface}}=M_{6}^{4}\int\textrm{d}^{4}x\int_{0}^{2\pi}\textrm{d}\varphi\,\left[\sqrt{|g|}g^{55}g^{\mu\nu}g_{\mu\nu,5}\right]_{r=0}^{r=L},\label{eq:Disk-S-surface-explicit}\end{equation}
can be also be removed by choosing appropriate boundary conditions
at~$r=0$ and~$r=L$. In addition, the lower limit would vanishes
because of~$\sqrt{|g|}g^{55}\propto r$ as long as~$g_{\mu\nu}$
does not diverge at~$r=0$.

Furthermore, we write the mass term action in a simplified form given
by\begin{equation}
S_{\textrm{mass}}=M_{6}^{4}\int\textrm{d}^{6}x\sqrt{|g|}\sum_{c=5,6}\Big[-\frac{1}{4}g^{cc}g_{\mu\nu,c}(g^{\mu\nu}g^{\alpha\beta}-g^{\mu\alpha}g^{\nu\beta})g_{\alpha\beta,c}\Big].\label{eq:Disk-S-mass-explicit}\end{equation}
Note that the naively discretised version of this equation would become
in the 5D setup of Sec.~\ref{sec:DiGr} similar to Eq.~(\ref{eq:DiGr-Link-action}),
which will become more obvious in Sec.~\ref{sec:Disk-Graviton-Spectrum}.

\section{Solving Einstein's Equations\label{sec:Disk-Einstein-Eqs}}

In this section we will find a simple solution of Einstein's equation
for the 6D metric given by Eq.~(\ref{eq:Disk-6D-metric}). For this
purpose let us write the gravitational part of the 6D action~(\ref{eq:Disk-6D-EH-matter})
in the form\[
S=M_{6}^{4}\int\textrm{d}^{6}x\,\sqrt{|g|}\left[R-2\lambda\cdot(g^{AB}n_{AB})\right],\]
where we have introduced the tensor\[
n_{AB}:=\textrm{diag}(0,0,0,0,\frac{1}{2}g_{55},\frac{1}{2}g_{66}),\]
given in the coordinates of the previous section. The term involving~$\lambda$
is a modified cosmological term in the following sense. As long as
we are working only with the action~$S$ as a number we can evaluate
the expression~$g^{AB}n_{AB}=n_{A}^{\,\,\, A}=1$ and thus obtain
standard 6D gravity with a 6D cosmological constant~$\lambda$. However,
if we want to derive Einstein's equations from the action functional~$S[g_{AB}]$
we have to leave~$g^{AB}n_{AB}$ unevaluated. The variational principle~$\delta S/\delta g^{AB}=0$
then gives 6D Einstein's equations\[
G_{MN}+g_{\mu\nu}\lambda=R_{MN}-\frac{1}{2}g_{MN}R+g_{\mu\nu}\lambda=0\]
with a CC~$\lambda$, that is located only within the whole 4D subspace
described by the metric\[
g_{\mu\nu}:=g_{MN}-2n_{MN}=\textrm{diag}(g_{00},g_{11},g_{22},g_{33},0,0).\]
 To calculate this we have used the relations~$\frac{\delta(\sqrt{|g|}R)}{\delta g^{MN}}=\sqrt{|g|}G_{MN}$
and respectively\begin{eqnarray*}
\frac{\delta(-2\lambda\sqrt{|g|}g^{AB}n_{A}n_{B})}{\delta g^{MN}} & = & -2\lambda(-\frac{1}{2}\sqrt{|g|}g_{MN}g^{AB}n_{AB}+\sqrt{|g|}\delta_{M}^{A}\delta_{N}^{B}n_{AB})\\
 & = & \lambda(g_{MN}-2n_{MN})=g_{\mu\nu}\lambda.\end{eqnarray*}
We now apply the simple ansatz~$g_{\mu\nu}=\eta_{\mu\nu}=\textrm{diag}(1,-1,-1,-1)$
leading to\[
G_{MN}+g_{\mu\nu}\lambda=\left[\begin{array}{ccc}
(-e+\lambda)\eta_{\mu\nu}\\
 & 0\\
 &  & 0\end{array}\right],\]
which becomes a solution of Einstein's equations~$G_{MN}+g_{\mu\nu}\lambda=0$
by setting~$\lambda=e$. In this case~$\lambda$ explicitly cancels
the cosmological term in Eq.~(\ref{eq:Disk-S-4D}). In the next section
we will investigate tensor fluctuations (gravitons) around the 4D
Minkowski metric~$\eta_{\mu\nu}$, which we have found to be a solution
of Einstein's equations.

\section{Massive 4D Gravitons\label{sec:Disk-Massive-Gravitons}}

In Sec.~\ref{sec:Disk} we have found the terms~(\ref{eq:Disk-S-mass-explicit})
that will become, as we show now, the mass terms for 4D gravitons.
Let us first write~$S_{\textrm{mass}}$ in the form\begin{eqnarray*}
S_{\textrm{mass}} & = & M_{6}^{4}\int\textrm{d}^{4}x\,\int\textrm{d}\varphi\,\textrm{d}r\,\sqrt{|g_{4}|}\,\Big[+\frac{1}{4}r\sqrt{1-er^{2}}\partial_{r}g_{\mu\nu}(g^{\mu\nu}g^{\alpha\beta}-g^{\mu\alpha}g^{\nu\beta})\partial_{r}g_{\alpha\beta}\\
 &  & +\frac{1}{4}\frac{1}{r\sqrt{1-er^{2}}}\partial_{\varphi}g_{\mu\nu}(g^{\mu\nu}g^{\alpha\beta}-g^{\mu\alpha}g^{\nu\beta})\partial_{\varphi}g_{\alpha\beta}\Big],\end{eqnarray*}
where we have used~$\sqrt{|g|}=\sqrt{|g_{4}|}\sqrt{g_{55}g_{66}}$
in the second line. Then we introduce graviton fields~$h_{\mu\nu}$
on a flat Minkowski metric~$\eta_{\mu\nu}$ for the 4D subspace by
the replacement\[
g_{\mu\nu}\rightarrow\eta_{\mu\nu}+h_{\mu\nu}.\]
Note that in our approach we ignore graviphoton and radion excitations,
which could result from the~$g_{5M}$ and~$g_{6M}$ components of
the metric. Since~$\eta_{\mu\nu}$ is constant we have~$g_{\mu\nu,A}\rightarrow h_{\mu\nu,A}$
and by expanding~$S_{\textrm{mass}}$ in second order in~$h_{\mu\nu}$
we find\begin{eqnarray*}
S_{\textrm{mass}} & \rightarrow & M_{6}^{4}\int\textrm{d}^{4}x\,\int\textrm{d}\varphi\,\textrm{d}r\,\Big[+\frac{1}{4}r\sqrt{1-er^{2}}\partial_{r}h_{\mu\nu}(\eta^{\mu\nu}\eta^{\alpha\beta}-\eta^{\mu\alpha}\eta^{\nu\beta})\partial_{r}h_{\alpha\beta}\\
 &  & +\frac{1}{4}\frac{1}{r\sqrt{1-er^{2}}}\partial_{\varphi}h_{\mu\nu}(\eta^{\mu\nu}\eta^{\alpha\beta}-\eta^{\mu\alpha}\eta^{\nu\beta})\partial_{\varphi}h_{\alpha\beta}\Big].\end{eqnarray*}

To obtain the discrete disk as described at the beginning of this
chapter we perform the discretisation of the disk in the following
way (Fig.~\ref{fig:Disk-Simple-Disk}). We put~$N$ points on the
boundary and one point in centre of the disk so that only two points
are lying in radial direction. The coordinate distance between the
centre point and every point on the boundary is given by the coordinate
radius~$L$ of the disk, which in general differs from its proper
radius. On the boundary we denote the graviton fields by~$h_{\mu\nu}^{i}$
with~$i=1\dots N$, and the position is given by~$\varphi^{i}=i\cdot\Delta\varphi$,
where $\Delta\varphi=2\pi/N$ is the angular lattice spacing. The
graviton field~$h_{\mu\nu}^{0}$ in the centre carries the index~$0$,
and the lattice spacing in radial direction is given by~$\Delta r=L$.
Explicitly, we discretise the disk by applying the following replacements
to~$S_{\textrm{mass}}$:\begin{eqnarray}
\partial_{r}h(\varphi^{i}) & \rightarrow & \frac{(h^{i}-h^{0})}{\Delta r},\nonumber \\
\partial_{\varphi}h(\varphi^{i}) & \rightarrow & \frac{(h^{i+1}-h^{i})}{\Delta\varphi},\nonumber \\
\int\textrm{d}r\, f(r) & \rightarrow & \sum_{r=L}\Delta r\cdot f(L)=\Delta r\cdot f(L),\label{eq:Disk-Def-discrete-integral}\\
\int\textrm{d}\varphi\, f(\varphi) & \rightarrow & \sum_{i=1}^{N}\Delta\varphi\cdot f(\varphi^{i}).\nonumber \end{eqnarray}
 Note that the integral~$\int\textrm{d}r$ is replaced by just one
summation region of length~$L$, where the summand is evaluated at
the position~$r=L$ to avoid problems with the derivative~$\partial_{\varphi}$
at~$r=0$. Thus we obtain for the mass terms on the discretised disk\begin{eqnarray*}
S_{\textrm{mass}} & \rightarrow & M_{6}^{4}\int\textrm{d}^{4}x\,\sum_{i=1}^{N}\Delta\varphi\Delta r\,\\
 & \times & \left[+\frac{1}{4}L\sqrt{1-eL^{2}}\cdot\frac{h_{\mu\nu}^{i}-h_{\mu\nu}^{0}}{\Delta r}(\eta^{\mu\nu}\eta^{\alpha\beta}-\eta^{\mu\alpha}\eta^{\nu\beta})\frac{h_{\alpha\beta}^{i}-h_{\alpha\beta}^{0}}{\Delta r}\right.\\
 &  & \,\,\,\left.+\frac{1}{4}\frac{1}{L\sqrt{1-eL^{2}}}\cdot\frac{h_{\mu\nu}^{i+1}-h_{\mu\nu}^{i}}{\Delta\varphi}(\eta^{\mu\nu}\eta^{\alpha\beta}-\eta^{\mu\alpha}\eta^{\nu\beta})\frac{h_{\alpha\beta}^{i+1}-h_{\alpha\beta}^{i}}{\Delta\varphi}\right],\end{eqnarray*}
which become more clear in the form \begin{eqnarray}
S_{\textrm{mass}} & \rightarrow & M_{4}^{2}\int\textrm{d}^{4}x\,\sum_{i=1}^{N}\,\nonumber \\
 & \times & \left[+m_{\star}^{2}\cdot(h_{\mu\nu}^{i}-h_{\mu\nu}^{0})(\eta^{\mu\nu}\eta^{\alpha\beta}-\eta^{\mu\alpha}\eta^{\nu\beta})(h_{\alpha\beta}^{i}-h_{\alpha\beta}^{0})\right.\nonumber \\
 &  & \,\,\left.+m^{2}\cdot(h_{\mu\nu}^{i+1}-h_{\mu\nu}^{i})(\eta^{\mu\nu}\eta^{\alpha\beta}-\eta^{\mu\alpha}\eta^{\nu\beta})(h_{\alpha\beta}^{i+1}-h_{\alpha\beta}^{i})\right].\label{eq:Disk-S-graviton-mass}\end{eqnarray}
From the last line we observe immediately the relation to Eq.~(\ref{eq:DiGr-Link-action})
that corresponds to the 5D case. Note that the actual graviton mass
scale from the radial links,~$m_{\star}$, and respectively from
the angular links,~$m$, depend on the 4D Planck mass~$M_{4}$ of
the observer's site (brane):\begin{equation}
m_{\star}^{2}:=\frac{M_{6}^{4}}{M_{4}^{2}}\cdot\frac{1}{4}\cdot\frac{2\pi}{N}\cdot\sqrt{1-eL^{2}},\,\,\,\, m^{2}:=\frac{M_{6}^{4}}{M_{4}^{2}}\cdot\frac{1}{4}\cdot\frac{N}{2\pi}\cdot\frac{1}{\sqrt{1-eL^{2}}}.\label{eq:Disk-mstar-m}\end{equation}
However, the ratio of masses is independent of the Planck scales,\begin{equation}
\frac{m_{\star}^{2}}{m^{2}}=\frac{(2\pi)^{2}}{N^{2}}(1-eL^{2}),\label{eq:Disk-Graviton-Mass-Ratio}\end{equation}
which also shows that arbitrarily large hierarchies between~$m_{\star}$
and~$m$ are possible by choosing~$eL^{2}$ and~$N$ appropriately.

Therefore the 6D action of the curved disk with~$\Lambda=e$, \[
S_{6\textrm{D}}=M_{6}^{4}\int\textrm{d}^{6}x\,\sqrt{|g|}\left[R-2\Lambda(g^{AB}n_{AB})\right],\]
 becomes with the discretisation procedure from above and the expansion~$g_{\mu\nu}\rightarrow\eta_{\mu\nu}+h_{\mu\nu}$\[
S_{6\textrm{D}}\rightarrow M_{6}^{4}\int\textrm{d}^{6}x\sqrt{|g|}R_{4\textrm{D}}+S_{\textrm{mass}},\]
where~$S_{\textrm{mass}}$ is now given by Eq.~(\ref{eq:Disk-S-graviton-mass}).

\section{4D Planck Scale on the Sites\label{sec:Disk-M4}}

\begin{figure}[t]
\begin{centering}\includegraphics[clip,width=0.7\columnwidth,keepaspectratio]{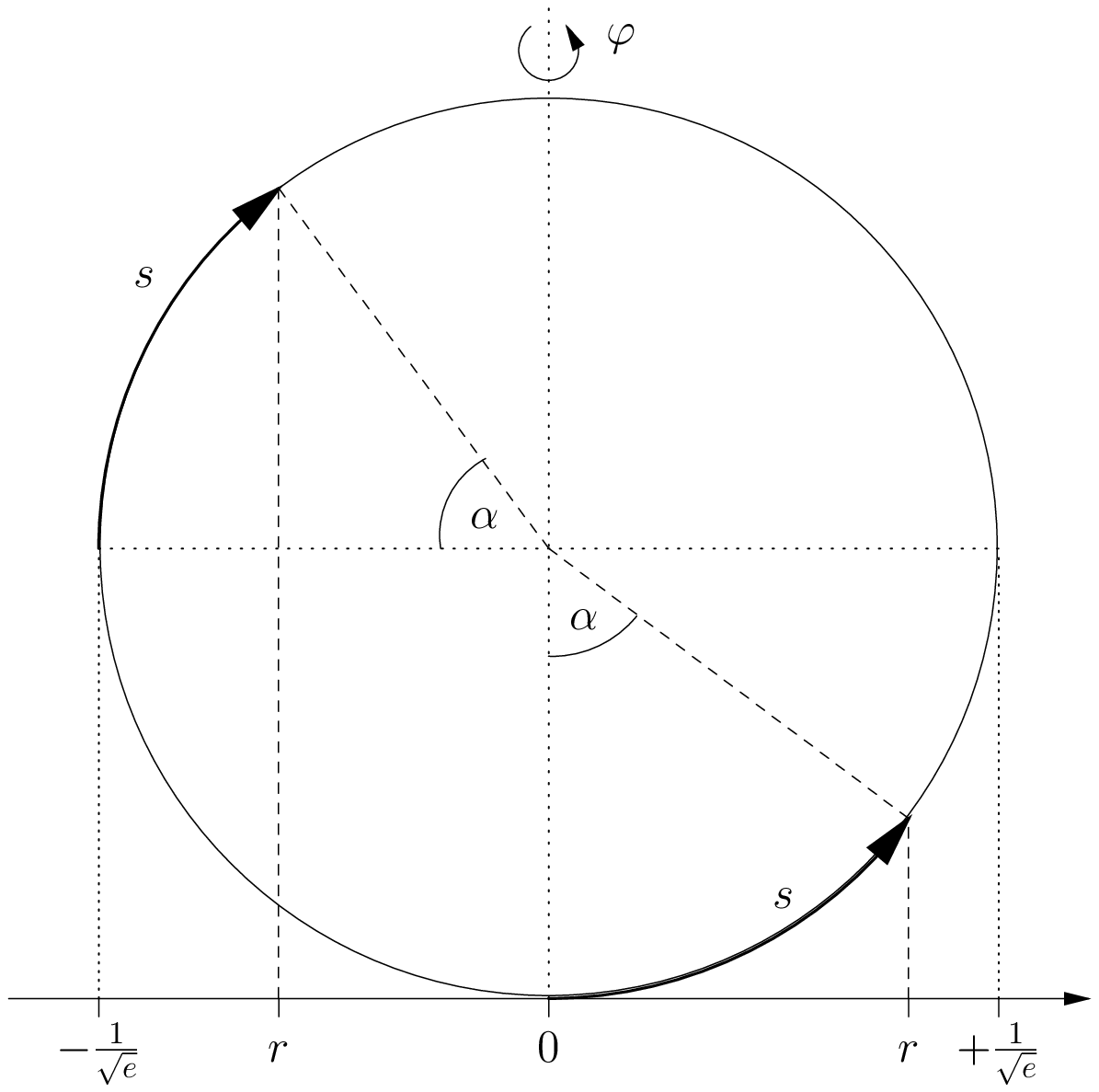}\par\end{centering}

\caption{\label{fig:Disk-sphere-arc-length}Illustration of the relation between
the coordinate~$r$ and the corresponding proper length~$s$ for
the spherically curved disk with~$e>0$. The first half of the sphere
is described by the coordinate range~$r\in[0,1/\sqrt{e}]$, which
covers the lower part of the sphere. The length~$s$ of the arc is
given by~$\alpha/\sqrt{e}$ with~$\alpha=\arcsin(\sqrt{e}r)$. The
upper part is covered by the range~$r\in[-1/\sqrt{e},0]$, where~$\alpha=\arcsin(\sqrt{e}r)+\pi/2$
and~$r\le0$.}
\end{figure}

Let us now turn to the determination of the 4D Planck mass~$M_{4}$
on the sites. For this purpose we need to know the proper area of
the curved disk that is given by\begin{equation}
A:=\int_{0}^{2\pi}\textrm{d}\varphi\int_{0}^{L}\textrm{d}r\sqrt{|g_{55}g_{66}|}=2\pi\int_{0}^{L}\textrm{d}r\frac{r}{\sqrt{1-er^{2}}}=\begin{cases}
\frac{2\pi}{e}(1-\sqrt{1-eL^{2}}), & e>0\\
\pi L^{2}, & e=0\\
\frac{2\pi}{|e|}(\sqrt{1+|e|L^{2}}-1), & e<0.\end{cases}\label{eq:Disk-proper-area}\end{equation}
For~$|e|L^{2}\ll1$ we find that the area scales like~$A=\pi L^{2}+\mathcal{O}(eL^{4})$
and for~$e<0$ the area grows approximately linearly with~$L$ for
large values of~$eL^{2}$: $A\approx2\pi L/\sqrt{|e|}$. In the case~$e>0$
the coordinate range~$r\in[0,1/\sqrt{e}]$ covers one half of a 2-sphere
with radius~$1/\sqrt{e}$, thereby~$r=1/\sqrt{e}$ corresponds to
the equator. The other half of the sphere can be described formally
by the same metric and the coordinate range~$r\in[-1/\sqrt{e},0]$,
where~$r\rightarrow0$ with~$r<0$ means approaching the antipode.
Thus the whole sphere has the usual area~$4\pi/e$. Moreover, on
the first half of the sphere the relation between the proper length~$s$
and the radial coordinate~$r$ is given by\[
s=\int_{0}^{r}\textrm{d}r^{\prime}\,\sqrt{|g_{55}|}=\int_{0}^{r}\textrm{d}r^{\prime}\,\frac{1}{\sqrt{1-er^{2}}}\stackrel{e>0}{=}\left[\frac{1}{\sqrt{e}}\textrm{arcsin}(\sqrt{e}r)\right]_{0}^{r}=\frac{1}{\sqrt{e}}\arcsin(\sqrt{e}r),\]
which is equal to the length of the arc on the right-hand side of
Fig.~\ref{fig:Disk-sphere-arc-length}.%

For completeness we remark that in the hyperbolic disk case~($e<0$)
the proper distance~$s$ from the centre in radial direction is given
by\[
s=\int_{0}^{r}\textrm{d}r^{\prime}\,\frac{1}{\sqrt{1-er^{2}}}\stackrel{e<0}{=}\left[\frac{1}{\sqrt{|e|}}\textrm{arsinh}(\sqrt{|e|}r)\right]_{0}^{r}=\frac{1}{\sqrt{|e|}}\textrm{arsinh}(\sqrt{|e|}r),\]
whereas for the flat case,~$e=0$, we simply find~$s=r$.

We now proceed to discretise the term~$M_{6}^{4}\int\textrm{d}^{6}x\sqrt{|g|}R_{4\textrm{D}}$
from the modified 4D action~(\ref{eq:Disk-S-4D}). Since the extra-dimensional
disk has a constant curvature it seems well motivated that also the
4D Planck scales~$M_{4}$ should be constant and universal on all
sites. Thus the~$R_{4\textrm{D}}$ terms of all~$N+1$ sites have
to take on the form\begin{equation}
M_{4}^{2}\sum_{i=0}^{N}\int\textrm{d}^{4}x\,\left.\sqrt{|g_{4}|}R_{4\textrm{D}}\right|_{\textrm{site }i}.\end{equation}
Since we cannot allow~$R_{4\textrm{D}}$ to depend on~$r$ and~$\varphi$,
we thus have~\[
\left.\sqrt{|g_{4}|}R_{4\textrm{D}}\right|_{\textrm{site }i}=\sqrt{|g_{4}|}R_{4\textrm{D}}\]
 for all sites~$i$. 

Now putting these parts together one finds that~$M_{6}^{4}\int\textrm{d}^{6}x\sqrt{|g|}R_{4\textrm{D}}$
has to be discretised as\begin{eqnarray}
M_{6}^{4}\underbrace{\int\textrm{d}\varphi\:\textrm{d}r\,\sqrt{|g_{55}g_{66}|}}_{\textrm{proper area }A}\int\textrm{d}^{4}x\,\sqrt{|g_{4}|}R_{4\textrm{D}} & \rightarrow & M_{4}^{2}\sum_{i=0}^{N}\int\textrm{d}^{4}x\,\left.\sqrt{|g_{4}|}R_{4\textrm{D}}\right|_{\textrm{site }i}\\
 & = & M_{4}^{2}\sum_{i=0}^{N}\int\textrm{d}^{4}x\,\sqrt{|g_{4}|}R_{4\textrm{D}}\\
 & = & M_{4}^{2}(N+1)\int\textrm{d}^{4}x\,\sqrt{|g_{4}|}R_{4\textrm{D}}.\end{eqnarray}
By comparing the left-hand side with the right-hand side the universal
4D site Planck scale is fixed,\begin{equation}
M_{4}^{2}=\frac{M_{6}^{4}A}{N+1},\label{eq:Disk-M4}\end{equation}
where the proper disk area~$A$ as given in Eq.~(\ref{eq:Disk-proper-area})
is equally divided upon all sites. However, it is important to note
that~$M_{4}$ is not the (reduced) Planck scale $M_{\textrm{Pl}}=1/(8\pi G)\sim10^{18}\:\textrm{GeV}$
that couples gravity to 4D matter. But~$M_{\textrm{Pl}}$ is determined
by integrating out the EDs in the continuum, which means~\[
M_{6}^{4}\int\textrm{d}^{6}x\:\sqrt{|g|}R_{4\textrm{D}}=M_{\textrm{Pl}}^{2}\int\textrm{d}^{4}x\:\sqrt{|g_{4}|}R_{4\textrm{D}}.\]
From this we find as in the discretised 5D case of Sec.~\ref{sec:DiGr}
the relation \begin{equation}
M_{\textrm{Pl}}^{2}=M_{6}^{4}A=(N+1)M_{4}^{2}.\label{eq:Disk-MPl}\end{equation}

In this section we have desisted from naively discretising the area
integral~$\int\textrm{d}\varphi\:\textrm{d}r\,\sqrt{|g_{55}g_{66}|}$
in~$S_{4\textrm{D}}$ since~$\sqrt{g_{55}g_{66}}\propto r$ vanishes
on the centre site and consequently there would not be a gravitational
action. On the other hand, this strategy could be useful when considering
a model, where the centre site just represents an auxiliary construction
without physical relevance. However, we will not pursue this idea
here any further. 

Instead, we mention another possibility, where one could obtain a
hierarchy between the centre and boundary sites. Let us consider for
this purpose the spherically~($e>0$) curved disk with~$L$ equal
to its maximal value on both hemispheres of the sphere. In this case,
we have a closed 2-sphere, where the centre sits on one point and
all the boundary points sit on the antipode. This means effectively
that there are only two points left. It might therefore appear {}``natural''
to assign both points (centre and boundary) one half of the total
area~$2A$, which leads to\[
S=M_{6}^{4}\int\textrm{d}^{6}x\sqrt{|g|}R_{4\textrm{D}}\rightarrow M_{6}^{4}\left[A\int\textrm{d}^{4}x\,\sqrt{|g_{4}|}R_{0}+\sum_{i=1}^{N}\frac{A}{N}\int\textrm{d}^{4}x\,\left.\sqrt{|g_{4}|}R_{4\textrm{D}}\right|_{\textrm{site }i}\right].\]
 We thus see that the squared 4D Planck mass on one boundary site
would just be one~$N^{\textrm{th}}$ of the squared 4D Planck mass
in the centre.

\section{Graviton Mass Spectrum\label{sec:Disk-Graviton-Spectrum}}

Up to now we have determined the mass term action for 4D gravitons
that follows from the discrete 6D model under consideration. Let us
now determine the actual mass spectrum. First we apply the graviton
expansion~$g_{\mu\nu}=\eta_{\mu\nu}+h_{\mu\nu}$ of the 4D metric~$g_{\mu\nu}$
also in the 4D curvature scalar~$R_{4\textrm{D}}$ that occurs in
the gravitational terms $M_{4}^{2}\sum_{i=0}^{N}\int\textrm{d}^{4}x\,\sqrt{|g_{4}|}R_{4\textrm{D}}$
for all sites. This leads to the kinetic terms for each graviton and
was found to have the form~\cite{Fierz:1939ix,'tHooft:1974bx}\begin{equation}
\sqrt{|g|}R_{4D}=\sqrt{|g_{55}\, g_{66}|}\frac{1}{4}(\partial^{\mu}h^{\nu\kappa}\partial_{\mu}h_{\nu\kappa}-\partial^{\mu}h\partial_{\mu}h-2h^{\mu}h_{\mu}+2h^{\mu}\partial_{\mu}h)+\mathcal{O}(h_{\mu\nu}^{3}),\label{eq:Disk-Kinetic-Graviton}\end{equation}
 where the following relations have been used:\begin{eqnarray*}
 & h:={h^{\mu}}_{\mu},\,\,\,\, h_{\nu}:=\partial^{\mu}h_{\mu\nu},\,\,\,\, h^{\mu\nu}:=\eta^{\mu\alpha}\eta^{\nu\beta}h_{\alpha\beta},\\
 & g^{\mu\nu}=\eta^{\mu\nu}-h^{\mu\nu}+h^{\mu\kappa}h_{\kappa}^{\nu}+\mathcal{O}(h_{\mu\nu}^{3}).\end{eqnarray*}
Up to second order in~$h_{\mu\nu}$ we find that the whole graviton
action, following from Eqs.~(\ref{eq:Disk-Kinetic-Graviton}) and
(\ref{eq:Disk-S-graviton-mass}), is given by\begin{eqnarray}
S_{\textrm{graviton}} & = & M_{4}^{2}\sum_{i=0}^{N}\,\int\textrm{d}^{4}x\,\frac{1}{4}(\partial^{\mu}h^{i\nu\kappa}\partial_{\mu}h_{\nu\kappa}^{i}-\partial^{\mu}h^{i}\partial_{\mu}h^{i}-2h^{i\mu}h_{\mu}^{i}+2h^{i\mu}\partial_{\mu}h^{i})\nonumber \\
 & + & M_{4}^{2}\sum_{i=1}^{N}\,\int\textrm{d}^{4}x\,\nonumber \\
 & \times & \left[+m_{\star}^{2}\cdot(h_{\mu\nu}^{i}-h_{\mu\nu}^{0})(\eta^{\mu\nu}\eta^{\alpha\beta}-\eta^{\mu\alpha}\eta^{\nu\beta})(h_{\alpha\beta}^{i}-h_{\alpha\beta}^{0})\right.\nonumber \\
 &  & \,\,\left.+m^{2}\cdot(h_{\mu\nu}^{i+1}-h_{\mu\nu}^{i})(\eta^{\mu\nu}\eta^{\alpha\beta}-\eta^{\mu\alpha}\eta^{\nu\beta})(h_{\alpha\beta}^{i+1}-h_{\alpha\beta}^{i})\right],\label{eq:Disk-S-graviton-full}\end{eqnarray}
where the mass scales~$m_{\star}$ and~$m$ have been determined
in Eq.~(\ref{eq:Disk-mstar-m}). 

Next we have to diagonalise the graviton mass terms by a unitary transformation.
As mentioned in Sec.~\ref{sec:Disk-M4} we choose the site Planck
scales~$M_{4}$ to be universal and fixed as in Eq.~(\ref{eq:Disk-M4}).
Then the squared $(N+1)$-dimensional mass matrix~$M^{2}$ in the
basis~$(h_{\mu\nu}^{0},h_{\mu\nu}^{1},\dots,h_{\mu\nu}^{N})$ has
the following form \begin{equation}
M^{2}=m_{\star}^{2}\left(\begin{matrix}N & -1 & -1 & \cdots & -1\\
-1 & 1\\
-1 &  & 1\\
\vdots &  &  & \ddots\\
-1 &  &  &  & 1\end{matrix}\right)+m^{2}\left(\begin{matrix}0 & 0 & 0 & \dots & 0\\
0 & 2 & -1 &  & -1\\
0 & -1 & 2 & \ddots\\
\vdots &  & \ddots & \ddots & -1\\
0 & -1 &  & -1 & 2\end{matrix}\right),\label{eq:Disk-Graviton-MassMatrix}\end{equation}
 where the first matrix proportional to~$m_{\star}$ originates from
the interactions in radial direction. The second matrix comes from
the angular interactions and is very similar to the gauge boson mass
matrix described by Eq.~(\ref{eq:VacD-A-Massterms}). Fortunately,
both matrices can be diagonalised simultaneously by a unitary transformation.
When we denote the graviton mass eigenstates by~$H_{\mu\nu}^{n}$
corresponding to the mass~$M_{n}$, we find the following relations
for the eigenvectors:\begin{eqnarray}
H_{\mu\nu}^{0} & = & \frac{1}{\sqrt{N+1}}\sum_{i=0}^{N}h_{\mu\nu}^{i},\label{eq:Disk-EV-H0}\\
H_{\mu\nu}^{p} & = & \frac{1}{\sqrt{N}}\sum_{i=1}^{N}\left[\sin(2\pi i\kl{\frac{p}{N}})+\cos(2\pi i\kl{\frac{p}{N}})\right]\cdot h_{\mu\nu}^{i},\label{eq:Disk-EV-Hp}\\
H_{\mu\nu}^{N} & = & \frac{1}{\sqrt{N(N+1)}}\left[-N\cdot h_{\mu\nu}^{0}+\sum_{i=1}^{N}h_{\mu\nu}^{i}\right],\label{eq:Disk-EV-HN}\end{eqnarray}
where~$p=1\dots N-1$. The eigenvalues~$M_{n}$ are respectively
given by \begin{eqnarray}
M_{0}^{2} & = & 0\\
M_{p}^{2} & = & m_{\star}^{2}+4m^{2}\textrm{sin}^{2}\frac{\pi p}{N},\label{eq:Disk-Graviton-Hp-Mass}\\
M_{N}^{2} & = & (N+1)m_{\star}^{2}.\label{eq:Disk-Graviton-Masses}\end{eqnarray}
From these results we observe that the zero-mode~$H_{\mu\nu}^{0}$
has a flat profile and is equally located on all sites, whereas the
mode~$H_{\mu\nu}^{N}$ with squared mass~$(N+1)m_{\star}^{2}$ is
peaked on the centre site with equal support on the boundary sites.
The~$N-1$ modes~$H_{\mu\nu}^{p}$ with~$p=1\dots N-1$ are located
only on the boundary with a typical discrete KK mass spectrum like
Eq.~(\ref{eq:DiED-Scalar-Spectrum}) that has been shifted by~$m_{\star}^{2}$.
In the limit~$m\ll m_{\star}$ the masses of the states~$H_{\mu\nu}^{p}$
from Eq.~(\ref{eq:Disk-EV-Hp}) become degenerate, and for~$N\gg1$
the mode~$H_{\mu\nu}^{N}$ becomes very heavy. 

Finally, we remark that a scenario related to ours has been discussed
recently in the context of multi-throat geometries~\cite{Kim:2005aa}.
It was shown that large EDs can be hidden in the sense that the occurrence
of massive KK modes is shifted to energies much higher than the compactification
scale of the ED, which helps evading limits on KK particles and other
bounds~\cite{Adelberger:2003zx,Cullen:1999hc,Hannestad:2001xi}.
This behaviour can be observed here for the modes~$H_{\mu\nu}^{n>0}$
in the limit~$m_{\star}\gg m$, too.

\section{Fermions on the Disk\label{sec:Disk-Fermions}}

Let us now investigate the incorporation of Dirac fermions into the
discretised disk model of Sec.~\ref{sec:Disk}. As in the graviton
case we start with a 6D Dirac fermion~$\Psi$ in the continuum. Using
the vielbein formalism~\cite{vierbein}, the corresponding action~$S$
on the curved disk reads~\cite{QFTonCurve-1}\begin{equation}
S=\int\textrm{d}^{6}x\,\sqrt{|g|}\left[\frac{1}{2}i\left(\overline{\Psi}G^{A}V_{A}^{M}\nabla_{M}\Psi-\overline{\nabla_{M}\Psi}V_{A}^{M}G^{A}\Psi\right)\right],\label{eq:Disk-6D-Fermion-Action}\end{equation}
where we denote 6D Lorentz indices by~$A,B,\dots$ and general coordinate
indices by~$M,N,\dots$, respectively. Moreover, we call~$G^{A}$
the 6D Dirac matrices, and for the barred spinor~$\overline{\Psi}$
we use the abbreviation $\overline{\Psi}=\Psi^{\dag}G^{0}$ . The
vielbein components~$V_{A}^{M}(x^{N})$ follow from the relation~$g_{MN}=V_{M}^{A}V_{N}^{B}\eta_{AB}$,
which connects the Lorentz coordinate system with the general coordinate
system. For the diagonal metric given by Eq.~(\ref{eq:Disk-6D-metric}),\begin{equation}
\textrm{d}s^{2}=g_{\mu\nu}(x^{M})\textrm{d}x^{\mu}\textrm{d}x^{\nu}-\frac{1}{1-er^{2}}\textrm{d}r^{2}-r^{2}\textrm{d}\varphi^{2},\end{equation}
we find~$V_{M}^{A}=\delta_{M}^{A}$ with the exceptions~$V_{M=5}^{A=5}=\sqrt{|g_{55}|}$
and~$V_{M=6}^{A=6}=\sqrt{|g_{66}|}$. To avoid confusion with indices
we denote the inverse vielbein components by\[
V_{5}:=V_{A=5}^{M=5}=\sqrt{|g^{55}|}\,\,\,\,\textrm{and}\,\,\,\, V_{6}:=V_{A=6}^{M=6}=\sqrt{|g^{66}|},\]
respectively. On a curved space-time the covariant derivative~$\nabla_{M}=\partial_{M}+\Gamma_{M}$
for spinors contains in addition to the usual partial derivative~$\partial_{M}$
also the spin connection\begin{equation}
\Gamma_{M}=\frac{1}{8}[G^{A},G^{B}]V_{A}^{N}V_{BN;M}.\label{eq:Disk-Spin-Connection}\end{equation}
To determine the form of the 6D $\gamma$-matrices~\cite{Pilaftsis:1999jk}
let us first look at the 4D case, where the $\gamma$-matrices are
given by \[
\gamma^{0}=\left[\begin{array}{cc}
0 & 1_{2}\\
1_{2} & 0\end{array}\right],\,\,\gamma^{k}=\left[\begin{array}{cc}
0 & \sigma^{k}\\
-\sigma^{k} & 0\end{array}\right],\,\,\gamma^{5}=i\gamma_{0}\gamma_{1}\gamma_{2}\gamma_{3}=\left[\begin{array}{cc}
-1 & 0\\
0 & +1\end{array}\right],\]
where~$k=1\dots3$ and~$\sigma^{k}$ denote the Pauli matrices.
This set of matrices has the properties\[
(\gamma^{0})^{\dag}=\gamma^{0},\,\,(\gamma^{k})^{\dag}=-\gamma^{k},\,\,(\gamma^{5})^{\dag}=\gamma^{5},\,\,\{\gamma^{5},\gamma^{\mu}\}=0\]
with the anti-commutator~$\{\square,\square\}$. In five dimensions
the number of spinor components is still four and the corresponding
$\gamma$-matrices are simply given by~$\Gamma^{0}=\gamma^{0}$,
$\Gamma^{k}=\gamma^{k}$ and~$\Gamma^{5}=i\gamma^{5}=-(\Gamma^{5})^{\dag}$.

In six dimensions, however, the Dirac algebra is 8-dimensional. Here,
we use the following set of $\gamma$-matrices\begin{eqnarray*}
G^{0} & = & \left[\begin{array}{cc}
0 & 1_{4}\\
1_{4} & 0\end{array}\right]=(G^{0})^{\dag},\\
G^{n} & = & \left[\begin{array}{cc}
0 & \Gamma^{0}\Gamma^{n}\\
-\Gamma^{0}\Gamma^{n} & 0\end{array}\right]=-(G^{n})^{\dag},\,\,\,\, n=1,2,3,5,\\
G^{6} & = & \left[\begin{array}{cc}
0 & \Gamma^{0}\\
-\Gamma^{0} & 0\end{array}\right]=-(G^{6})^{\dag},\end{eqnarray*}
which fulfil the Cifford algebra~$\{ G^{A},G^{B}\}=2\eta^{AB}\cdot1_{8}$.
Furthermore, in 6D one can also define a chirality matrix~$G_{c}$
by\[
G_{c}=G^{0}G^{1}G^{2}G^{3}G^{5}G^{6}=\left[\begin{array}{cc}
-1_{4} & 0\\
0 & 1_{4}\end{array}\right]=G_{c}^{\dag},\]
which satisfies~$G_{c}^{2}=1$ and~$\{ G_{c},G^{A}\}=0$.

From the form of the vielbeins and the $\gamma$-matrices it follows
that the terms~$V_{BN;M}$ that occur in the spin connection%
\footnote{Note that we use subscript indices for the spin connection~$\Gamma_{M}$
and superscript indices for the 5D $\gamma$-matrices~$\Gamma^{A}$.%
}~$\Gamma_{M}$ vanish except for~$V_{(B=6,N=5);6}=1$ and~$V_{(B=5,N=6);6}=-r\sqrt{1-er^{2}}$,
which leads to\begin{eqnarray*}
\Gamma_{6} & = & \frac{1}{8}[G^{5},G^{6}]V_{5}V_{B=6,N=5;6}+\frac{1}{8}[G^{6},G^{5}]V_{6}V_{B=5,N=6;6}\\
 & = & \frac{1}{4}[G^{5},G^{6}]\sqrt{1-er^{2}}=\frac{1}{2}iV_{5}\left[\begin{array}{cc}
\gamma^{5} & 0\\
0 & \gamma^{5}\end{array}\right]=-\Gamma_{6}^{\dag},\end{eqnarray*}
while all other components~$\Gamma_{M\neq6}$ vanish. Finally, we
observe that in the action~(\ref{eq:Disk-6D-Fermion-Action}) the
term involving~$\Gamma_{6}$, \[
\frac{1}{2}i\left(\overline{\Psi}G^{6}V_{6}\Gamma_{6}\Psi-\overline{\Gamma_{6}\Psi}G^{6}V_{6}\Psi\right),\]
vanishes because of~$\overline{\Gamma_{6}\Psi}G^{6}=\Psi^{\dag}\Gamma_{6}^{\dag}G_{0}G_{6}=\overline{\Psi}G^{6}\Gamma_{6}$. 

Let us now diagonalise the action by the substitution~$\Psi:=G^{6}\Phi$,
which yields~$\overline{\Psi}=-\Phi^{\dag}G^{6}G^{0}$ and respectively\begin{eqnarray*}
i\overline{\Psi}G^{A}V_{A}^{M}\nabla_{M}\Psi & = & i\Phi^{\dag}\left[\begin{array}{cc}
-\Gamma^{0} & 0\\
0 & \Gamma^{0}\end{array}\right]\left[G^{0}\partial_{0}+G^{k}\partial_{k}+V_{5}G^{5}\partial_{5}+V_{6}G^{6}\partial_{6}\right]G^{6}\Phi\end{eqnarray*}
in the action~(\ref{eq:Disk-6D-Fermion-Action}). It turns out that
all the products~$G^{A}G^{6}$ are block diagonal and if we decompose
the eight-component spinor~$\Phi=(\Phi_{a},\Phi_{b})^{\textrm{T}}$
into two four-component spinors~$\Phi_{a}$, $\Phi_{b}$ the last
line reads\begin{eqnarray*}
 & i\left[\overline{\Phi_{a}},\overline{\Phi_{b}}\right]\!\!\! & \times\,\left[\left(\begin{array}{cc}
\gamma^{0} & 0\\
0 & \gamma^{0}\end{array}\right)\partial_{0}+\left(\begin{array}{cc}
-\gamma^{k} & 0\\
0 & \gamma^{k}\end{array}\right)\partial_{k}\right.\\
 &  & \left.+\left(\begin{array}{cc}
-i\gamma^{5} & 0\\
0 & +i\gamma^{5}\end{array}\right)V_{5}\partial_{5}+\left(\begin{array}{cc}
1 & 0\\
0 & -1\end{array}\right)V_{6}\partial_{6}\right]\times\left[\begin{array}{c}
\Phi_{a}\\
\Phi_{b}\end{array}\right]\end{eqnarray*}
with~$\overline{\Phi_{a,b}}=\Phi_{a,b}^{\dag}\gamma^{0}$. From this
one can read off that~$\Phi_{a}$ corresponds to~$\Phi_{b}$ but
with negative energy, therefore we will work only with~$\Phi_{b}$
in the following. If we now denote the left- and right-handed components
of~$\Phi_{b}$ by~$\Phi_{\textrm{L,R}}:=\frac{1}{2}(1\mp\gamma^{5})\Phi_{b}$,
then the full action for~$\Phi_{b}$ can be written in the form\begin{eqnarray*}
S & = & \int\textrm{d}^{6}x\sqrt{|g|}\Big[\frac{1}{2}i\left(\overline{\Phi_{b}}\gamma^{\mu}\partial_{\mu}\Phi_{b}-\overline{\partial_{\mu}\Phi_{b}}\gamma^{\mu}\Phi_{b}\right)\\
 & - & V_{5}\frac{1}{2}\left(\overline{\Phi_{\textrm{L}}}\partial_{5}\Phi_{\textrm{R}}-\overline{\Phi_{\textrm{R}}}\partial_{5}\Phi_{\textrm{L}}-\overline{\partial_{5}\Phi_{\textrm{L}}}\Phi_{\textrm{R}}+\overline{\partial_{5}\Phi_{\textrm{R}}}\Phi_{\textrm{L}}\right)\\
 & - & iV_{6}\frac{1}{2}\left(\overline{\Phi_{\textrm{L}}}\partial_{6}\Phi_{\textrm{R}}+\overline{\Phi_{\textrm{R}}}\partial_{6}\Phi_{\textrm{L}}-\overline{\partial_{6}\Phi_{\textrm{L}}}\Phi_{\textrm{R}}-\overline{\partial_{6}\Phi_{\textrm{R}}}\Phi_{\textrm{L}}\right)\Big].\end{eqnarray*}
In the last line let us further transform both middle terms by an
integration by parts and apply periodicity conditions with respect
to the $\varphi$-coordinate:\[
\int_{0}^{2\pi}\textrm{d}\varphi\left(\overline{\Phi_{\textrm{R}}}\partial_{6}\Phi_{\textrm{L}}-\overline{\partial_{6}\Phi_{\textrm{L}}}\Phi_{\textrm{R}}\right)=\left[\overline{\Phi_{\textrm{R}}}\Phi_{\textrm{L}}-\overline{\Phi_{\textrm{L}}}\Phi_{\textrm{R}}\right]_{0}^{2\pi}-\int_{0}^{2\pi}\textrm{d}\varphi\left(\overline{\partial_{6}\Phi_{\textrm{R}}}\Phi_{\textrm{L}}-\overline{\Phi_{\textrm{L}}}\partial_{6}\Phi_{\textrm{R}}\right).\]
And for both middle terms in the second line we respectively obtain\begin{eqnarray*}
\int_{0}^{L}\textrm{d}r\sqrt{|g_{55}g_{66}|}V_{5}\left(-\overline{\Phi_{\textrm{R}}}\partial_{5}\Phi_{\textrm{L}}-\overline{\partial_{5}\Phi_{\textrm{L}}}\Phi_{\textrm{R}}\right) & = & \left[-r\overline{\Phi_{\textrm{R}}}\Phi_{\textrm{L}}-r\overline{\Phi_{\textrm{L}}}\Phi_{\textrm{R}}\right]_{0}^{L}\\
 & - & \int_{0}^{L}\textrm{d}r\left(-\overline{\Phi_{\textrm{R}}}\Phi_{\textrm{L}}-\overline{\Phi_{\textrm{L}}}\Phi_{\textrm{R}}\right)\\
 & - & \int_{0}^{L}\textrm{d}r\cdot\sqrt{|g_{66}|}\left(-\overline{\partial_{5}\Phi_{\textrm{R}}}\Phi_{\textrm{L}}-\overline{\Phi_{\textrm{L}}}\partial_{5}\Phi_{\textrm{R}}\right).\end{eqnarray*}
By using the same discretisation prescription \[
-\int_{0}^{L}\textrm{d}r\left(-\overline{\Phi_{\textrm{R}}}\Phi_{\textrm{L}}-\overline{\Phi_{\textrm{L}}}\Phi_{\textrm{R}}\right)\rightarrow L\left[\overline{\Phi_{\textrm{R}}}\Phi_{\textrm{L}}+\overline{\Phi_{\textrm{L}}}\Phi_{\textrm{R}}\right]_{r=L}\]
 as given in Eq.~(\ref{eq:Disk-Def-discrete-integral}) for the gravitons
in Sec.~\ref{sec:Disk-Massive-Gravitons}, we find that only the
terms in the last line will survive in the end.

Since 6D spinors have mass dimension~$\frac{5}{2}$ we have to rescale
them in order to obtain usual 4D spinors. As in the graviton case
we will integrate the kinetic terms over the EDs and also apply a
similar discretisation procedure as in Sec.~\ref{sec:Disk-M4}, which
means\[
\int\textrm{d}^{6}x\sqrt{|g|}\frac{1}{2}i\overline{\Phi_{b}}\gamma^{\mu}\partial_{\mu}\Phi_{b}\rightarrow\sum_{j=0}^{N}\frac{A}{N+1}\int\textrm{d}^{4}x\frac{1}{2}i\overline{\Phi_{b}^{j}}\gamma^{\mu}\partial_{\mu}\Phi_{b}^{j},\]
where~$A$ is the proper area of the EDs as given in Eq.~(\ref{eq:Disk-proper-area}).
Finally we absorb the factor~$A/(N+1)$ into the fermion fields $\chi:=\Phi_{b}\sqrt{A/(N+1)}$
and subsequently apply the discretisations\begin{eqnarray}
\partial_{5}\chi & \rightarrow & (\chi^{j}-\chi^{0})/\Delta r,\nonumber \\
\partial_{6}\chi & \rightarrow & (\chi^{j+1}-\chi^{j})/\Delta\varphi,\label{eq:Disk-Fermion-Diskretisierung}\\
\int\textrm{d}r\textrm{d}\varphi\cdot f & \rightarrow & \sum_{j=1}^{N}\Delta r\Delta\varphi\cdot f^{j},\nonumber \end{eqnarray}
where we also use~$\Delta r=L$ and~$\Delta\varphi=2\pi/N$. Note
that in radial direction we cannot use a symmetric derivative as in
Sec.~\ref{sec:DiED-Casimir-Fermion} since there are only two lattice
points available. For the derivative in radial direction we will later
on consider a more symmetric form. As a result we obtain the action
for~$N+1$ 4D fermions,\begin{eqnarray*}
S & = & \sum_{j=0}^{N}\int\textrm{d}^{4}x\frac{1}{2}i\left(\overline{\chi^{j}}\gamma^{\mu}\partial_{\mu}\chi^{j}-\overline{\partial_{\mu}\chi^{j}}\gamma^{\mu}\chi^{j}\right)\\
 & - & \sum_{j=1}^{N}\int\textrm{d}^{4}x\cdot m_{\star}\left(\overline{\chi_{\textrm{L}}^{j}}(\chi_{\textrm{R}}^{j}-\chi_{\textrm{R}}^{0})+(\overline{\chi_{\textrm{R}}^{j}}-\overline{\chi_{\textrm{R}}^{0}})\chi_{\textrm{L}}^{j}\right)\\
 & - & \sum_{j=1}^{N}\int\textrm{d}^{4}x\cdot i\cdot m\left(\overline{\chi_{\textrm{L}}^{j}}(\chi_{\textrm{R}}^{j+1}-\chi_{\textrm{R}}^{j})-(\overline{\chi_{\textrm{R}}^{j+1}}-\overline{\chi_{\textrm{R}}^{j}})\chi_{\textrm{L}}^{j}\right),\end{eqnarray*}
where the mass scales read\begin{eqnarray*}
m_{\star} & := & \Delta r\Delta\varphi\sqrt{|g_{55}g_{66}|}V_{5}\frac{N+1}{A}\frac{1}{\Delta r}=\frac{2\pi(N+1)L}{NA},\\
m & := & \Delta r\Delta\varphi\sqrt{|g_{55}g_{66}|}V_{6}\frac{N+1}{A}\frac{1}{\Delta\varphi}=\frac{(N+1)L}{A\sqrt{1-eL^{2}}}.\end{eqnarray*}
Note that the ratio~$m_{\star}/m$ is the same as in Eq.~(\ref{eq:Disk-Graviton-Mass-Ratio})
for the gravitons. However, the mass terms are different and can be
written symbolically in the form \[
M=m_{\star}\cdot\left[\begin{array}{ccccc}
 & \chi_{\textrm{R}}^{0} & \chi_{\textrm{R}}^{1} & \chi_{\textrm{R}}^{2} & \cdots\\
\overline{\chi_{\textrm{L}}^{0}} & 0 & 0 & 0\\
\overline{\chi_{\textrm{L}}^{1}} & -1 & 1\\
\overline{\chi_{\textrm{L}}^{2}} & -1 &  & 1\\
\vdots & \vdots &  &  & \ddots\end{array}\right]+im\cdot\left[\begin{array}{ccccc}
 & \chi_{\textrm{R}}^{0} & \chi_{\textrm{R}}^{1} & \chi_{\textrm{R}}^{2} & \cdots\\
\overline{\chi_{\textrm{L}}^{0}} & 0 & 0 & 0\\
\overline{\chi_{\textrm{L}}^{1}} & 0 & -1 & 1\\
\overline{\chi_{\textrm{L}}^{2}} & 0 &  & -1 & 1\\
\vdots & \vdots &  &  & \ddots\end{array}\right].\]
In order to find the mass eigenvalues we apply a bi-unitary transformation
meaning that left- and right-handed fields transform differently.
Thus the states~$\chi$ and the mass eigenstates~$\psi$ are related
by\begin{eqnarray}
\overline{\chi_{\textrm{L}}^{0}} & = & \overline{\psi_{\textrm{L}}^{0}}\nonumber \\
\overline{\chi_{\textrm{L}}^{j}} & = & \frac{1}{\sqrt{N}}\sum_{n=1}^{N}\exp(+2\pi i\cdot j\frac{n}{N})\overline{\psi_{\textrm{L}}^{n}}\nonumber \\
\chi_{\textrm{R}}^{0} & = & \frac{1}{\sqrt{N+1}}\psi_{\textrm{R}}^{0}-\frac{N}{\sqrt{N(N+1)}}\psi_{\textrm{R}}^{N}\label{eq:Disk-Fermion-Trafo}\\
\chi_{\textrm{R}}^{j} & = & \frac{1}{\sqrt{N}}\sum_{n=1}^{N-1}\exp(-2\pi i\cdot j\frac{n}{N})\psi_{\textrm{R}}^{n}+\frac{1}{\sqrt{N+1}}\psi_{\textrm{R}}^{0}+\frac{1}{\sqrt{N(N+1)}}\psi_{\textrm{R}}^{N}.\nonumber \end{eqnarray}
Expressed by the mass eigenstates~$\psi$ the action~$S$ for~$N+1$
4D Dirac fermions reads \begin{eqnarray*}
S & = & \sum_{j=0}^{N}\int\textrm{d}^{4}x\,\Big[\frac{1}{2}i\left(\overline{\psi^{j}}\gamma^{\alpha}\partial_{\alpha}\psi^{j}-\overline{\partial_{\alpha}\psi^{j}}\gamma^{\alpha}\psi^{j}\right)\\
 & - & \sum_{j=1}^{N-1}\int\textrm{d}^{4}x\cdot\overline{\psi_{\textrm{L}}^{j}}\psi_{\textrm{R}}^{j}[m_{\star}+im(e^{-2\pi i\frac{n}{N}}-1)]+\overline{\psi_{\textrm{R}}^{j}}\psi_{\textrm{L}}^{j}[m_{\star}-im(e^{2\pi i\frac{n}{N}}-1)]\\
 & - & m_{\star}\sqrt{N+1}\left(\overline{\psi_{\textrm{L}}^{N}}\psi_{\textrm{R}}^{N}+\overline{\psi_{\textrm{R}}^{N}}\psi_{\textrm{L}}^{N}\right)\Big].\end{eqnarray*}
From this result we read off that there is one massless fermion~$\psi^{0}$,
one heavy fermion~$\psi^{N}$ with mass~$m_{\star}\sqrt{N+1}$ and~$N-1$
fermions~$\psi^{1},\dots,\psi^{N-1}$ with complex masses, whose
squared absolute values read\[
|m_{\star}+im(e^{-2\pi i\frac{n}{N}}-1)|^{2}=m_{\star}^{2}+4m^{2}\sin^{2}(\frac{\pi n}{N})+2m_{\star}m\sin(\frac{2\pi n}{N}).\]
In contrast to the graviton case from Eq.~(\ref{eq:Disk-Graviton-Hp-Mass}),
here we find an additional interference term~$2m_{\star}m\sin(\frac{2\pi n}{N})$
in the mass spectrum. Since this looks a bit strange we propose a
slightly modified discretisation procedure for the angular direction.
Instead of~$\partial_{6}\chi\rightarrow(\chi^{j+1}-\chi^{j})/\Delta\varphi$
from Eq.~(\ref{eq:Disk-Fermion-Diskretisierung}) we choose the following
prescription on a formal level\[
\partial_{6}\rightarrow i\cdot\frac{\chi^{j+\frac{1}{2}}-\chi^{j-\frac{1}{2}}}{\Delta\varphi}.\]
This does not change anything with the zero mode or the heavy mode,
but the transformations in Eqs.~(\ref{eq:Disk-Fermion-Trafo}) lead
to modified complex masses for the modes~$\psi^{1}\dots\psi^{N-1}$.
Their squared absolute values read in this case\[
|m_{\star}+im(2\sin\frac{\pi n}{N})|^{2}=m_{\star}^{2}+4m^{2}\sin^{2}(\frac{\pi n}{N}).\]
Here we see that the interference terms from above do not exist and
the fermion mass spectrum has exactly the same structure as the gravitons.

\section{Small Fermion Masses}

Here, we finish this chapter with a simple application, where the
results for the fermions on the discretised curved disk can be applied
directly to generate small fermion masses. For this purpose let us
assume that our known 4D world is located on the centre site of the
disk. At this place the particle standard model may couple to the
4D component~$\chi_{\textrm{R}}^{0}$ of the 6D Dirac field via a
Yukawa coupling schematically given by~$\overline{L}\langle H\rangle\chi_{\textrm{R}}^{0}$,
where~$\overline{L}$ is a left-handed lepton doublet and~$\langle H\rangle$
is the VEV of the Higgs doublet, respectively. In addition we consider
a large number~$N$ of lattice sites so that the heaviest mode~$\Phi^{N}$
decouples due to its large mass~$m_{\star}\sqrt{N+1}$. From Eq.~(\ref{eq:Disk-Fermion-Trafo})
it follows that the right-handed fermion~$\chi_{\textrm{R}}^{0}$
on the centre site essentially consists only of the zero-mode~$\psi_{\textrm{R}}^{0}$
with a weight factor~$1/\sqrt{N+1}$. Thus the Yukawa coupling of
the left-handed SM doublet~$L$ with the right-handed fermion~$\chi_{\textrm{R}}^{0}$
leads to a suppression of the SM neutrino mass,\[
\overline{L}\langle H\rangle\chi_{\textrm{R}}^{0}\rightarrow\frac{1}{\sqrt{N+1}}\nu_{\textrm{L}}\langle H\rangle\psi_{\textrm{R}}^{0}.\]
For example, the number~$N\sim10^{24}$ might explain the lightness
of the SM neutrinos~\cite{Bauer:2006xxx}. For related applications
in this context see also Ref.~\cite{Hallgren:2004mw}. Finally, we
mention that this mass suppression mechanism can be considered as
a discrete version of the wave function suppression mechanism in continuous
higher dimensions from Ref.~\cite{Arkani-Hamed:1998vp}. 

\clearpage{\pagestyle{empty}\cleardoublepage}

\chapter{Summary and Conclusions\label{cha:Conclusions}}

In this thesis we have investigated the dark energy problem from two
different perspectives that both involve vacuum energy, or equivalently
the CC, as the major dark energy candidate. In the first part we have
started with scaling laws for the CC originating from RGEs in quantum
field theory and quantum gravity. However, since these theories do
not fix the physical interpretation of the corresponding renormalisation
scale, some typical scales occurring in the context of cosmology have
been chosen to fit into this role. In this sense we have considered
the Hubble rate in addition to the scales characterised by the cosmological
event and particle horizons. In the end, this leads to a time-dependent
CC implying a non-trivial scaling of the matter energy density with
the cosmic scale factor. Solving Einstein's equations therefore becomes
more complicated, but the solutions we found exhibit some very interesting
features.

In the late-time epoch of the cosmological evolution we have found,
apart from the well known de~Sitter final states, also power-law
and super-exponential solutions for the scale factor. Additionally,
we have observed in some cases future singularities of the big rip
and big crunch type, which usually appear only in more exotic dark
energy models. In detail we have discussed in this analysis all combinations
of scaling laws and renormalisation scale identifications and solved
Einstein's equations analytically and numerically. As a result we
have determined the fate of the universe for each case and studied
the influence of the parameters on the solutions. By doing this it
happened that some final states occur only for certain scaling laws
or scale interpretations. Finally, the running of Newton's constant
was taken into account in one case, which we treated numerically in
more detail. In conclusion, the results found in this thesis feature
several very different solutions for the cosmological late-time evolution.
Based on this outcome the analysis may help to discriminate between
different combinations of scaling laws and scales.

The second part of the work consists of an investigation of vacuum
energy in higher dimensions, where we make use of the fact that the
zero-point energy of quantum fields depends on boundary conditions.
Starting from the Casimir effect in a continuous higher-dimensional
space-time, where the Casimir energy density was determined and compared
with results from the literature, we have adjusted the calculations
to the case of discrete EDs. Within a detailed analysis we have discussed
the influence of boundary conditions and the number of lattice sites
on the value of the Casimir energy density, which contributes to the
effective 4D CC. Special emphasis was placed on the effect of bulk
field masses, which open up the possibility to considerably suppress
the corresponding vacuum energy contribution. In addition, we have
discussed the differences between the bosonic and fermionic case and
observed some typical lattice artefacts for fermions.

A nice motivation for discrete EDs comes from the model building sector
in the form of deconstruction. Representing its most important property,
this framework describes discretised EDs on the basis of 4D quantum
fields thereby avoiding problems that appear in continuous higher
dimensions. As quantum fields in this setup are not subject to boundary
conditions their zero-point energy is a priori completely unconstrained.
At this stage we have proposed a prescription to solve this problem
by employing the correspondence between the discrete 5D setup and
deconstruction. Therefore the vacuum energy of all 4D fields in deconstruction
has been identified with the finite Casimir energy of the corresponding
5D quantum field. Within a specific model we have demonstrated this
idea on a formal level for bosons and fermions. By using the suppression
mechanism of large field masses in this setup, we have also shown
that the vacuum energy contributions lie below or around the observed
value of the CC.

Even if the strong condition of tiny CC contributions might be satisfied
for arbitrary small EDs by implementing sufficiently large bulk masses,
we have found that this might not be true for discrete gravitational
EDs. In these models there usually exists a strong coupling scale
that depends on the size of the ED thereby implying an upper limit
for the mass scales in the theory. Due to this UV/IR connection one
has lost the ability to arbitrarily suppress the Casimir energy by
considering large bulk masses. In this context, a lower limit on the
ED size emerges. By applying our results for the Casimir effect we
have found the minimal size~$R_{\textrm{min}}$ to be roughly in
the range~$(10^{12}\,\textrm{GeV})^{-1}\dots(10^{7}\,\textrm{GeV})^{-1}$.
As a consequence, one can exclude, e.g., Planck scale sized EDs in
this framework.

Motivated by the appearance of massive gravitons in discrete gravitational
dimensions our last subject of investigation consists of a 6D model,
where both EDs form a discretised two-dimensional curved disk. After
discussing the disk in the continuum, we have applied a special discretisation
procedure that features 4D gravitons with an interesting mass spectrum.
The explicitly performed calculation of the masses has led, for instance,
to the observation of a gap between the zero mode and the finite KK
tower. As the main result of this subject, both mass scales appearing
in this spectrum were found to be completely adjustable by the parameters
of the curved disk. In contrast to a flat disk, here it is the curvature
of the disk that makes this flexibility possible. In the future, it
will be interesting to explore some applications and the strong coupling
scale in this setup or more generalised scenarios~\cite{Bauer:2006xxx}.
As a foretaste, we have finally shown in this work the explicit implementation
of fermions, which could be directly applied to generate small fermion
masses.

As our final conclusion let us mention that, despite many possible
candidates for the source of the accelerated expansion of the universe,
the CC should be the first one to look at. We have seen in this thesis
that quantum effects lead not only to a dependence on boundary conditions
as given by discrete EDs, but they may also induce a time-dependence.
By investigating both subjects we have found new insights for cosmology
and the structure of our space-time.

\chapter*{Acknowledgements}

At first, I would like to thank Manfred Lindner for great support,
collaboration and the freedom in doing my research. Furthermore, I
wish to thank my collaborators Marc-Thomas Eisele, Mathias Garny,
Tomas Hällgren and Gerhart Seidl for excellent teamwork and fruitful
discussions. I also thank Markus Michael Müller for accepting the
sunshine in our office and for the nice coffee discussions. Moreover,
I want to express my appreciation to all members and guests of our
group for the excellent atmosphere during the last years. Also, I
acknowledge financial support in terms of a doctorate grant by the
Freistaat Bayern. Finally, my biggest thanks go to my parents for
their great support.

\clearpage{\pagestyle{empty}\cleardoublepage}

\end{document}